\documentclass[reprint,
superscriptaddress,
showpacs,
nofootinbib,
prd,
aps,
floatfix,
]{revtex4-1}

\usepackage{amsmath,amssymb,amsfonts,amsthm}
\usepackage[english]{babel}
\usepackage{url}
\usepackage{bm}
\usepackage[latin1]{inputenc}
\usepackage{graphicx,rotating}
\usepackage[colorlinks=true,linkcolor=black, citecolor=blue]{hyperref}
\usepackage{csvsimple}
\usepackage{booktabs}
\usepackage{multirow, array}
\usepackage{slashed}
\usepackage{xcolor}
\usepackage[export]{adjustbox}

\usepackage{natbib}

\usepackage[normalem]{ulem}

\begin{document}
\title{Precision pulse shape simulation for proton detection at the Nab experiment}
\author{Leendert Hayen}
\email[Corresponding author: ]{lmhayen@ncsu.edu}
\author{Jin Ha Choi}
\author{Dustin Combs}
\author{R.J. Taylor}
\affiliation{Department of Physics, North Carolina State University, Raleigh, 27695 North Carolina, USA}
\affiliation{Triangle Universities Nuclear Laboratory, Durham, 27710 North Carolina, USA}

\author{Stefan Bae{\ss}ler}
\affiliation{Oak Ridge National Laboratory, Oak Ridge, TN 37831, USA}
\affiliation{University of Virginia, Charlottesville, VA 22904, USA}
\author{Noah Birge}
\affiliation{University of Tennessee, Knoxville, TN 37996, USA}
\author{Leah J. Broussard}
\altaffiliation[This manuscript has been authored in part by UT-Battelle, LLC, under contract DE-AC05-00OR22725 with the US Department of Energy (DOE). The publisher acknowledges the US government license to provide public access under the DOE Public Access Plan (http://energy.gov/downloads/doe-public-access-plan)]{Oak Ridge National Laboratory, Oak Ridge, TN 37831, USA}
\author{Christopher B. Crawford}
\affiliation{University of Kentucky, Lexington, KY 40506, USA}
\author{Nadia Fomin}
\affiliation{University of Tennessee, Knoxville, TN 37996, USA}
\author{Michael Gericke}
\affiliation{University of Manitoba, Winnipeg, MB R3T 2N2, Canada}
\author{Francisco Gonzalez}
\affiliation{Oak Ridge National Laboratory, Oak Ridge, TN 37831, USA}
\author{Aaron Jezghani}
\affiliation{University of Kentucky, Lexington, KY 40506, USA}
\author{Nick Macsai}
\affiliation{University of Manitoba, Winnipeg, MB R3T 2N2, Canada}
\author{Mark Makela}
\affiliation{Los Alamos National Laboratory, Los Alamos, NM 87545, USA}
\author{David G. Mathews}
\affiliation{University of Kentucky, Lexington, KY 40506, USA}
\author{Russell Mammei}
\affiliation{University of Winnipeg, Winnipeg, MB R3B 2E9, Canada}
\author{Mark McCrea}
\affiliation{University of Winnipeg, Winnipeg, MB R3B 2E9, Canada}
\author{August Mendelsohn}
\affiliation{University of Manitoba, Winnipeg, MB R3T 2N2, Canada}
\author{Austin Nelsen}
\affiliation{University of Kentucky, Lexington, KY 40506, USA}

\author{Grant Riley}
\affiliation{Los Alamos National Laboratory, Los Alamos, NM 87545, USA}
\author{Tom Shelton}
\affiliation{University of Kentucky, Lexington, KY 40506, USA}
\author{Sky Sjue}
\affiliation{Los Alamos National Laboratory, Los Alamos, NM 87545, USA}
\author{Erick Smith}
\affiliation{Los Alamos National Laboratory, Los Alamos, NM 87545, USA}

\author{Albert R. Young}
\affiliation{Department of Physics, North Carolina State University, Raleigh, 27695 North Carolina, USA}
\affiliation{Triangle Universities Nuclear Laboratory, Durham, 27710 North Carolina, USA}
\author{Bryan Zeck}
\affiliation{Department of Physics, North Carolina State University, Raleigh, 27695 North Carolina, USA}
\affiliation{Triangle Universities Nuclear Laboratory, Durham, 27710 North Carolina, USA}
\affiliation{Los Alamos National Laboratory, Los Alamos, NM 87545, USA}

\date{\today}
\begin{abstract}
The Nab experiment at Oak Ridge National Laboratory, USA, aims to measure the beta-antineutrino angular correlation following neutron $\beta$ decay to an anticipated precision of approximately 0.1\%. The proton momentum is reconstructed through proton time-of-flight measurements, and potential systematic biases in the timing reconstruction due to detector effects must be controlled at the nanosecond level. We present a thorough and detailed semiconductor and quasiparticle transport simulation effort to provide precise pulse shapes, and report on relevant systematic effects and potential measurement schemes.
\end{abstract}

\maketitle


\section{Introduction}

Precise measurements of weak interaction effects in (nuclear) $\beta$ decay have been at the forefront of the Standard Model's (SM) development and continue to provide stringent tests of Beyond SM physics \cite{Donoghue1992, Holstein2014, Cirigliano2013a, Cirigliano2013b, Ramsey-Musolf2000, Erler2005, Vos2015, Severijns2006, Gonzalez-Alonso2018, Ramsey-Musolf2008, Falkowski2017}. The neutron, in particular, is an attractive system due to the absence of nuclear structure corrections, its role in cosmology and accessibility by lattice QCD techniques \cite{Dubbers2011, Abele2008, Nico2009a, Adelberger2011, Chang2018, Walker-Loud2020, Seng2020, Feng2020}. Within the Standard Model, the decay process can be completely determined through measurements of the neutron lifetime \cite{Snow2000, Pattie2018, Gonzalez2021, Yue2013, Arzumanov2000, Arzumanov2015, Serebrov2005}, and the admixture of vector and axial vector strengths in its decay through, e.g., angular correlation measurements \cite{Mund2013, Markisch2019, Mendenhall2013, Brown2018, Beck2019}. Stringent tests of the Cabibbo-Kobayashi-Maskawa (CKM) quark mixing matrix using these results are particularly powerful \cite{Czarnecki2018, Czarnecki2004} due to the absence of nuclear structure corrections that dominate the uncertainty on the global average on $V_{ud}$, the up-down matrix element \cite{Hardy2020}. The importance of the latter is amplified by the current tension in the top-row CKM unitarity requirement \cite{Falkowski2020, Cirigliano2022} (the so-called 'Cabibbo Angle Anomaly' \cite{Coutinho2020, Grossman2020, Crivellin2021, Crivellin2020}), and recent progress on electroweak radiative corrections \cite{Seng2018, Seng2019, Czarnecki2019, Hayen2021, Shiells2021, Seng2021c}. Additionally, spectral and angular correlation measurements have complementary sensitivity to exotic scalar and tensor currents in the weak interaction \cite{Naviliat-Cuncic2013, Wauters2010, Wauters2014, Bhattacharya2012,Pattie2013,Pattie2015}, including right-handed neutrino couplings \cite{Falkowski2020, Gonzalez-Alonso2018}.

The Nab experiment\cite{Pocanic2009} aims to measure the angular correlation between the emitted electron and the antineutrino following neutron $\beta$ decay at the per-mille level, resulting in a determination of the axial-to-vector coupling constant at the 0.04\% level. When combined with a measurement of the neutron lifetime at the 0.25~s (0.03\%) level\cite{Gonzalez2021}, this will enable a determination of $V_{ud}$ at precision levels comparable to the superallowed Fermi decay data set. Given the exceedingly small interaction cross section for antineutrinos, this correlation is typically measured by detecting the outgoing proton, possibly in coincidence with the outgoing electron. Because the proton emerges with a maximal kinetic energy of 751 eV, these are detected after post-acceleration using a variety of detector technologies \cite{Erozolimskii1991, Mostovoi2001, Beck2002, Soldner2004, Stratowa1978, Schumann2008a, Byrne2002}. The Nab experiment uses high-purity, thick silicon detectors \cite{Broussard2017, Salas-Bacci2014} which display excellent linearity over the full range of energies for electrons emerging from neutron $\beta$ decay. The proton momentum is reconstructed from the time of flight from the decay vertex to its detection after passing through a magnetic field-expansion region. As a consequence, for Nab to reach its anticipated precision, systematic effects in the timing reconstruction must be understood at the sub-nanosecond level. As such, detector-related effects that change the anticipated pulse shape must be sufficiently understood.

A detailed description of the response of high-purity silicon detectors under irradiation is a central pillar of much of nuclear and particle physics \cite{Knoll2010, Owens2019, Spieler2005, Lutz1999, Schenk1998}. In particular, the segmentation and extreme radiation conditions at colliders have driven substantial efforts for numerical simulation of device performance and signal prediction \cite{Richter1996, Brigida2004, Unno2013, Demaria2000, Petasecca2006, Piemonte2006, DanielElvira2017}. Extensive work has been performed to unlock position sensitivity in large Germanium detectors using pulse shape discrimination \cite{Bruyneel2006, Bruyneel2006a, Bruyneel2016, Korichi2017, Schmid1999, Paschalis2013, Eberth2001, Cooper2011} while the KATRIN collaboration developed custom simulation software to model low-energy electrons incident on Si detectors \cite{Renschler2011, Renschler2012}. Additionally, the advent of (cryogenic) semiconductor technology for dark matter searches \cite{Agnese2018, Armengaud2017} points to an increased need for precise descriptions of low-energy nuclear radiation interactions  \cite{Bonhomme2022, Ramanathan2020}. Even so, the operational regime for proton detection in the Nab experiment is virtually unexplored due to the low proton energy and stringent timing constraints. In this work, we describe a detailed model for pulse shape simulation of incoming protons, with results directly applicable to electrons as well.

The paper is organized as follows: Section \ref{sec:experiment_overview} describes an overview of the experiment, requirements on detector timing performance and accurate decay event reconstruction. Section \ref{sec:model_input} describes a number of general semiconductor inputs to the model and a critical literature study, which is used in the following sections. In Sec. \ref{sec:FieldSimulations} we describe detailed electric and weighting field simulations for our experimental configuration, which are used in Sec. \ref{sec:carrier_transport_sim} to perform Monte Carlo simulations of quasiparticle transport and investigate collective effects. Section \ref{sec:pulse_simulation} takes the preceding ingredients to perform detailed pulse shape simulation through Monte Carlo charged particle transport and electronics simulations. By varying internal parameters of the models and results from the preceding chapters, Sec. \ref{sec:sensitivity_study} studies Nab's sensitivity to a variety of observables and proposals of measurement schemes. Finally, Sec. \ref{sec:conclusion} provides a summary and outlook.

Features of note in this analysis of pulse shapes include: (1) the explicit incorporation of diffusion, plasma effects and Coulomb repulsion into pulse shape evolution, (2) an energy per quasi-particle tuned to reproduce empirical energy dependence data, (3) the capability to model undepleted material using Gunn's theorem, (4) dead layer models based on manufacturer's impurity density profiles near the rectifying junction, (5) detailed models of pixel isolation using p-stop and combined p-spray and p-spray geometries.

\section{Experiment overview}
\label{sec:experiment_overview}
The goal of the Nab experiment is to measure the $\beta$-$\nu$ angular correlation following neutron $\beta$ decay. The Standard Model differential decay rate for neutron decay rate is known to high precision and can be written as \cite{Jackson1957,Hayen2018, Hayen2020a}
\begin{align}
\dfrac{d\Gamma}{dE_e d\Omega_e d\Omega_\nu} &\propto |V_{ud}|^2G_F^2\,p_eE_e(E_0-E_e)^2 F(Z, E_e) \nonumber \\
&\times\left\{1 + b_F \frac{m_e}{E_e}+a\frac{\bm{p}_e\cdot \bm{p}_\nu}{E_eE_\nu} \right. \nonumber \\
&\left.+ \bm{\sigma}_n \cdot \left[A \frac{\bm{p}_e}{E_e} + B\frac{\bm{p}_\nu}{E_\nu} +\ldots \right] + \ldots \right\}
\label{eq:diff_decay_rate}
\end{align}
where ellipses represent higher-order terms that vanish in the measurement scheme in Nab, $E_e (\bm{p}_e)$ is the electron energy (momentum), $F$ is the traditional Fermi function, $\bm{\sigma}_n$ is the initial neutron polarization and we omitted a number of multiplicative small corrections for notational clarity \cite{Hayen2018}. For an unpolarized neutron beam, terms proportional to $\bm{\sigma}_n$ average to zero and one is left with the $\beta$-$\nu$ angular correlation, denoted $a$, and the Fierz interference term, $b_F$. The former can be written in terms of the vector and axial coupling constants as
\begin{equation}
    a \stackrel{LO}{=} \frac{1-\lambda^2}{1+3\lambda^2}
    \label{eq:a_LO}
\end{equation}
where $\lambda \equiv g_A/g_V$ is the ratio of coupling constants. The Particle Data Group (PDG) reports the value of  the correlation parameter  $a$ already corrected for higher order effects, making Eq. (\ref{eq:a_LO}) rigorously correct to first order in recoil corrections. Using the current PDG \cite{Workman2022} average, $\lambda = 1.2754(13)$, Eq. (\ref{eq:a_LO}) resolves to $a^\mathrm{PDG} = -0.10657(38)$ implying a substantial cancellation in Eq. (\ref{eq:a_LO}). As a result, while such a cancellation precipitates increased sensitivity to $\lambda$ it obtains larger fractional changes originating from higher-order corrections. These have been recently reevaluated \cite{Hayen2020a}, however, and are adequately understood. The Fierz interference term, on the other hand, is sensitive only to Beyond Standard Model currents (assuming left-handed neutrinos)
\begin{align}
    &b_F = 2 \frac{\sqrt{1-\alpha^2}}{1+3\lambda^2} \nonumber \\
    &\times \text{Re}\left\{\frac{g_S\epsilon_S}{g_V(1+\epsilon_L+\epsilon_R)}- \lambda \frac{12g_T\epsilon_T}{g_V(1+\epsilon_L - \epsilon_R)} \right\},
    \label{eq:bF}
\end{align}
where $\alpha$ is the fine-structure constant and $\epsilon_X \sim (246\, \mathrm{GeV}/\Lambda_{X})$ are exotic couplings appearing due to loop effects from new physics at a scale $\Lambda_X$ \cite{Bhattacharya2012, Cirigliano2013, Falkowski2020}. Similar exotic currents appear in $a$ but appear only at quadratic order. Sensitivity therefore originates predominantly from an (effective) Fierz measurement, or CKM unitarity tests through the joint determination of $\lambda$ and the neutron lifetime \cite{Gonzalez2021}.

\subsection{Measurement principle}
\label{sec:Measurement_principle}
Neglecting radiative decay (i.e. $n\to pe\Bar{\nu}\gamma$), conservation of three-momentum implies
\begin{equation}
    p_p^2 = p_e^2 + 2 p_e p_\nu \cos \theta_{e\nu} + p_\nu^2
\end{equation}
where $\theta_{e\nu}$ is the angle between electron and anti-neutrino three-momentum. The latter is the same as that of Eq. (\ref{eq:diff_decay_rate}) and the main quantity of interest. The large proton mass means its kinetic energy contribution can be neglected when compared to that of the antineutrino, so that one can set $p_\nu = (E_0-E_e)$. Determining both proton and electron momentum then allows one to determine $\theta_{e\nu}$ on an event-by-event basis. The term proportional to $a$ in the differential decay rate of Eq. (\ref{eq:diff_decay_rate}) becomes a linear function of $p_p ^2$
\begin{equation}
    \frac{d\Gamma}{dE_edp_p^2} \propto 1+a\beta\frac{p_p^2-p_e^2-(E_0-E_e)^2}{2E_e(E_0-E_e)}
\end{equation}
when $(p_p^2-p_e^2-p_\nu^2)/2p_ep_\nu \leq 1$ and zero otherwise. Here, $\beta = p_e/E_e = v/c$ is the electron velocity. For a constant electron energy, the slope of the decay distribution is proportional to $a$. Performing this procedure at various different electron energies allows one to disentangle systematic effects related to electron spectroscopy and particle transport dynamics.

\begin{figure}[ht]
    \centering
    \includegraphics[width=0.48\textwidth]{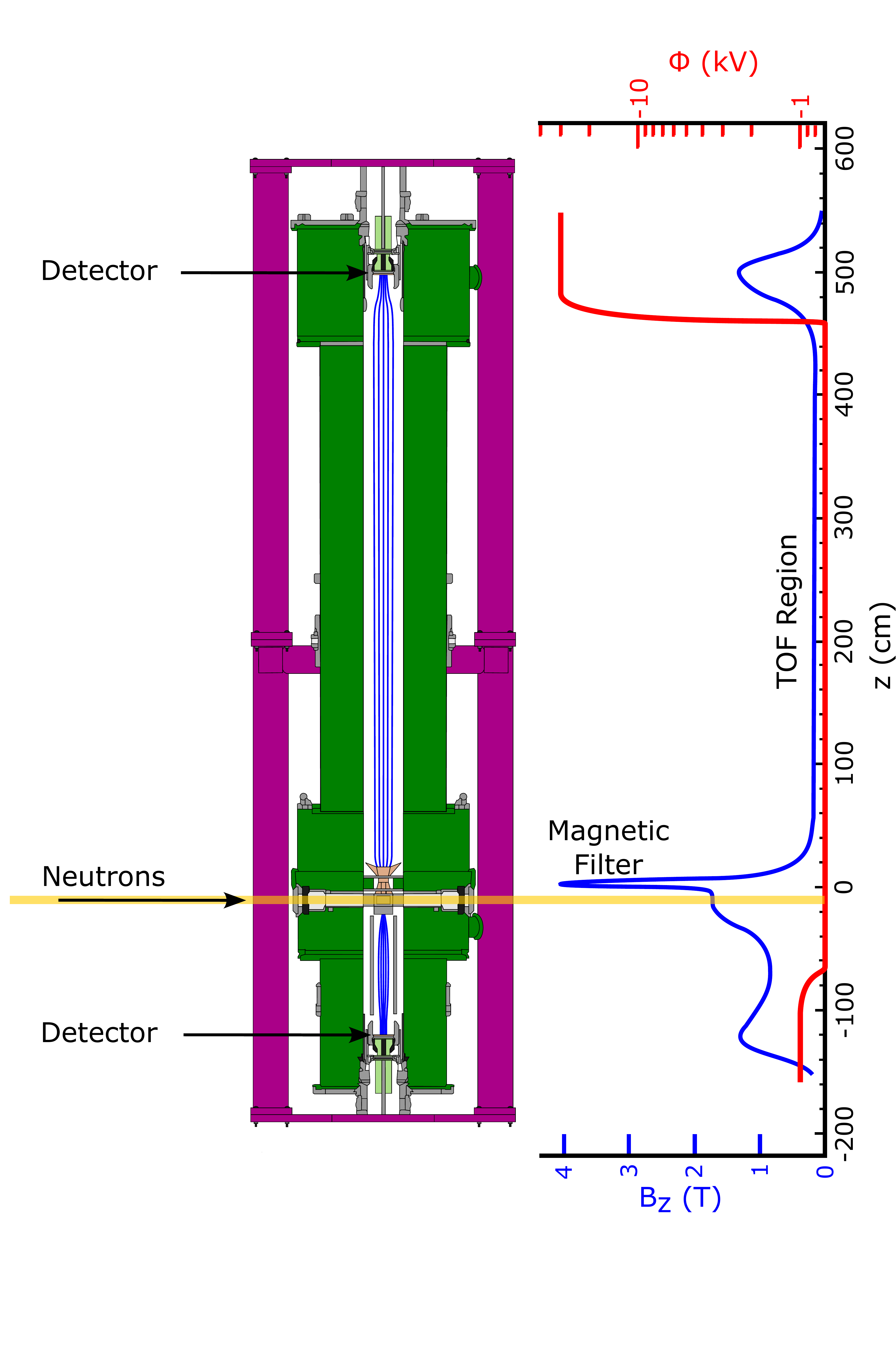}
    \caption{Overview of the Nab apparatus and electromagnetic fields in the `$a$' configuration. Cold neutrons from the Fundamental Neutron Physics beam line at Oak Ridge National Laboratory decay in flight. Emitted charged particles can only overcome the rapid increase in magnetic field (denoted Magnetic Filter) if their momentum is sufficiently longitudinal. The upper detector is floated at -30 kV to accelerate protons above the detection threshold.}
    \label{fig:nab_setup_overview}
\end{figure}

Figure \ref{fig:nab_setup_overview} shows an overview of the Nab apparatus and electromagnetic field arrangement for optimal sensitivity to an $a_{\beta\nu}$ measurement. A cold neutron beam passes through a decay volume inside a magnetic spectrometer with segmented silicon detectors placed on either end in an asymmetric fashion (see Refs. \cite{Pocanic2009, Broussard2017} for a more complete discussion). Electrons and protons emerging from the decay volume can move either to a detector located about 1\,m below beam height or move through a 6\,m low-field region into an upper detector. 

The Nab experiment requires a coincidence signal of both particles, with the electron detected in either detector and the proton in the upper detector. The proton momentum is reconstructed through the time difference between the relativistic electron trigger and the proton time of flight in the low-field region. In order to narrow the proton momentum reconstruction function, the Nab experiment implements an angular filter for particles to reach the upper detector. Particles with an upwards velocity component encounter a substantial magnetic field increase that acts as an angular cut, such that only particles with angles larger than $\theta_\text{min} = \cos^{-1}\sqrt{1-B_0/B_\text{max}}$ will make it to the top detector. A substantial magnetic field decrease after the angular selection serves to {(nearly) adiabatically} longitudinalize the momentum along the flight direction. For protons, with a maximum of 751\,eV of kinetic energy, transport to the upper detector takes at least about 10\,$\mu$s (compared to 10's of nanoseconds for electrons) so that to first order the time difference between electron and proton hits is simply the proton travel time. In the ideal (but unphysical) scenario, this would result in the trivial relationship
\begin{equation}
    p_p = \frac{m_pL}{t_p}
    \label{eq:pp_tof_ideal}
\end{equation}
where $L$ is the path length and $t_p$ the proton time of flight. In reality, the random initial emission angle smears the transport time from the decay volume to the magnetic field maximum, resulting in a broadening of the proton momentum extraction. 
Taking into account several additional complications, the time-of-flight distribution can instead be written as
\begin{equation}
    P_t\left(\frac{1}{t_p^2}\right) = \int dp_p^2\int d\Omega P_p(p_p^2)\phi\left(\frac{1}{t_p^2},p_p^2, \Omega\right)
\end{equation}
where $\phi$ is the spectrometer response function and $\Omega$ encodes further broadening effects. In the idealized case of Eq. (\ref{eq:pp_tof_ideal}) we may simply write $\phi(t_p^{-2}, p_p^2) = \delta(t_p^{-2}-p_p^2/(m_pL)^2)$. A realistic assessment of $\phi$ is one of the main targets of the Nab experiment and can be addressed in a number of complementary ways. There are a number of effects, however, that are not easily accessible without dedicated study and requiring input from simulation.

\subsection{Timing requirement}
\label{sec:timing_requirement}

A major concern is the appearance of a timing mismatch between what is predicted from spectrometer transport and what is extracted from the detector response. The latter is a three-step process, as it comprises physical transport from the decay volume to the detector, the transport time of quasiparticles and induced charge on the electrodes, and a reconstruction of the impact time from the saved waveform after analog and digital filtering. In this work we will be concerned mainly with the latter two mechanisms, as we explore how different processes introduce timing offsets and wave form variability depending on internal detector parameters. 

We may estimate the relevant scale for these offsets through a dimensional analysis, as a change in $a$ resulting from an offset $\Delta$ will be proportional to $\Delta/t_p$. Using $t_p \sim 10\mu$s for a proton's 5\,m transport time, a $1$\,ns unaccounted offset results in a false offset in $a$ at the $10^{-4}$ level. Relative to the Standard Model prediction, $a_{\beta\nu}^{\rm SM} \sim -0.1$, such an effect would constitute a relative $\mathcal{O}(0.1\%)$ systematic bias. A more elaborate way of performing an analytical estimate of the introduced timing bias results in a \textit{false} $a_\mathrm{false}$ as
\begin{equation}
    a_{\rm false} \approx \frac{2p_ep_\nu}{\beta p_{p, \rm{max}}^2} \frac{t_p}{t_{p, \rm{min}}^2} \Delta
\end{equation}
where $p_{p, \rm{max}}$ is the maximal proton momentum and $t_{p, \rm{min}}$ the minimal proton time of flight, in agreement with the dimensional analysis estimate. As Nab aims for a determination of $a_{\beta\nu}$ to 0.1\%, systematic timing offsets in the reconstruction must be controlled at the nanosecond or below level. This stringent requirement will be a common thread throughout this work as we study various effects of timing bias and pulse shape changes.



\subsection{Event reconstruction}
In order to achieve the physics goals of the Nab experiment, decays must be accurately reconstructed, i.e. a determination of proton and electron hit locations, the point in time when they are incident on a detector face, and the electron energy. As discussed, electrons generate a prompt ``start'' for an extraction of the proton time of flight (TOF), with any bias in this measurement being a critical parameter for the Nab experiment. The extracted TOF is potentially very strongly influenced by detector pulse-shape effects, motivating the development of an accurate model for detector response. The accurate binning of detected coincidences with respect to the electron energy is also important. Dominant effects for the electron energy binning are expected at roughly the percent level from bremsstrahlung losses. The interpretation of electron events can be complicated because they have a relatively high probability of scattering in the detector material, potentially resulting in one or more ``back-scattering'' events. When electrons back-scatter, they do not deposit their full energy in one interaction with the detector, but instead reverse their longitudinal momentum and re-emerge from the detector. They can subsequently either reflect from the magnetic ``pinch" (located at $z=0$ in Figure \ref{fig:nab_setup_overview}) or hit the opposite detector face. While, to zeroth order, all of the electron energy is deposited in either one detector or another, meaning the resultant energy errors are expected to be much smaller than those from bremsstrahlung, ``missed'' backscatters can result in a shifted ``start'' time and strong variations in pulse shape. A detailed treatment of these event topologies lies beyond the scope of the current manuscript and will be discussed in a follow-up work.

From the point of view of the detector response, several complicating factors occur that can potentially introduce bias in TOF and energy measurements: ($i$) the backscattering probability off the detector is slightly energy-dependent, and its threshold detection depends on the semiconductor junction structure; ($ii$) the quasiparticle transport time can have strong local dependencies due to local impurity density variations affecting the electric field; ($iii$) the time dependence of the induced charge on an electrode will be strongly deformed near pixel boundaries due to geometrical effects; ($iv$) non-ionizing energy losses (NIEL) depend on the proton momentum, which translates into an energy-dependent sub-threshold event fraction; ($v$) quasiparticle creation and transport are strongly temperature dependent. While several of these can be addressed in part through calibration, high quality model input is required to disentangle experimental results and train analysis extraction scripts and apply ``benchmark'' calibration data to the global beta decay data set (over all possible particle energy combinations and initial emission angles).

Finally, to avoid events where decay particles interact with the spectrometer boundary surfaces and to ensure each proton event is paired with a physically reasonable electron coincidence, the fiducial volume for allowed decays must be unambiguously defined. The segmentation of the dectector provides this capability, but also introduces charge sharing effects near pixel boundaries and rather large pulse shape effects as a function of the position for a given event incident on a given pixel. Defining the fiducial volume also plays a role in measuring backgrounds produced when the neutron beam is present (making a neutron ``beam off'' measurement not possible). These backgrounds will be determined using the events detected in pixels where particles originating in the neutron beam (entrained in the spectrometer fields) can not reach, and are outside the fiducial volume for neutron decay events. An accurate detector model can use the expected pulse shapes and event histories to strongly constrain events near pixel boundaries and account for charge sharing effects.

\section{Model input}
\label{sec:model_input}
In this section we summarize the model input and discuss consequences of its individual components. Some components are specific to the Nab apparatus, such as the detector geometry (Sec. \ref{sec:detector_geometry}) and doping (Sec. \ref{subsection: DopingProfile}), whereas the carrier transport (Sec. \ref{sec:charge_carrier_transport}), charge collection efficiency (Sec. \ref{sec:charge_collection_efficiency}) and pair creation energy (Sec. \ref{sec:pair_creation_energy}) are more generally applicable. The combination of these ingredients will combine with detailed simulation results discussed in the following sections to provide the most precise description of pulse shapes in ultrapure silicon detectors discussed in Sec. \ref{sec:pulse_simulation}.

\subsection{Detector geometry}
\label{sec:detector_geometry}
\begin{figure}[ht]
    \centering
    \includegraphics[width=0.23\textwidth]{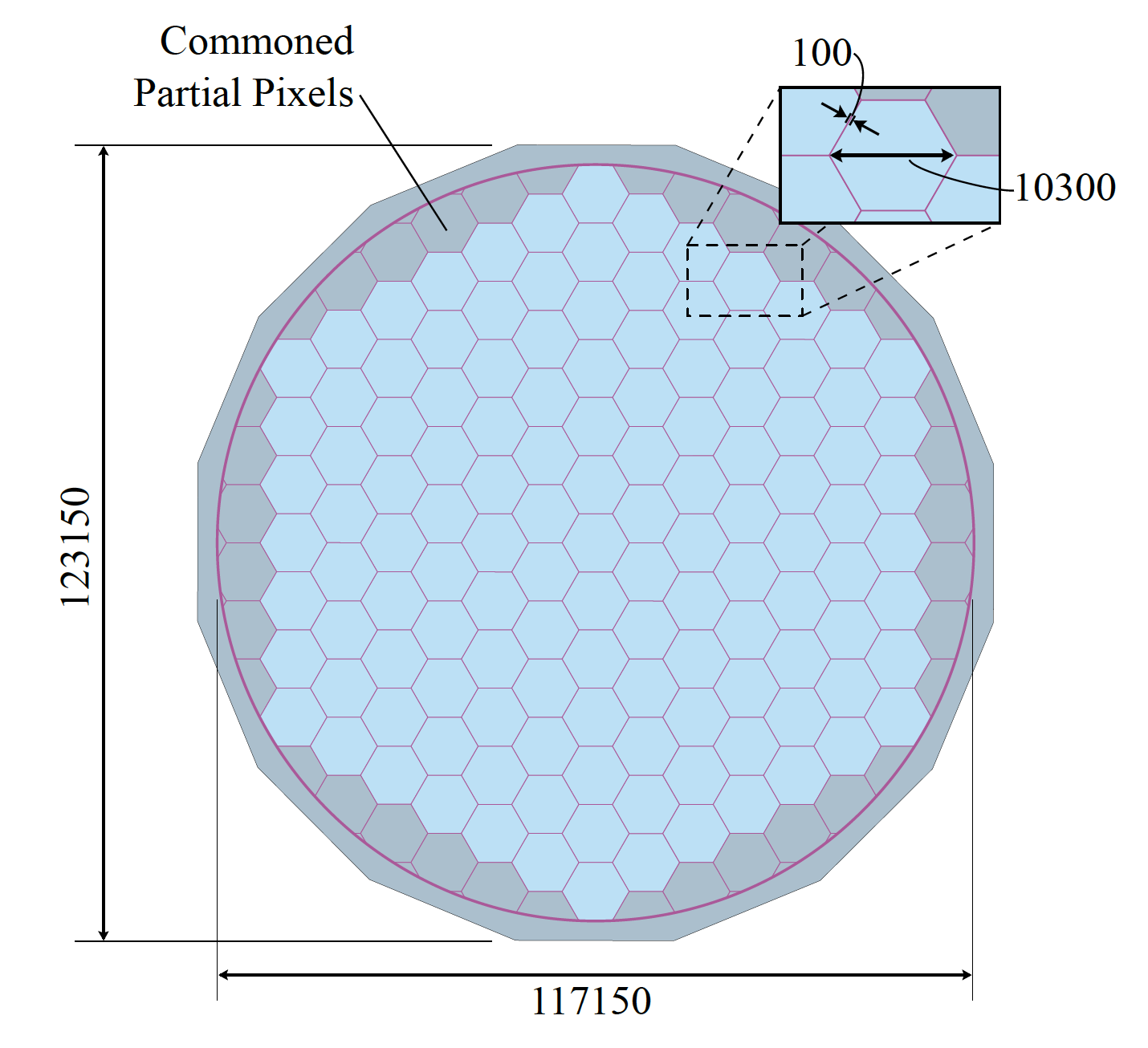}
    \includegraphics[width=0.23\textwidth]{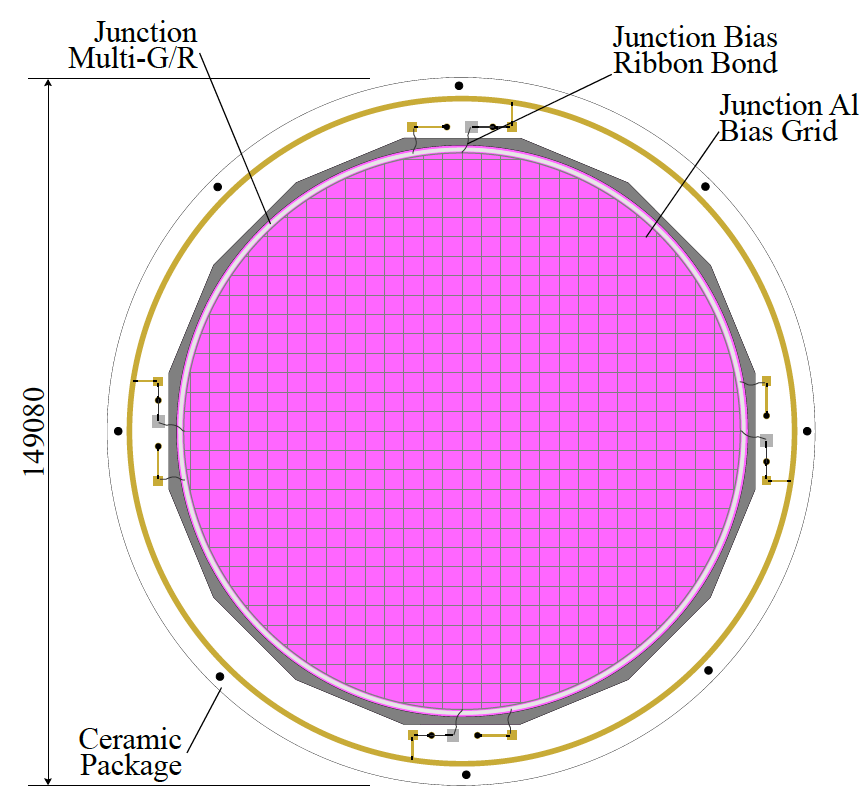}
    \includegraphics[width=0.48\textwidth]{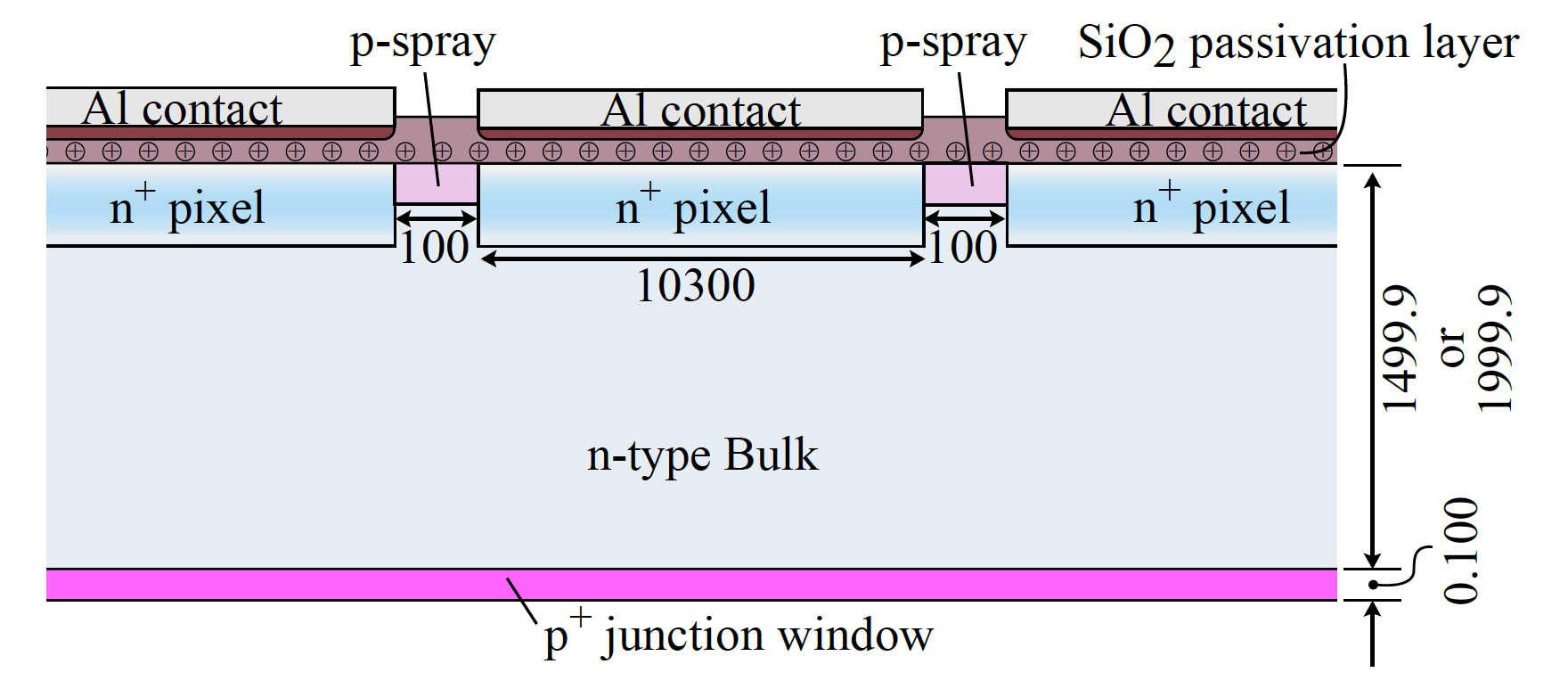}
    \caption{Overview of the Nab silicon detector geometry \cite{Jezghani2019}. (Top) The ohmic side is segmented into 127 individual pixels each with a total area of 70 mm$^2$. The junction side is featureless by design with the exception of an Aluminum biasing grid with 4 mm separation. (Bottom) Cross-section showing schematic impurity density profiles and inter-pixel isolation.}
    \label{fig:detector_schematic}
\end{figure}

Nab's choice of detector technology was guided by a number of constraints: ($i$) due to the sensitivity to the electron energy, detectors should be highly linear; ($ii$) for accelerated protons to be detected with high efficiency, incomplete charge collection in the entrance window should be minimal ($iii$) to reduce backgrounds and force topologically consistent coincidence events, the detector should be highly segmented. The result is a 127-pixel, 1.5\,mm or 2\,mm thick ultra-pure silicon detector with an implanted, sub-100\,nm p-type entrance window. A schematic overview is shown in Fig. \ref{fig:detector_schematic}.

The Nab detectors are made from high-purity Si with slab thicknesses of 1.5 and 2.0 mm and outer diameter of 13.5\,cm. The front face is a rectifying contact made through Boron implantation with an estimated thickness of 100 nm, overlaid with a square Al grid for biasing and charge restoration. The latter covers about 0.4\% of the surface and is effectively opaque to protons. The back side is highly segmented via 127 individual Ohmic contacts through Aluminum coating to form individual hexagonal pixels with a surface of about $A=70$ mm$^2$ and separated from neighbouring pixels by a $100 \mu$m gap. The Si bulk and Al coating are separated by an oxide passivation layer of a few nanometers thick. The choice of hexagonal pixels has a number of benefits: ($i$) a planar surface can be efficiently filled; ($ii$) each corner connects only three pixels, thereby limiting charge sharing effects; ($iii$) most of the inner surface is quasi-cylindrically symmetric.

\subsection{Doping profile}
\label{subsection: DopingProfile}
\subsubsection{Bulk}

The Nab detectors are constructed using ultrapure silicon grown along the $\langle 100 \rangle$ crystal axis. The detectors should be capable of fully stopping 782 keV (the neutron decay $Q_\beta$ value) electrons. The minimal thickness must therefore be at least 1.5 mm, which implies that the required crystal purity approaches that of intrinsic silicon. As such, crystals are grown using the float-zone technique, where a polycrystalline feed rod is made molten using high-power radiofrequency coils to create a liquid interface with the seed crystal \cite{Dietze1981}. Because of the large diameter of the crystal required, the needle-eye technique was used, where the induction coils have a much smaller diameter than the boule to ensure homogeneous heating. The subsequent widening of the melt means the connection to the target rod and its crystallization process are a complex interplay between a variety of local and environmental conditions such as temperature, pressure and impurity concentration. On average, the bulk resistivity for the Nab detectors is estimated to be at least 25 k$\Omega \cdot$cm, with depletion studies pointing towards an effective impurity density of $3-6\times 10^{10}$ cm$^{-3}$. Local deviations introduce a position-dependence on the pulse shapes for physics events, however, and require additional scrutiny.

As impurities (i.e. metals, but also oxygen, carbon, and others) have a higher diffusivity in the silicon liquid phase, the monocrystal can be made substantially more pure than the polycrystalline feed rod \cite{Burton1953, Burton1953a, Christenserr2003}. On the other hand, float-zone silicon is susceptible to variations in the impurity concentration. The dopant concentration along the length of the boule, for example, varies due to the buildup of impurities in the melt as the process proceeds up the feed rod. Depending on the type of impurity, however, axial concentration gradients can be minimized through one or more passes \cite{Sze2007}. Radial gradients, on the other hand, present a much more substantial issue for larger crystals ($>$100 mm diameter) \cite{Schroder2001}. Unlike the impurities present in the polycrystalline feed rod, dopants such as phosphorus for $n$-type silicon are typically introduced through a vapour inside the chamber and therefore follow the flow of the silicon melt. Due to the large diameter difference between the needle-eye feed rod-melt interface and the melt-target rod interface, the dynamics of the melt is determined by substantial temperature gradients along with gravity following the Navier-Stokes equation \cite{Ratnieks2008}. 

In a simplified picture, the flow of the liquid silicon is determined by the Marangoni force, electromagnetic forces from the induction coils and buoyancy forces inside the melt due to temperature and density gradients \cite{Ratnieks2008}. Whereas the former two (partially) cancel, buoyancy forces create convection cells inside the melt. Detailed simulations of the float-zone process \cite{Ratnieks2008, Sabanskis2017, Han2020, Han2020a} show crystals with diameters larger than 100 mm having two or more convection cells along the radial direction. The consequence is that, as impurities and dopants will follow the flow inside the melt before crystallizing, an accumulation of dopants occurs at the confluence of these convection cells. These result in a reduced resistivity band concentric with the boule axis, and more generally a complex radial impurity density profile. Relative differences in impurity concentration can exceed 50-100\%, in agreement with experimental observations \cite{Quaranta1970, Dietze1981}.

The presence of radial gradients in the bulk impurity density profile will have profound effects on the electronic pulse shape of physical events and their timing reconstruction, and will be studied in detail in Secs. \ref{sec:FieldSimulations} and \ref{sec:RiseTimeDistributions}.

\subsubsection{Junction and contact implantation}

\begin{figure}[ht]
    \centering
    \includegraphics[width=0.48\textwidth]{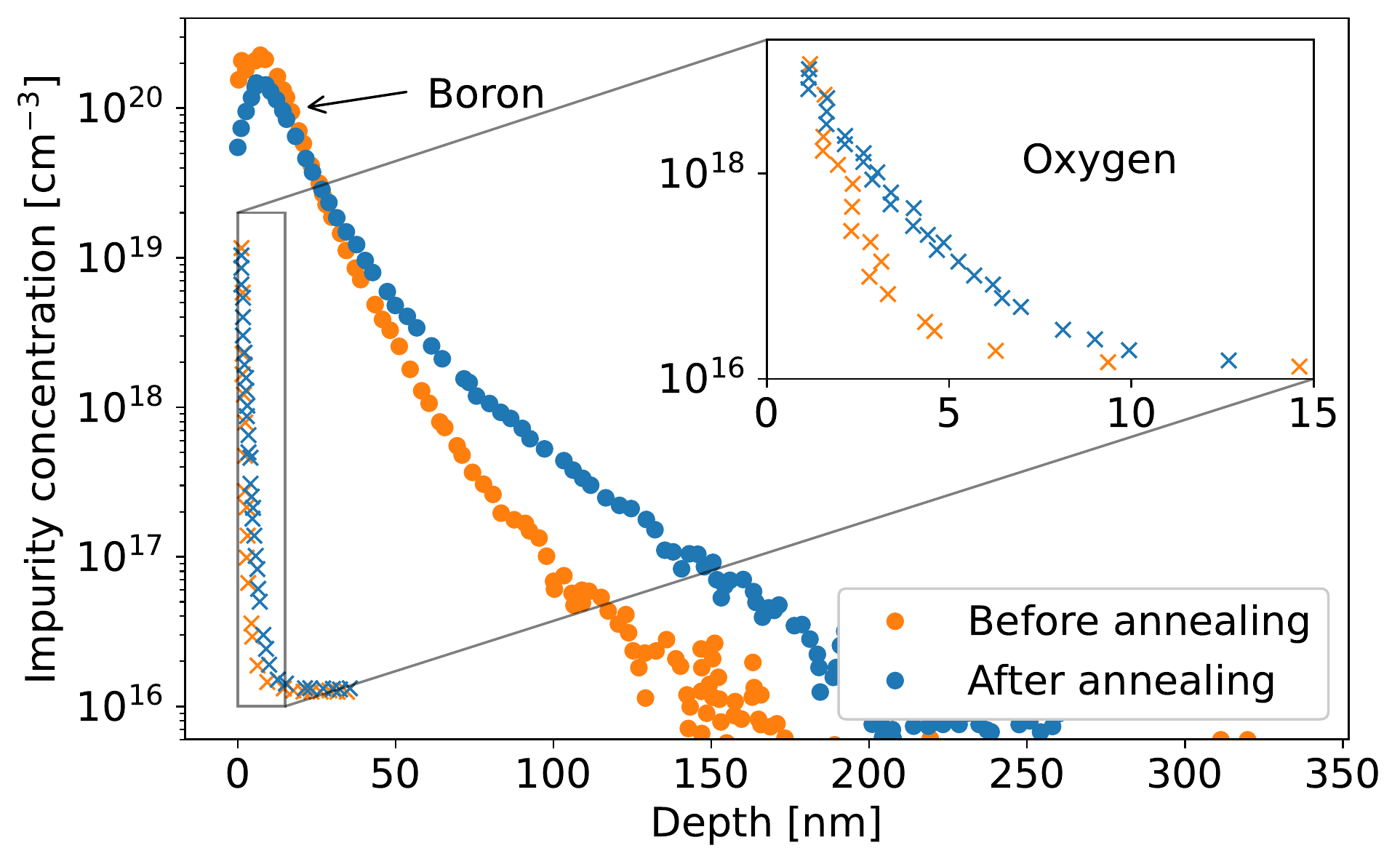}
    \caption{Secondary Ion Mass Spectroscopy results using an Oxygen beam for the Boron and Oxygen concentration on the front face of the detector before and after annealing.}
    \label{fig:sims_data}
\end{figure}

Following the schematic representation of Fig. \ref{fig:detector_schematic}, the bulk material as described above undergoes a number of implantation steps to create the diode junction and pixel contacts. The former is a $p^+$ junction created via Boron implantation with a penetration depth of $\mathcal{O}(100)$ nm. As this junction results in poor charge collection (see Sec. \ref{sec:charge_trapping_dl}) and crystal damage, it is imperative that this be layer be as thin as possible and remain so even after annealing. Results from secondary ion mass spectroscopy (SIMS) \cite{Benninghoven1987, Francois-Saint-Cyr2001, Svensson1990} using a primary Oxygen beam are shown in Fig. \ref{fig:sims_data} before and after annealing. The latter results in a general loss of Boron by 28\% and thermal diffusion increases the depth by which the concentration reaches $1\times 10^{16}$ cm$^{-3}$ by 40\%. Similar changes occur for the thermally grown oxide layer at the front face, where the concentration decreases by an order of magnitude over the space of a nanometer. The concentration bottoms out at $1\times 10^{16}$ cm$^{-3}$ due to implantation of the primary Oxygen beam. The crystal structure after annealing should resolve many of the defects introduced after implantation. Even so, we neglect effects due to channeling in this work when discussing charge transport in Sec. \ref{sec:MC_sim}.

The readout geometry, on the other hand, is defined via $n^+$ implantation on the backside of the crystal, creating an $n^+$-on-$n$ Ohmic pixel. Together with the Aluminum metal contact, a thin (several nm) thermally grown SiO$_2$ layer and monocrystalline silicon, these form a MOS junction. Due to ultra-thin insulating oxide layer, the tunneling process is exponentially enhanced even though multiple phenomena contribute at different temperatures \cite{Sze2007, Lutz1999}. The oxide layer experiences strain at the interface with the bulk silicon, however, and additional static positive charges are always present in the oxide layer even after annealing \cite{Fowkes1969, Goetzberger1967, Nicollian1982}. The latter causes an accumulation layer of electrons, effectively creating a conductive channel between n$^+$ implants. 

\begin{figure}[ht]
    \centering
    \includegraphics[width=0.48\textwidth]{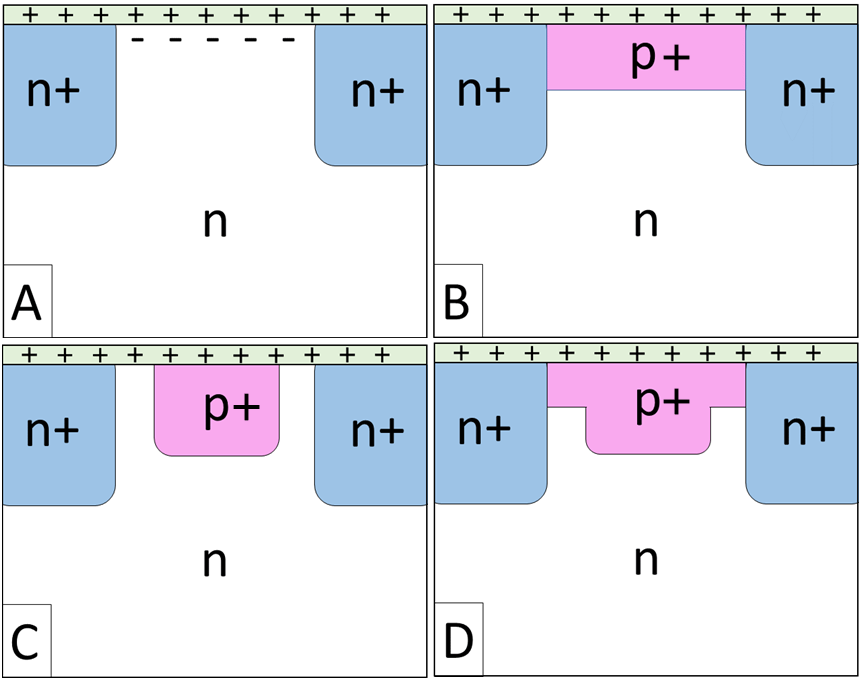}
    \caption{Overview of pixel isolation strategies: (A) No isolation, (B) p-spray, (C) p-stop, (D) p-stop and p-spray. The two Nab configurations consist of pure p-spray (B) and a combination of p-stop and p-spray (D).}
    \label{fig:pixel_isolation}
\end{figure}

One way of interrupting this channel is by introducing a large $p$-type doping in between pixels, using either p-stop or p-spray technologies \cite{Batignani1989, Matheson1995, Richter1996, Gorelov2002, Unno2013, Piemonte2006, Gorfine2001}. In the case of p-spray, the full surface is implanted with Boron which is then overcompensated through Phosphorus implantation to create the n$^+$ regions. For p-stop, meanwhile, the $p$-type doping is introduced via implantation using an additional mask. Due to the sharp features of the implant on the edges (see Fig. \ref{fig:pixel_isolation}), however, high electric fields that may cause breakdown are sometimes observed. As a remedy, moderated p-spray \cite{Gorelov2002} is sometimes preferred, where the centre of the inter-pixel gap contains a higher dopant concentration, as shown in Fig. \ref{fig:pixel_isolation}. The Nab detectors come in two varieties, with some produced using p-stop implants and the others using moderated p-spray.

In Sec. \ref{sec:pixel_isolation}, we perform detailed simulations of the electric and weighting fields close to the p-type implantation for both types of pixel isolation technology. These results are used in Sec. \ref{sec:chargeSharing} to study charge sharing effects for events occurring close to pixel boundaries.

\subsection{Pair creation energy}
\label{sec:pair_creation_energy}

Charged particles entering the detector lose energy through a variety of channels, some of which cause the creation of particle-hole pairs. In a typical ionization event, a bound electron inside the valence band is promoted to the conduction band leaving behind a hole. As such, the minimal energy required is the size of the bandgap, which for silicon is $E_{g} = 1.12$ eV at 300\,K. Experimentally, however, the mean energy required for the creation of a particle-hole pair is substantially higher than $E_\mathrm{gap}$ - a feature that is observed in all semiconductors. Additionally, the variation in the number of created particle-hole pairs is non-zero, but substantially lower than expected from independent Poisson processes. One defines the following semiconductor-specific pair-creation energy (PCE) and Fano factor,
\begin{subequations}
\begin{align}
    \epsilon_{ph} &= \frac{E}{\langle N(E)\rangle} \\
    F &= \frac{\langle N^2(E) \rangle - \langle N(E) \rangle^2}{\langle N(E) \rangle}
\end{align}
\label{eq:epsilon_Fano}
\end{subequations}
where $E$ is the incoming particle energy, $N$ the number of particle-hole pairs and $\langle \ldots\rangle$ denotes the average value. The Fano factor, $F$, reduces the variance through $\sigma^2 = F N$ and is experimentally found to be close to 0.11, while for the average particle-hole creation energy one finds $\epsilon_{ph} \sim 3.6$ eV. The threefold increase of the latter over the bandgap energy is typically understood via the creation of optical phonons and population of final state energies below $E_g$ \cite{Shockley1961}. As the primary electron slows down its conversion into optical phonons becomes more  efficient and the average energy needed for a particle-hole pair increases. Both the bandgap energy and phonon population depend on the detector temperature, so that thermal changes in $\epsilon_{ph}$ set a constraint for the required thermal stability of the Nab detectors. 

\begin{figure}[ht]
    \centering
    \includegraphics[width=0.48\textwidth]{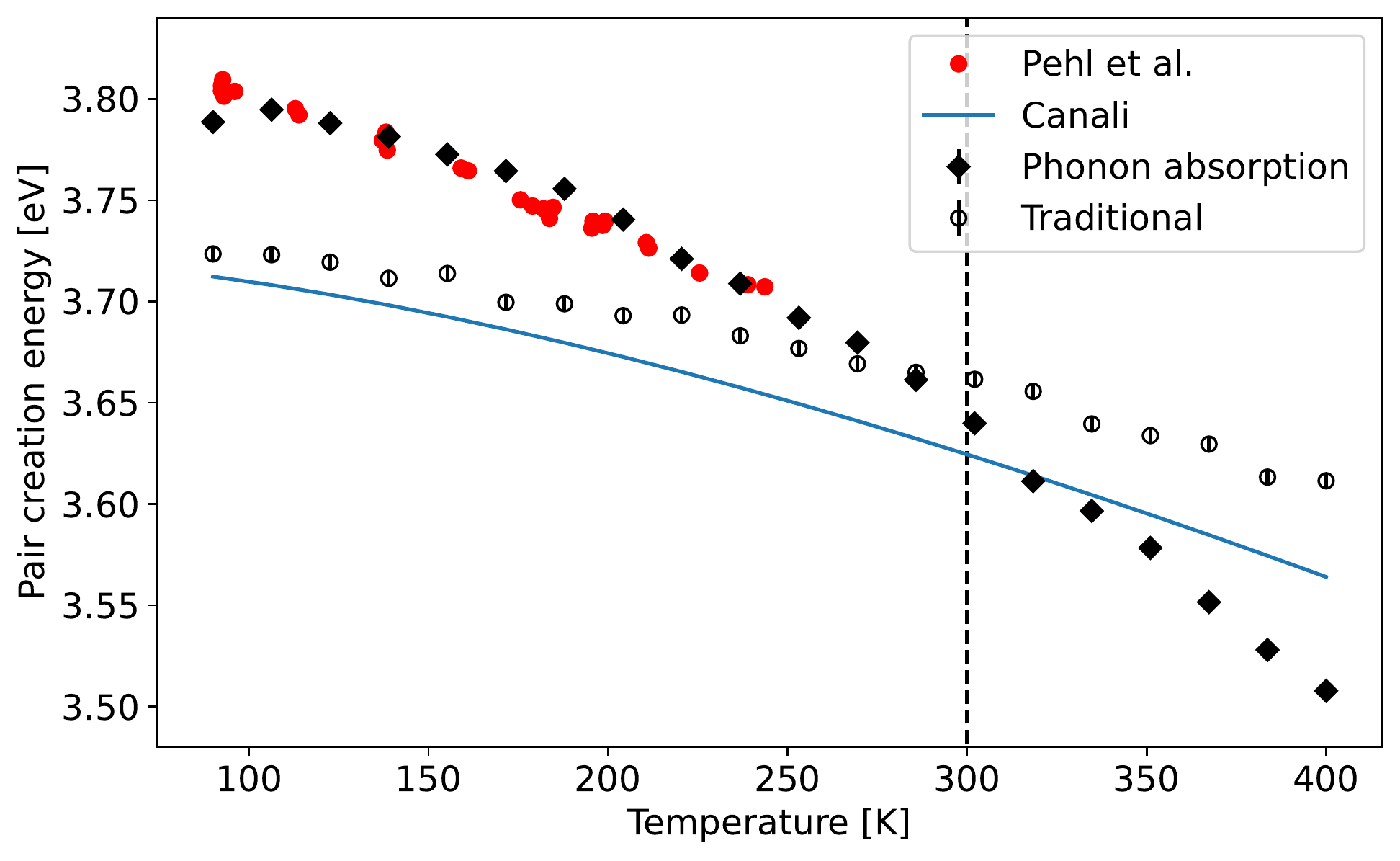}
    \caption{Behaviour of the pair-creation energy as a function of temperature for electrons and $\gamma$ rays. Here, Canali refers to Eq. (\ref{eq:Canali_PCE_vs_T}), the data are by Pehl et al. \cite{Pehl1968}, and the simulation results are discussed in a follow-up work. A vertical line is shown at room temperature, whereas the typical Nab detector operating temperature lies around 110 K.}
    \label{fig:eph_vs_T}
\end{figure}

Temperature effects on $\epsilon_{ph}$ have been studied by a number of authors in the past, although results do not unequivocally agree \cite{Pehl1968, Emery1965, Bussolati1964, Canali1972}. Early theoretical arguments pointed towards a linear relationship between $\epsilon_{ph}$ and $E_g$, where all temperature dependence is assumed to come from that of the bandgap energy \cite{Alig1980, Chang1985, Ramanathan2020, Balkanski1983}. The latter is typically written as
\begin{equation}
    E_g(T) = E_g(0) - \frac{aT^2}{T+b}
\end{equation}
where $E_g(0) = 1.1692$~eV, $a=(4.9\pm 0.2)\times 10^{-4}$ eV, and $b = 655 \pm 40$ K. This results in $E_g(300\,K) = 1.12$ eV and $E_g(120\,K) = 1.16$ eV. Following earlier partitions of $\epsilon_{ph}$ into phonon and ionization contributions, Canali et al. \cite{Canali1972} find
\begin{equation}
    \epsilon_{ph}(T) = 2.15 E_g(T) + 1.2\,\mathrm{eV}
    \label{eq:Canali_PCE_vs_T}
\end{equation}
but is generally in poor agreement with the precise data of Pehl et al. \cite{Pehl1968}. While several theoretical descriptions have been performed in a variety of semiconductor compounds at room temperature, little focus has been dedicated to study its temperature dependence. For this work we take the experimentally determined pair production energy as empirical input for our detector model. By taking into account a more sophisticated treatment of phonon creation and absorption, discussed in more detail in a follow-up work, we are able to recover the behaviour found by Pehl et al, shown in Fig. \ref{fig:eph_vs_T}.

\subsection{Charge carrier transport}
\label{sec:charge_carrier_transport}
Once liberated, the movement of free charge carriers is determined completely by a set of coupled equations, the first being the drift-diffusion equation
\begin{subequations}
\begin{align}
    \frac{\bm{J}}{-q} &= -D\nabla n -n \mu_n \bm{E} \label{eq:net_current_drift_diffusion} \\
    \frac{\partial n}{\partial t} &= -\nabla \cdot \bm{J} + R
\end{align}
\label{eq:drift_diffusion}
\end{subequations}
where $q=e$ is the absolute electron charge, $n$ is the free electron concentration, $\mu_n$ is the mobility, $D$ is the diffusion coefficient, $\bm{J}$ is the particle current and $R$ is a generation-recombination coefficient. The electric field $\bm{E}$ is determined as the solution to the Poisson equation $\nabla\cdot \bm{E} = q(p-n)$. While the sudden liberation of charge carriers due to a traversing charged particle can be interpreted as a non-zero $R$, the effect of the additional charge is typically sufficiently small for it not to disturb the externally applied fields. An exception occurs when the density is sufficiently high, leading to plasma-like effects discussed in Sec. \ref{sec:self_repulsion_plasma}, and when the detector is not fully depleted (see Sec. \ref{sec:time_dependent_weighting_field}). We additionally note that in thermal equilibrium (i.e. $\bm{J} = 0$), Eq. (\ref{eq:net_current_drift_diffusion}) implies that a non-zero gradient in carrier density results in a similarly non-zero electric field even outside of the traditional depletion zone. This fact will become essential when discussing entrance window effects in Sec. \ref{sec:charge_trapping_dl}.

In the assumption that the charge injection is sufficiently small with respect to the doping concentration, the charge carrier motion is determined completely by drift and diffusion. The diffusion coefficient can be obtained from the Einstein relation
\begin{equation}
    D = \frac{k_B T}{q}\mu_n
    \label{eq:einstein_diffusion}
\end{equation}
where $k_B$ is Boltzmann's constant. Before we discuss the effects of diffusion in greater detail in Sec. \ref{sec:diffusion_sim}, it is worthwhile to provide an order of magnitude estimate of the relative effects of drift and diffusion as a function of time. The latter proceeds according to classical expectations, where the charge cloud expands over time to a Gaussian shape with a standard deviation set by $\sigma_D = \sqrt{2Dt}$ after an elapsed time $t$. After the same amount of time, drift along the electric field has propagated charges along a distance $x = \langle \mu E\rangle t$. As a consequence, the effects of diffusion are relevant predominantly at short time scales whereas drift determines the long-term motion. The turnover time can be simply estimated as $t = 2D/(\mu E)^2$, which for typical conditions in the Nab experiment results in a few picoseconds (discussed in more detail below). Diffusion at this timescale results in a charge cloud with a width on the order of a few hundred nanometers. The latter is negligible with respect to the thickness of the detector, so that transport through the detector occurs predominantly through drift. It is comparable, however, to the size of the junction entrance and will significantly influence the charge collection efficiency (see Sec. \ref{sec:charge_trapping_dl}).

\subsubsection{Mobility}
\label{sec:mobility}
Following the drift-diffusion equation (Eq. (\ref{eq:drift_diffusion})), charge carriers propagate along the local electric field through the mobiliy, a proportionality constant defined as
\begin{equation}
    \bm{v}_d = \mu \bm{E}
    \label{eq:mobility_definition}
\end{equation}
where $\bm{v}_d$ is the drift velocity. For low electric fields and high temperatures, the drift velocity behaviour is purely Ohmic and $\mu$ reduces to a constant $\mu_0$. At high fields, saturation occurs due to interaction with the lattice as the phonon production rate vastly outpaces the absorption cross section as the carrier energy increases \cite{Jacoboni1977, Fischetti2019, Lombardi1988}. In these circumstances, the effective temperature of the charge carriers exceeds that of the lattice so that these are typically referred to as `hot electrons' \cite{Reggiani1985}.

While \textit{ab initio} treatments of electron-phonon interactions are evolving rapidly \cite{Giustino2017, Yoder1993, Asche1981, Murphy-Armando2008, Zhou2021}, similar treatments of the mobility reach good qualitative agreement at room temperature \cite{Ponce2018, Ponce2020, Desai2021, Restrepo2009} but are underexplored at low temperatures. In these cases, one must fall back on semi-classical Monte Carlo treatments \cite{Jacoboni1983, Pop2004} using phenomenological models of phonon scattering \cite{Herring1956, Mitin1985}. There is a vast library of experimental work available, however, including high quality data sets for ultrapure silicon at cryogenic temperatures by Canali and collaborators \cite{Canali1975, Canali1973, Ottaviani1975, Canali1971, Canali1975a, Jacoboni1977}. Drift velocity measurements are performed along different crystalline axis, and can generally be described well using an empirical function \cite{Caughey1967, Knoll2010}
\begin{equation}
    \bm{v}_d = \frac{\mu_0 \bm{E}}{[1+(\mu_0E/v_s)^{1/\beta}]^\beta} \label{eq:drift_velocity}
\end{equation}
where $v_s$ is a saturation velocity and $\beta$ a fit coefficient. In the case of Germanium, an additional term is often added to account for the Gunn effect \cite{Mihailescu2000} but this is not applicable to silicon.

The Nab detectors contain both extremes in impurity density, since the implanted junction region has extremely high dopant concentrations (see Fig. \ref{fig:sims_data}), whereas the bulk is very pure. Traditionally, Klaassen's model \cite{Klaassen1992, Klaassen1992a} describes the Ohmic moblity, $\mu_0$, as a combination of different mobilities through Matthiesen's rule, i.e. $\mu_0^{-1} = \sum_i \mu_i^{-1}$ where each $\mu_i$ corresponds to a scattering mechanism
\begin{equation}
    \mu_0^{-1} = \mu_L^{-1} + \left[\mu_{i, N}\left(\frac{N_{ref, 1}}{N_I}\right)^{\alpha_1} + \mu_{i,c}\right]^{-1}
\end{equation}
where $N_I$ is the impurity density and parameters are defined in Ref. \cite{Klaassen1992}. Since then, a number of modifications have been proposed \cite{Schindler2014, Dhillon2022} to extend or improve the agreement with data but do not change our conclusions.

For bulk transport, the mobility is limited only by electron-phonon interactions and the saturation velocity can be described by a phenomenological fit function \cite{Jacoboni1977}
\begin{equation}
    v_s = \frac{v^*}{1+C\exp(T/\Theta)}
\end{equation}
where $v^* = 2.4 \times 10^7$ cm/$s$, $C=0.8$ and $\Theta = 600$ K. The Ohmic mobility, on the other hand, was measured by a number of different authors in the limit of ultrapure samples. The precise measurements along the $\langle 100 \rangle$ axis by Refs. \cite{Norton1973, Logan1960, Canali1975} can be summarized by a power-law fit
\begin{equation}
    \mu_L^{\langle 100\rangle } = 1521(216) \left(\frac{300 K}{T} \right)^{2.01(12)}
    \label{eq:power_law_ohmic_mobility}
\end{equation}
where the uncertainty is due to the spread in literature values.


At cryogenic temperatures, however, features show up in experimental mobility measurements which are not covered by Klaassen's model and which require extra attention. In particular, at medium strength electric fields ($10-10^3$ V/cm) below 77 K one observes experimentally a negative differential mobility \cite{Canali1973, Nougier1975, Nougier1976, Jorgensen1972} where the drift velocity saturates before increasing again at high electric fields. This is qualitatively understood as a consequence of intervalley scattering 
, but theoretical efforts never obtained a better than 10\% level agreement with experimental data. As such, in order to describe the mobility in this region we use empirical fits to the data by Canali \cite{Canali1975}.


\subsection{Charge collection efficiency}
\label{sec:charge_collection_efficiency}

As free charge carriers are created after ionizing energy losses throughout the material, propagation along electric field lines towards electrodes can be interrupted through (temporary) capture via a number of different mechanisms. For most high-speed applications, even brief interruptions in charge transport will result in net loss of signal strength as typical holding times are significantly longer than those of the shaping electronics. For high-purity silicon detectors as those used in the Nab experiment, there are two different regimes of charge loss throughout the detector: ($i$) continuous bulk losses via residual impurities with energy levels close to the middle of the band gap; ($ii$) inside the front face $p^+$ implantation region, also known as the `dead layer'.

In the bulk, dopant concentrations are exceedingly low and loss of charge carriers occurs predominantly via trapping in deep trapping centers close to the middle of the bandgap described by Shockley-Reed-Hall (SRH) statistics \cite{Sze2007}. Typical contaminants such as oxygen and gold trap charges with an average release time longer than the integration time of the current pulse, leading to effectively lost charges. If their concentration is constant throughout the bulk, one can instead define a mean carrier lifetime, $\tau_c$, to obtain Hecht's equation in a constant electric field \cite{Hecht1932}. This equation simply states that the signal size is proportional to $\exp(-t/\tau_c)$ where $t$ is the transport time. We instead use a trivial generalization to linearly dependent electric fields (see Eq. (\ref{eq:E_undepl_depl})) \cite{Zanichelli2012}
\begin{align}
    |Q(t)| &= \frac{q}{L}\frac{|\mu\tau_c a|(x_0+b/a)}{\mu\tau_c|a|-1}H(d-x_0) \nonumber \\
    &\times \left[1-\exp\left(-\frac{1-\mu\tau|a|}{\tau_c}\right)\right]
\end{align}
where $x_0$ is the starting location, the electric field can be written as $E(x) = ax+b$, $d$ is the depletion thickness discussed below, and $t$ is understood to be less than the signal collection time.

The highly doped, implanted part of the p$^+$-n junction, on the other hand, contains a variety of loss mechanisms for charge carriers. Often, charge collection is considered negligible in this region and it is commonly referred to as a `dead layer'. Obtaining a minimal thickness for this layer is crucial, and alluded to in Sec. \ref{subsection: DopingProfile}. In the nuclear physics community, its thickness is typically studied using $\alpha$ spectroscopy under varying incidence angles and assuming
\begin{equation}
    {\rm CCE}(x) = \left\{\begin{array}{lr}
        0 & x < t_{\rm dead} \\
        1 & x > t_{\rm dead}
    \end{array} \right.
    \label{eq:cce_hard_dead}
\end{equation}
where CCE$(x)$ is the charge collection efficiency, with the aim of extracting $t_{\rm dead}$. Work by the KATRIN collaboration \cite{Wall2014} noted that effects from diffusive processes originating in this layer can be transported into the active volume before loss. Several phenomenological parametrizations have been proposed in the literature \cite{Beck2019, Popp2000}, such as that in recent work \cite{Gugiatti2020}
\begin{equation}
    {\rm CCE}(x) = \left\{\begin{array}{lr}
        p_0 & x < t_{\rm ox} \\
        1+(p_1-1)\exp\left(-\dfrac{x-t_{\rm ox}}{\lambda}\right) & x > t_{\rm ox}
    \end{array} \right.
    \label{eq:cce_soft_dead}
\end{equation}
where $t_{\rm ox}, p_0, p_1$ and $\lambda$ are free fit parameters representing an initial oxide layer thickness, $t_{\rm ox}$, with constant efficiency, $p_0$, and a region beyond with maximal efficiency $p_1$ achieved over a length scale $\lambda$. In Sec. \ref{sec:charge_trapping_dl} we perform a novel, detailed Monte Carlo simulation of the Nab entrance window collection efficiency, where we will compare both descriptions to simulation results.

\section{Field simulations}
\label{sec:FieldSimulations}
The components of the model discussed in the previous section often depend on the electric field inside the material, particularly for the drift motion of the free charge carriers. The latter move according to the electric field through Eq. (\ref{eq:mobility_definition}), and therefore determine the transit time throughout the active detector volume. In turn, the electric field depends on the local impurity density and geometry. In this section, we report on detailed simulations of electric fields and weighting potentials, defined below. In particular, we study the effects of radial gradients in the impurity density profile and edge effects near pixel boundaries. 

The detailed simulations are to be compared to the standard description in textbooks on particle and nuclear physics \cite{Knoll2010, Leo1987}. There, the induced current in the physical electrodes is written in terms of a weighting potential in conjunction with the Shockley-Ramo theorem \cite{Ramo1939, Shockley1938}. The latter states that the induced current on electrode $k$  due to the movement of a single charge carrier is
\begin{equation}
    I_k^{\rm SR} = q \bm{v}_d \cdot \nabla W_k(x)
    \label{eq:I_SR}
\end{equation}
where $W_k$ is the weighting potential obtained by solving the Laplace equation after setting electrode $k$ to unit potential and grounding all others. The gradient of the weighting potential is typically referred to as the weighting field, which will be discussed in greater detail below.

For a simple parallel plate detector one finds $\nabla W(x) = (1/L) \hat{\bm{z}}$ where $L$ is the spacing between the plates and $\hat{\bm{z}}$ a unit vector perpendicular to the plates. The induced current then simply depends on the electric field and mobility as the charge carriers move through the geometry. The electric field for a simple planar $p^+$-$n$ geometry with a homogeneous impurity density is
\begin{equation}
    |\bm{E}(z)| = \left\{ \begin{array}{lr}
        \sqrt{\dfrac{2VNq}{\varepsilon}}-\dfrac{Nq}{\varepsilon}z & \text{undepleted} \\
        \dfrac{V}{L}+\dfrac{NqL}{2\varepsilon}-\dfrac{Nq}{\varepsilon}z & \text{depleted}
    \end{array} \right.
    \label{eq:E_undepl_depl}
\end{equation}
where $V$ is the potential difference between the plates (called the bias voltage below), $N$ is the impurity density in the $n$ region and $\varepsilon$ is the dielectric strength of the material. We included in Eq. (\ref{eq:E_undepl_depl}) also the case of an undepleted material, i.e. where $V < V_d$ with $V_d$ the depletion voltage
\begin{equation}
    V_d = \frac{L^2Nq}{2\varepsilon}.
\end{equation}
In this case, the electric field is non-zero only for a thickness $d < L$ as given by Eq. (\ref{eq:E_undepl_depl}). The undepleted region behaves like a high-resistivity conductor which gives rise to interesting time-dependent behaviour during charge transport (see Sec. Sec. \ref{sec:time_dependent_weighting_field}), but can otherwise be considered an extension of the electrode. More importantly, however, the weighting field is not $|\nabla W| = 1/L$ as the full bias potential drop occurs over a smaller thickness. Specifically, Eq. (\ref{eq:I_SR}) is valid only for a fully depleted detector, i.e. when the material between the electrodes is linear. The generalization is usually written as Gunn's theorem, which states that \cite{Gunn1964, Vittone2004, Hamel2008}
\begin{equation}
    I_k^{\rm Gunn} = q \bm{v}_d \cdot \left.\frac{\partial \bm{E}}{\partial V}\right\vert_{V_{\rm op}^k},
    \label{eq:I_Gunns}
\end{equation}
where $V_{\rm op}^k$ is the actual operating voltage of electrode $k$. The partial derivative of the electric field with respect to the operating voltage takes the place of the weighting field and is valid even for non-linear media. 

While detectors are typically run overdepleted (i.e. $V > V_d$), the electric fields inside the junction window and in p-stop and p-spray regions can be treated correctly only using Eq. (\ref{eq:I_Gunns}). For a partially depleted parallel plate detector, Gunn's theorem can be applied analytically and correctly reproduces the weighting field $\partial \bm{E} / \partial V = (1/d) \hat{\bm{z}}$, where $d \leq L$ is the depletion thickness. For more complex geometries, weighting fields must be calculated numerically as discussed below.

\subsection{Pixel weighting potential}
\label{sec:pixel_weighting_potential}
For pixels of finite spatial extent, the weighting potential will deviate from the parallel plate result, $W(z) = z/L$, even in the case of full depletion. In particular, edge effects will strongly distort the weighting potential as it must accommodate the sudden change in boundary conditions at the detector back face from the main pixel ($W_k = 1$) to an adjacent one ($W_k = 0$). Throughout the detector, the weighting potential for a pixel will be non-zero even outside of its canonical volume so that an induced current appears across all adjacent pixels. In order to study this effect, we present an analytical result for circular pixels and afterwards discuss numerical results for hexagons.

\subsubsection{Analytical approximation}
\label{sec:pixel_weighting_analytical}
The behaviour of the weighting potential close to the edge of a pixel changes dramatically for small changes in displacement. As such, numerical methods are typically employed to perform the standard Shockley-Ramo procedure (see Eq. (\ref{eq:I_SR})). These are typically computationally costly, however, and here we present a closed-form analytical result that is able to capture most of the behaviour.

For this scenario, we use a simplified geometry and consider only a circular pixel of radius $R$ situated a distance $L$ away from a grounded electrode. We assume the circular pixel is surrounded completely by other electrodes on the back face, which for the purpose of the weighting potential calculation are set to ground together with the front face electrode. In other words, we impose Dirichlet boundary conditions on the weighting potential $W(\rho, z)$ so that $W(\rho, 0) = 0$, $W(S, z) = 0$ and $W(\rho, t) = V(\rho)$ where $S$ is the radius of the cylindrical shroud and $V(\rho) = 1$ when $\rho < R$ and 0 otherwise. Due to the cylindrical symmetry, the weighting potential may be expanded using only a lowest-order Bessel function \cite{Jackson1999}
\begin{equation}
    W(\rho, z) = \int_0^\infty J_0(k\rho) \sinh{kz}B_0(k),
\end{equation}
where we let $S \to \infty$ to remove effects from the finite shroud radius. Using the orthogonality of the Bessel functions we can determine the form of $B_0(k)$,
\begin{align}
    B_0(k) &= \frac{k}{\sinh{kL}}\int_0^\infty d\rho J_0(k\rho) V(\rho) \rho \nonumber \\
    &= \frac{R}{\sinh{kL}}J_1(kR).
\end{align}
 The weighting potential for a circular pixel is then
\begin{equation}
    W(\rho, z) = \int_0^\infty dk R \frac{\sinh{kz}}{\sinh{kL}}J_0(k\rho)J_1(kR).
    \label{eq:weighting_pot_circ_pixel}
\end{equation}
While there exists no analytical solution for this integral equation, it is straightforwardly to numerically integrate. When $kL \ll 1$, the integrand is proportional to $(R^2/L)$ so that the result will depend strongly on the pixel aspect ratio, $R/L$, as intuitively expected. Similarly, for high $k$ the ratio of hyperbolic sine functions approaches $\exp[k(z-L)]$, implying the potential depends strongly on $\rho/R$ as anticipated close to the edge. Using $J_1(x) =- d J_0(x)/dx$ and performing integration by parts we can write Eq. (\ref{eq:weighting_pot_circ_pixel}) as
\begin{equation}
    W(\rho, z) = \frac{z}{L} + \int_0^\infty dk J_0(kR)\frac{\partial}{\partial k}\left[\frac{\sinh kz}{\sinh kL}J_0(k\rho) \right].
\end{equation}
Taking either $\rho = 0$ or $R\to \infty$ it is then trivial to see that it reduces to the infinite parallel plate result.

\begin{figure}[ht]
    \centering
    \includegraphics[width=0.48\textwidth]{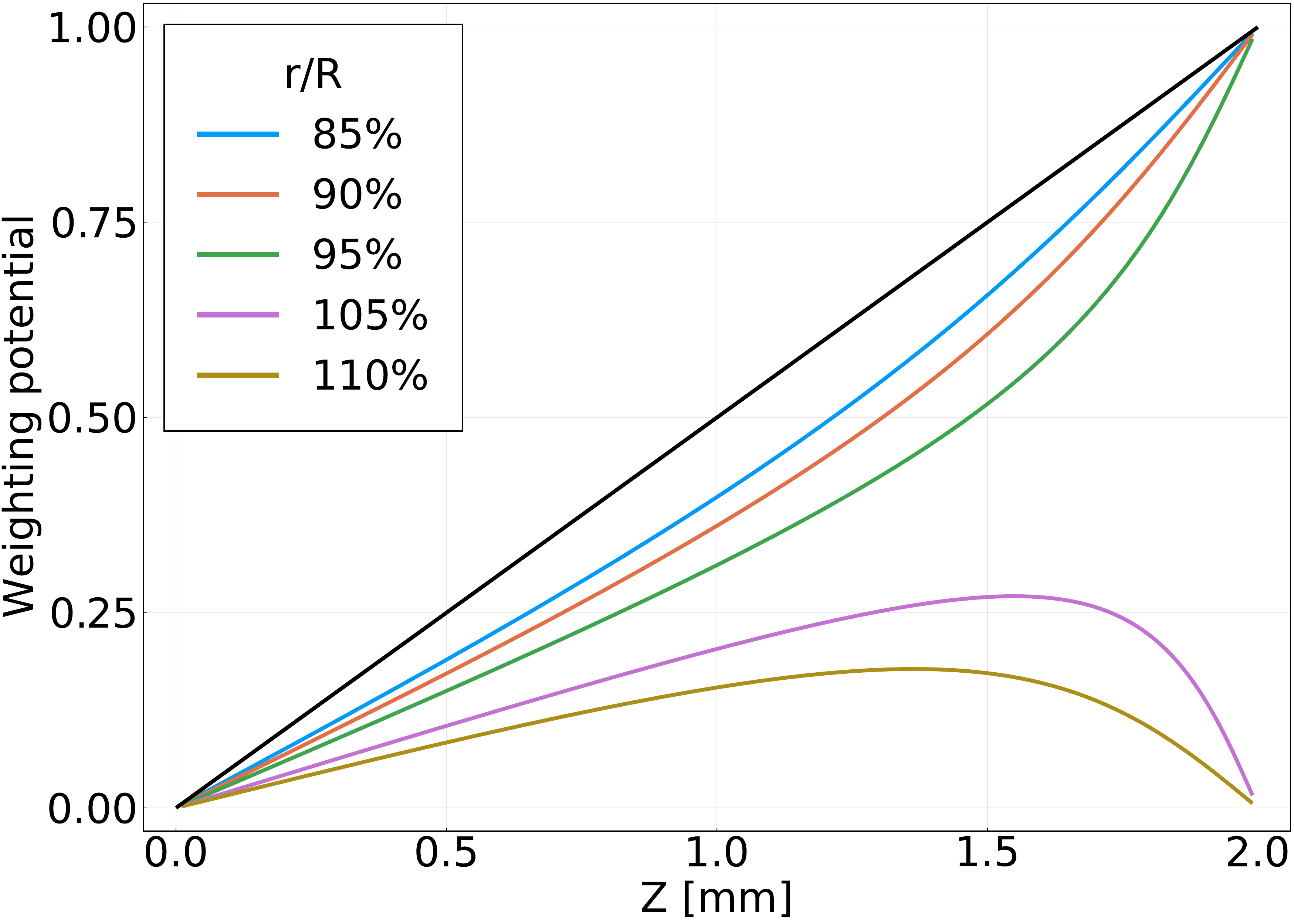}
    \caption{Analytical results for the static weighting potential close to the pixel boundary. The legend refers to radial distance relative to edge distance. Black line shows a perfectly linear relationship which applies near the center of the pixel. Inside the pixel, the weighting potential always reaches unity implying full charge collection, whereas it goes to zero outside of the pixel resulting in bipolar current pulses.}
    \label{fig:weighting_pixel_analytical}
\end{figure}

Results of  the numerical integration of Eq. (\ref{eq:weighting_pot_circ_pixel}) for $R = 5.15$ mm and $L = 2$ mm are shown in Fig. \ref{fig:weighting_pixel_analytical}. Close to the pixel boundary, the weighting potential differs significantly from the infinite parallel plate result, $W(z) = z/L$. Charge moving close to the pixel boundary will be collected more slowly during the first part of its transit and will increase to its total value more swiftly as it approaches the contact. Outside the pixel, the weighting potential is non-zero throughout the volume but vanishes for $z=L$ in accordance with the boundary condition. Note that the analytical result is valid also for underdepleted detector geometries, as one may consider the undepleted region to be simply an extension of the conductive contact. The correct result is then obtained simply by changing $L$ to correspond to the depletion thickness, $d < L$.

Following Eq. (\ref{eq:I_SR}), charge transport outside the canonical volume of the pixel will induce a bipolar current pulse typically denoted as differential cross-talk. In an idealized situation, the total charge collected on neighbouring pixels will resolve to zero, however, as it ends up on a neighbouring electrode. Finite pixel-to-pixel capacitances and charge sharing due to carrier diffusion will give rise to finite amounts of charge collected, however, and are known as integral cross-talk and charge sharing, respectively, and are discussed later (Sec. \ref{sec:chargeSharing}).

\subsubsection{Hexagonal simulation}
\label{subsec:SSD_sim}
In the Nab experiment, hexagonal pixels are employed due to a variety of benefits as mentioned in the introduction. While the analytical results of the previous section can be expected to work well for the flat edge of the hexagon, differences due to the sharp corners result in substantial changes. The latter requires the use of numerical solvers, for which we initially use the open-source \texttt{SolidStateDetectors.jl} package \cite{Abt2021}. This Julia package solves the Poisson equation on a rectangular grid with a user-specified grid size. Results in this section are obtained using a grid spacing of 0.05 mm, which is small enough to accurately probe the hexagonal geometry, but large enough to maintain good performance. Strong local potential changes due to pixel isolation (p-stop/p-spray) are not observable with this technique and instead we treat these in more detail in Sec. \ref{sec:pixel_isolation}.

\begin{figure}[ht]
    \centering
    \includegraphics[width=0.48\textwidth]{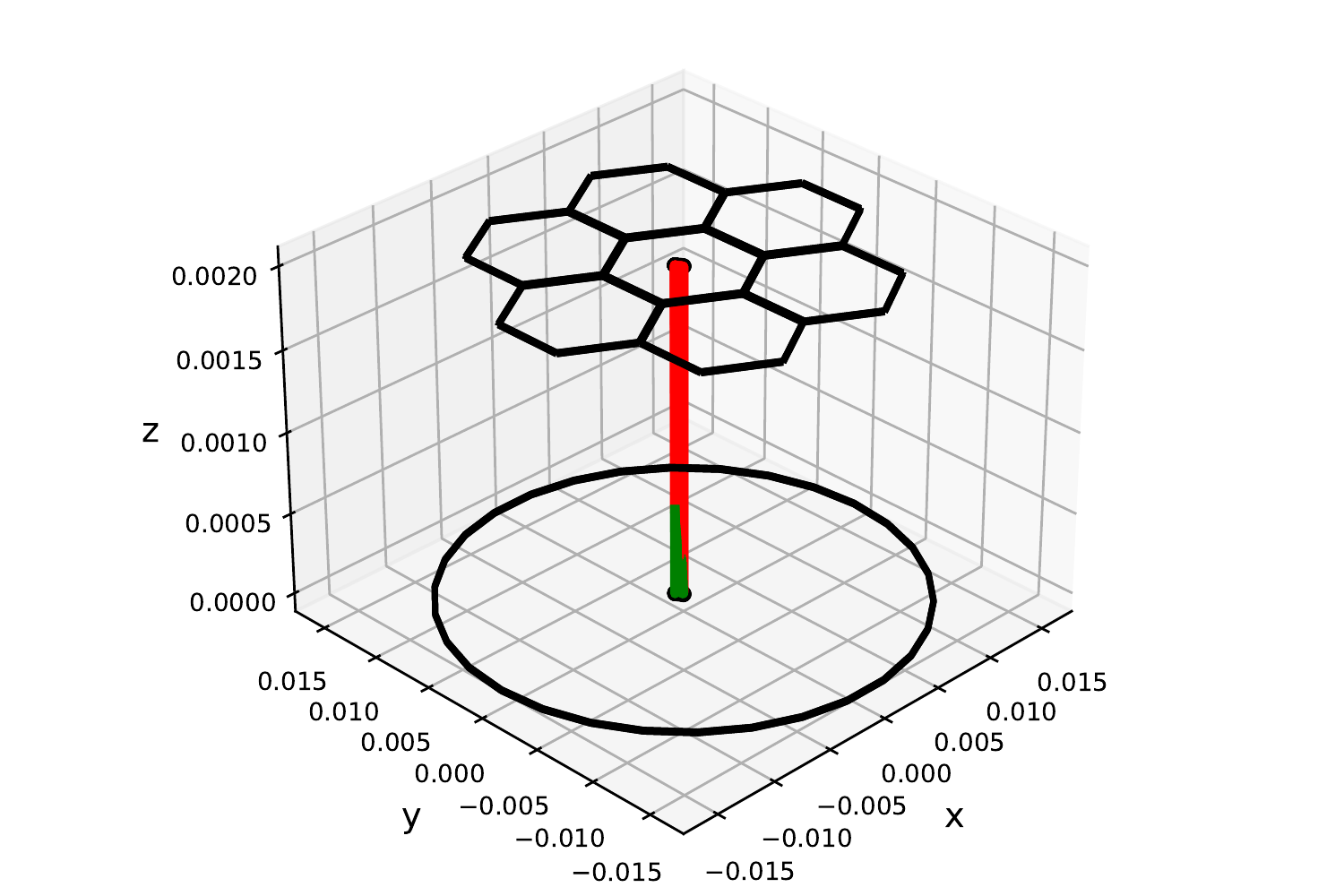}
    \caption{Drawing of contacts of a 7 pixel detector simulation. 500 keV electron event is shown with electron and hole drift paths in red and green, respectively.}
    \label{fig:7pixelEvent}
\end{figure}

We define a simplified detector geometry in \texttt{SolidStateDetectors.jl} with 7 hexagonal contacts arranged in a ring around a central contact, shown in Figure \ref{fig:7pixelEvent}. The hexagonal contacts are positioned 2 mm away from a single circular contact with a radius of 16.26 mm (see Fig. \ref{fig:detector_schematic}) and are spaced 0.1 $\mu$m from each other. This geometry is an accurate model for contacts with 6 neighboring contacts, representing all but the outer contacts in the Nab detectors.

\begin{figure}[ht]
    \centering
    \includegraphics[width=0.48\textwidth]{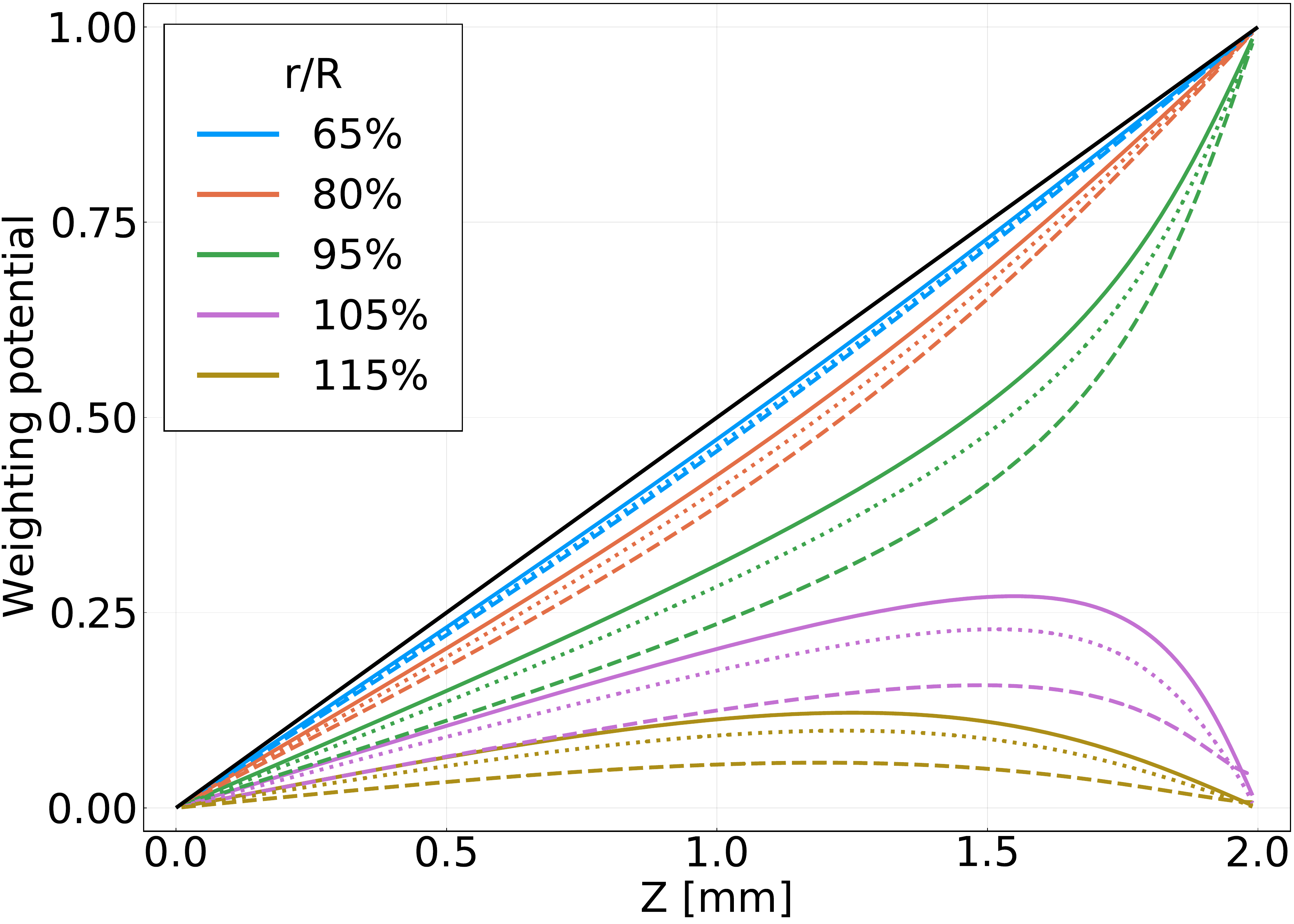}
    \caption{Simulation results for the weighting potential of a central hexagonal contact at the edge and corner compared to theoretical results from a circular contact. Represented by solid, dashed, and dotted lines respectively. Solid line represents perfectly linear weighting potential. Different colors show the relative distance from the contact boundary.}
    \label{fig:EdgeWP}
\end{figure}

Figure \ref{fig:EdgeWP} shows the weighting potential as a function of distance from the circular front face contact at different starting radii relative to the hexagonal geometry. As an example, for a hexagonal contact with a corner at $(x,y)=(5.15,0)$, 80\% refers to the weighting potential at (4.12,0) along the $z$-axis. The dashed and solid lines represent weighting potentials measured along the line extending from the center to the corner ($\theta=0$) and to the edge ($\theta=30$) of the contact respectively. The dotted line represents the analytical results from Equation \ref{eq:weighting_pot_circ_pixel} for $L=2$ mm and $R=5.15$ mm, where $L$ and $R$ are defined in Section \ref{sec:pixel_weighting_analytical}. At the center of the pixel the weighting potential changes linearly with $z$ analogously to the infinite parallel plate capacitor result. Away from the center the weighting potential is slightly nonlinear as seen with the 65\% and 80\% lines in blue and orange. As we cross the edge the potential becomes increasingly nonlinear, until we are under the grounded contact and the potential now returns to zero for $z =2$ mm. 

The analytical solution to the weighting potential for a circular pixel (Eq. (\ref{eq:weighting_pot_circ_pixel})) is in good agreement with that of a hexagon along its flat edge. Along a path towards a corner, however, differences appear when extending beyond 80\%. Therefore, about 70\% of the detector can be modeled well by a circular contact with appropriate parameter choices, whereas one must instead rely on numerical potentials near pixel edges. As the weighting potential determines the induced current on the contact from moving charges (see Eq. (\ref{eq:I_SR})), differences in the weighting potential will have a direct effect on the predicted pulse shape. We will investigate this in great detail in Sec. \ref{sec:RiseTimeDistributions}, but first consider the bulk and pixel isolation behaviour in the following sections.


\subsection{Bulk electric field}
\label{sec:electric_field}
The electric field in the detector causes the electrons and holes to drift to their respective contacts as discussed in Section \ref{sec:FieldSimulations}, which in turn induce a current in the electrodes according to Eq. (\ref{eq:I_Gunns}). The electric field is generally a combination of the applied bias voltage, the bulk impurity concentration and detector geometry as visible in Eq. (\ref{eq:E_undepl_depl}). As detailed in section \ref{subsection: DopingProfile}, the Nab detectors are made of high purity Si with impurity densities of $\mathcal{O}(10^{10}\mathrm{cm}^{-3})$. Both radial and longitudinal gradients affect the overall shape of the electric field throughout the entire crystal, however, and are not easily analytically tractable. 

Like in the previous section, we performed the numerical evaluation of electric fields with the \texttt{SolidStateDetectors.jl} package using the same contact design. We define radial gradients in the impurity density such that its value at any point in the bulk is $n=n_0+g_n\times r$, where $n_0$ is the base concentration and $g_n$ is the gradient.\footnote{The electric field results from the center pixel can be used to model any other interior pixel in the detector by stitching together the appropriate combination of gradients. For example, if $n=n_0+g_n\times |r|$ we can use $n_0-g_n\times r$ for $r=x<0$ and  $n_0+g_n\times r$ for $r=x>0$. In the same way we recreate any impurity density profile made of approximately linear segments.} While we are not sensitive to small-scale features below the grid size, we also define the p$^+$ window, the n$^+$ hexagonal pixels, and a p-spray layer as described in Section \ref{subsection: DopingProfile}.

\begin{figure}
    \centering
    \includegraphics[width=0.48\textwidth]{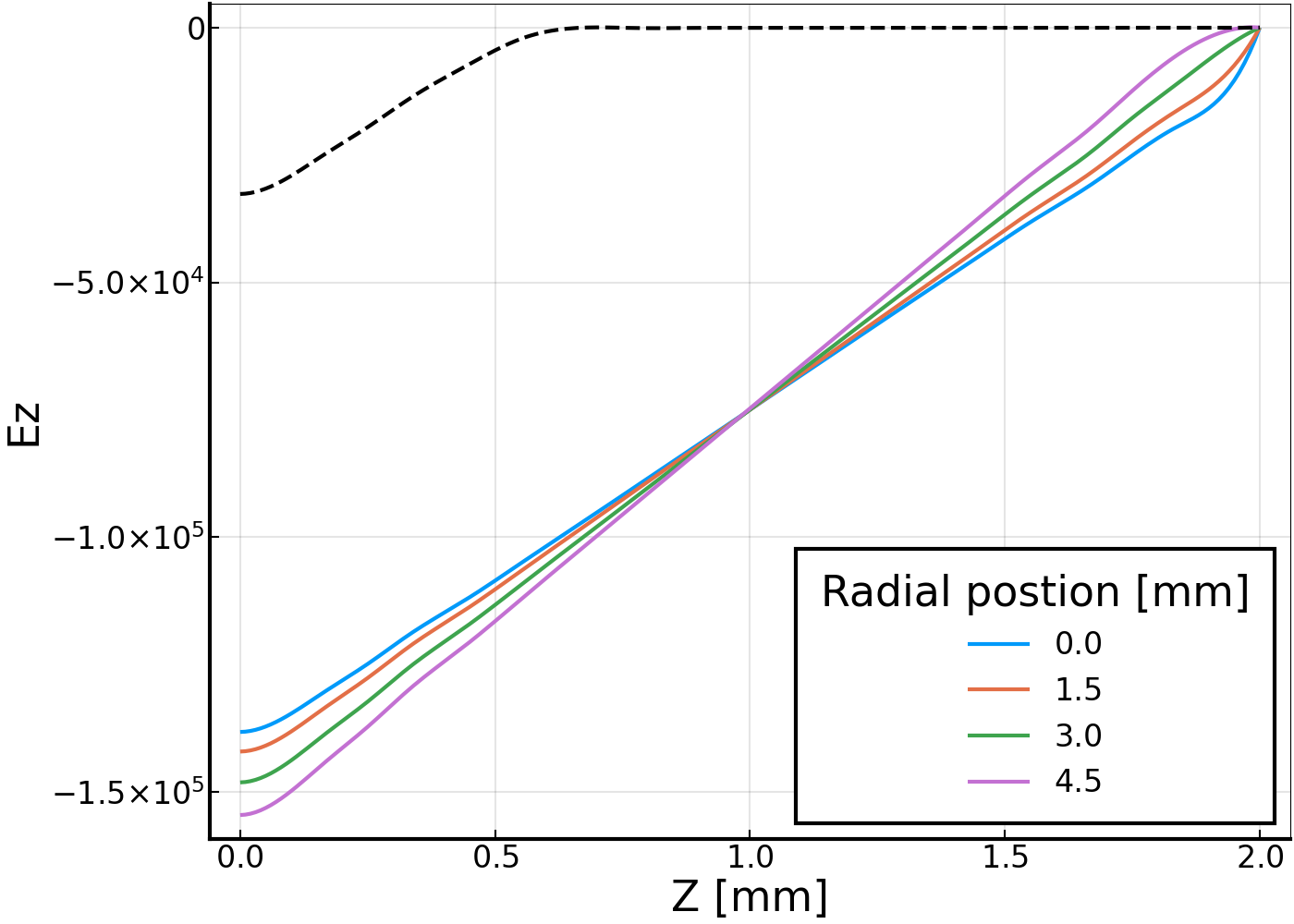}
    \caption{$z$ component of electric field (V/m) vs $z$ position (mm) at various radial positions. $4\times10^{10}$cm$^{-3}$ detector with $3\times10^{10}$cm$^{-4}$ gradient at 150V is shown with solid lines. The dotted line shows the results for an undepleted detector at 30V measured at the center of the pixel.}
    \label{fig:Efield_z}
\end{figure}

We simulate 7 pixels with radial impurity concentration gradients of $g_n=\pm(0,1, 2, 3) \times 10^{10} \mathrm{cm}^{-4}$. We simulate all these gradients for base concentrations of $n_0=\pm(1, 2, 3, 4, 5)10^{10}\times \mathrm{cm}^{-3}$. This combination covers a range of reasonable concentrations for large high-purity silicon detectors, as discussed in Section \ref{subsection: DopingProfile}. The depletion voltage for a $2$ mm detector is approximately $V_d\approx -30 n/ 1\times10^{11} $, and we simulate with a bias voltage of -30, -60, -90, -120, and -150 V to check for depletion and simulate signals for overdepleted detectors. The Nab detectors will typically be operated above depletion voltage, but impurity density gradients can cause regions of the detector to be undepleted. Depletion can be directly checked with \texttt{SolidStateDetectors.jl} by looking at the fields versus the $z$ direction. Undepleted sections of the detector still have mobile charges and so the material acts as a poor conductor. The electric field will therefore be zero in the undepleted region as seen in Figure \ref{fig:Efield_z}. The field for an undepleted detector is shown by a black dotted line and depleted detector fields are shown in solid colored lines. We see in this figure that the undepleted detector field reaches 0 V/m at about 0.5 mm so three quarters of the detector is undepleted.

When a radial impurity density gradient is present in the detector the electric field will have a radial dependence. The $z$ component of the electric field, for $g_n=3\times10^{10} \mathrm{cm}^{-4}$, at different radii is shown in Figure \ref{fig:Efield_z}. If there were no gradient there would be just the blue line, which is the center of the pixel. As the impurity concentration increases radially so does the maximum field strength and the slope of the fields. Near the junction contact all fields have a positive second derivative, but near the grounded contact the second derivative is more positive for smaller $r$ (the more over-depleted regions) and is slightly negative at $r=4.6$mm indicating that a small portion of the detector is undepleted. All the fields are zero at $2$ mm as that is inside the grounded contact in our simulation.

\begin{figure}
    \includegraphics[width=0.85\textwidth, left]{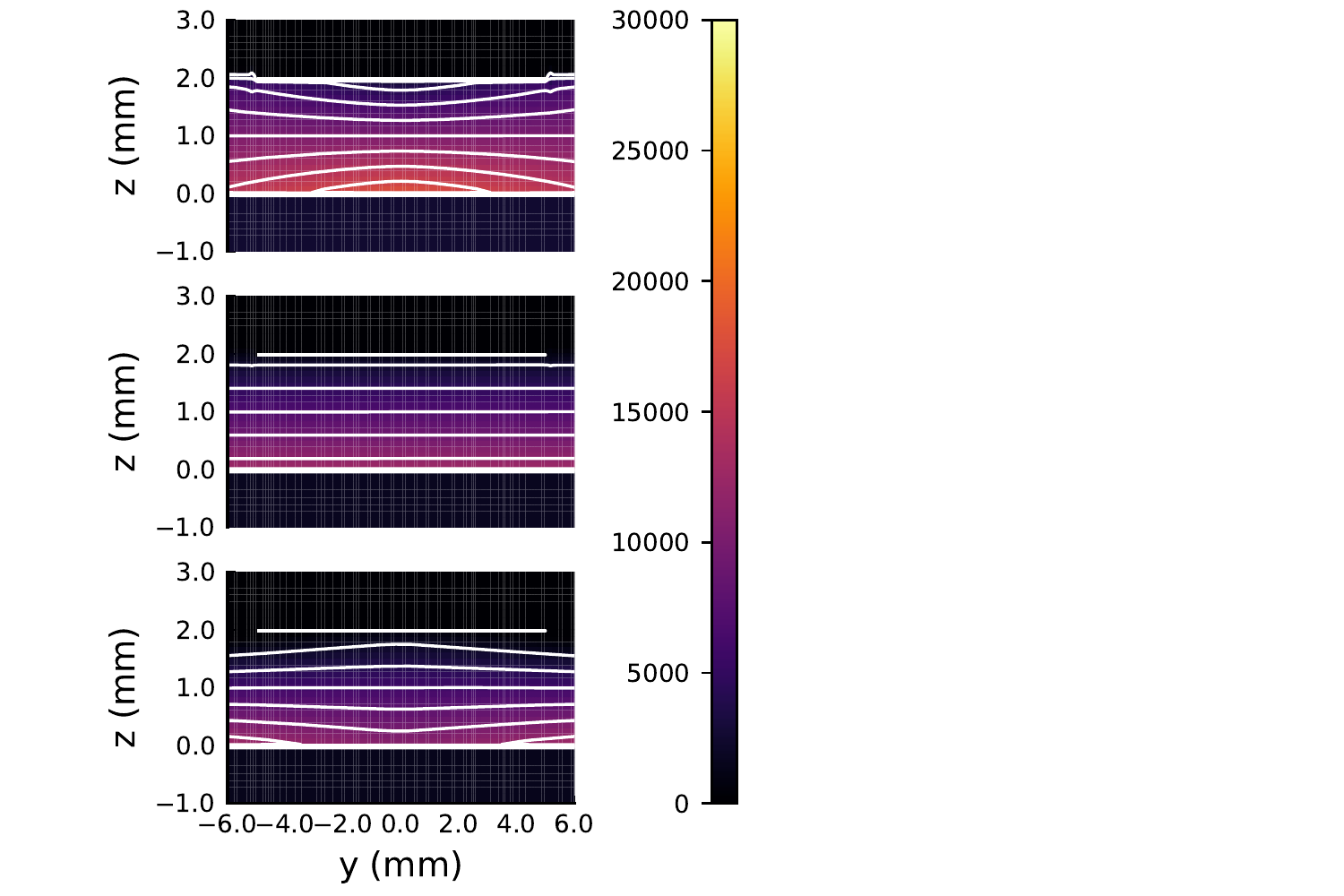}
    \caption{Electric field magnitude and equipotential lines for a $n_0 = 4\times10^{10}$cm$^{-3}$ concentration detector with $g_n = -3\times10^{10}$, 0, and $3\times10^{10}$ cm$^{-4}$ radial gradients from top to bottom. Color scale is in units of V/m.}
    \label{fig:fields}
\end{figure}

The total strength of the electric field and equipotential lines are shown in Figure \ref{fig:fields} for a detector with no impurity density gradient and for a positive and negative gradient. We see that the potential expands and contracts as the impurity concentration changes in the detector. Some edge effect are visible in the bulk of the detector for the negative gradient results, but not for the positive and no gradient fields; we will see those features more clearly in the next section. The differences in the fields seen in Figure \ref{fig:fields} have two effects: ($i$) the field magnitude is radially dependent which directly affects the magnitude of the electron drift velocity; ($ii$) the field lines curve towards or away from the center for negative and positive gradients, respectively. Both the varying field strengths and the field anisotropy will affect the drift velocity
. The bulk effects along with the weighting potentials found in the previous section will be used to investigate pulse shapes in Section \ref{sec:RiseTimeDistributions}. Small scale features of the electric field and weighting potential will be discussed in Section \ref{sec:pixel_isolation}.

\subsection{Pixel isolation simulation}
\label{sec:pixel_isolation}
While the previous section discussed large-scale behaviour of electric fields, substantial changes are expected to occur near pixel boundaries. Here, individual pixels are electrically isolated by manufacturing small depletion zones on pixel boundaries (see Sec. \ref{subsection: DopingProfile}). These local impurity depositions create significant distortions to the local electric field while keeping the bulk largely unchanged, and as such leave most of the charge carrier transport stable. Properties such as charge sharing (Sec. \ref{sec:chargeSharing}) and pulse shape close to the pixel separation (Sec. \ref{sec:pixel_weighting_potential}), on the other hand, depend critically on the local properties and feed into event reconstruction efficiency and timing extraction.

\subsubsection{Geometry}

We simulate a small cross section of the experimental geometry to focus on the pixel isolation features. Simulations are performed in COMSOL v5.2 using the Semiconductor module as a 2D geometry, with the third dimension along the inter-pixel separation. The bulk doping, p$^+$ junction implant and n$^+$ Ohmic contacts are implemented as described in Sec. \ref{sec:detector_geometry}, with a concentration of $1\times 10^{16}$ cm$^{-3}$ and extending 1 $\mu$m into the bulk using an error function fall-off. We implement an oxide charge layer of 10 nm thickness as observed in the SIMS data of Sec. \ref{subsection: DopingProfile} with a a static charge distribution of $10^{11}$ cm$^{-2}$ \cite{Richter1996}. For the pixel isolation technology, we consider both p-stop and p-spray geometries. For the former, we assume a surface implantation with a 50 $\mu$m width (i.e. covering half of the inter-pixel gap surface), 1$\times 10^{17}$ cm$^{-3}$ impurity concentration and an error function depth profile determined by a 100\,nm fall-off\footnote{These values are approximate and based on common literature values. Sensitivity due to deviations are discussed in the text.}. For p-spray, the deposit spans the entire width of the pixel separation and we assume an order of magnitude lower in impurity density with similar depth profiles. In all cases, the Poisson and drift-diffusion equations (see Eqs. (\ref{eq:drift_diffusion})) are solved at the same time, such that local carrier densities can be extracted. Due to the smallness of the region in the plane of the detector, we neglect any impurity density gradients in the bulk.

Boundary conditions (BC) are split up between metal contacts and other physical boundaries. All contacts are defined using Dirichlet BC, with both pixel contacts set to ground potential and the front face using a negative bias voltage. All other physical boundaries are set using von Neumann BC. Whereas the latter is evident for the boundaries of the bulk silicon on the sides, the choice is less trivial for the region between the metal contacts on the back. Specifically, significant differences between von Neumann and Dirichlet BC on the interface region were observed \cite{Richter1996}. For a clean, unirradiated surface a von Neumann BC, i.e. $d \phi / dx = 0$ for an electrostatic potential $\phi$, should hold, whereas any form of moisture or contamination can form a conductive channel between the metal contacts or build up a static potential difference. As the Nab detectors will not experience significant irradiation over their lifetime compared to running at, e.g., the Large Hadron Collider \cite{Richter1996, Gorelov2002}, we apply von Neumann BC on the insulating boundary.

The grid is a finite element mesh constructed using COMSOL and set to be very fine near the contacts where large differences in doping are present and less dense in the intermediate regions. The bias voltage is swept from zero to twice the depletion voltage, where the result of the previous calculation is used as a starting point for faster convergence. For each applied bias voltage on the bottom contact, the calculation is run several times with small differences to the voltage applied to individual contacts. Taking the bottom contact as an example, calculations at a bias voltage $V_{\rm bias}$ are performed twice with a small difference $\Delta V$ between them. The two data sets are used to calculate the weighting field numerically via Gunn's theorem (Eq. (\ref{eq:I_Gunns})), using the extracted electric fields at both voltages \cite{Riegler2019}. The procedure is repeated for one of the two contacts at the back, as both are symmetrically placed in the geometry.

\subsubsection{Results}

\begin{figure*}[ht]
    \centering
    \includegraphics[width=\textwidth]{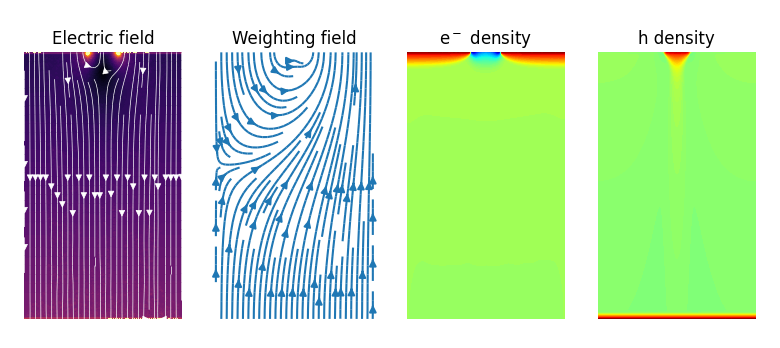}
    \caption{Overview of the local properties around a pixel boundary with a p-spray isolation. Shown, from left to right, are the local electric field in logarithmic scale, the weighting field for the pixel to the right of the boundary, and free quasiparticle electron and hole densities close to the depletion voltage in logarithmic color scale. For the latter, red areas indicate high density compared to blue areas for a density deficiency.}
    \label{fig:overview_pspray_COMSOL}
\end{figure*}

An example of the electric and weighting fields with the corresponding electron and hole densities can be found in Fig. \ref{fig:overview_pspray_COMSOL} for a p-spray configuration. The electric field at the physical boundaries of the electrodes is larger than that of the bulk by several orders of magnitude, as expected from the sudden change in impurity density. In the region around the pixel isolation, however, the electric field magnitude drops significantly over the full width of the pixel separation and extending several tens of $\mu$m into the crystal. This is the result of the large impurity deposition and free charge carriers diffusing into the bulk material. As a consequence, the simulation results for the p-stop configuration is very similar and has analogous consequences for the electric field shape and corresponding charge carrier movement.

The profile of the electric field is such that the local minimum in electric field strength is surrounded by what can be described as a `protective breakwater'. This can be understood by analysing the electric field lines, which in this configuration are pushed away from the pixel isolation and channel charge carriers to either pixel rather than the low field region in the isthmus. This will be a critical component of the waveform simulation and charge sharing analysis in Sec. \ref{sec:chargeSharing}.

The weighting field (see Eq. (\ref{eq:I_Gunns})) shows the large-scale anticipated behaviour. The change in direction when moving along a straight path for the left pixel implies the expected bipolar current structure, while the weighting field well inside the right pixel is a constant according to the analytical results of Eq. (\ref{eq:E_undepl_depl}). Following a path along the geometrical pixel isolation, however, reveals complicated dynamics as the weighting field becomes progressively more perpendicular to the electric field, implying that little charge is collected on either electrode despite moving closer to the pixels. As the total line integral must add to unity for a charge carrier collected on a pixel, this implies large, sudden changes to the pulse shape close to the collecting electrode. This can throw off the timing reconstruction and will be discussed in greater depth in Sec. \ref{sec:chargeSharing}.

The sensitivity of this behaviour to the impurity density and geometrical shape has been studied qualitatively and is found to be of little significance for proton and electron detection in the Nab experiment. This can be understood intuitively in a similar fashion to the simple p-n junction, where the spatial extent of the depletion zone into the heavily doped volume depends more upon its implantation profile than its density so long as the latter is significantly larger than for its junction partner. Whereas the `dead layer' profile is important for the Nab experiment due to the extremely local energy deposition of incident protons, the isolation structures at the back of the detector are irradiated only for background events due to low energy gamma or X-rays and Compton scatters from nearby materials.

Finally, we comment on the introduction of additional capacitance due to the pixel isolation structures. By adding additional depletion zones across all pixel boundaries, pixels are now explicitly capacitively coupled to all of their neighbours. As such, when signals are generated through charge collection on any one pixel, the total charge collection on any of its neighbours will not resolve to zero but instead be proportional to the mutual capacitance. This additional capacitance appears on top of the usual capacitive coupling when putting two conductors in close proximity, and contributes to so-called integral cross-talk \cite{Leviner2014}. Using the available 2D geometry, we then obtain a mutual capacitance along the pixel boundary of 0.5 pF/mm. For hexagonal pixels with ~10mm outer diameter, a naive estimate results in a mutual capacitance of 2.5 pF with each of a pixel's six neighbours. When compared to the simplified geometrical capacitance of a planar pixel ($C_{p0} = $ 3.8 pF for a 2mm thick detector), this represents a substantial increase in the total capacitance before connection the amplifier system.

\subsection{Time-dependent effects in underdepleted geometries}
\label{sec:time_dependent_weighting_field}
We have used Gunn's theorem (Eq. (\ref{eq:I_Gunns})) to derive the correct weighting potential for an undepleted detector and, more interestingly, the behaviour near the pixel isolation technology in the previous section. We can go one step further by taking into account the time-dependent response of the undepleted layer. Specifically, as this layer has some finite conductivity, it will respond to changes in the electric field due to moving charges in a time-dependent fashion. The movement of charges in the undepleted layer will equally affect the induced charges in the electrode, resulting in a delayed enhancement of the induced charge. Riegler \cite{Riegler2002, Riegler2004, Riegler2019} has treated this in some detail and introduces a time-dependent weighting field so that
\begin{equation}
    I_k = q \int_0^t \bm{v}_d(t)\cdot \bm{W}(x, t') dt'.
\end{equation}
where $W(x, t)$ now depends explicitly on time. For a simple parallel plate geometry of thickness $L$ and depletion thickness $d < L$, the weighting field can be derived as \cite{Riegler2019}
\begin{equation}
W(t) =
\left\{ \begin{array}{lr}
     \dfrac{1}{d}\left(\delta(t) + \dfrac{L-d}{d}\dfrac{1}{\tau} e^{-t/\tau} \right) & x < d \\
    \dfrac{1}{d}\left(\delta(t)-\dfrac{1}{\tau}e^{-t/\tau}\right) & d < x < L
\end{array}
\right.
    \label{eq:W_t_simple}
\end{equation}
where $\tau$ is a characteristic response time of the undepleted medium,
\begin{equation}
    \tau = \frac{\epsilon L}{d \sigma},
    \label{eq:t_undepleted}
\end{equation}
where $\sigma = q \mu n$. The delta function in Eq. (\ref{eq:W_t_simple}) represents the instantaneous induced current due to the charge movement inside the active region, whereas the second term is the reaction of the undepleted medium. The latter is positive for charges moving in the depleted volume, so that the charge movement in the undepleted material can be likened to an inertial effect.

\begin{figure}[ht]
    \centering
    \includegraphics[width=0.48\textwidth]{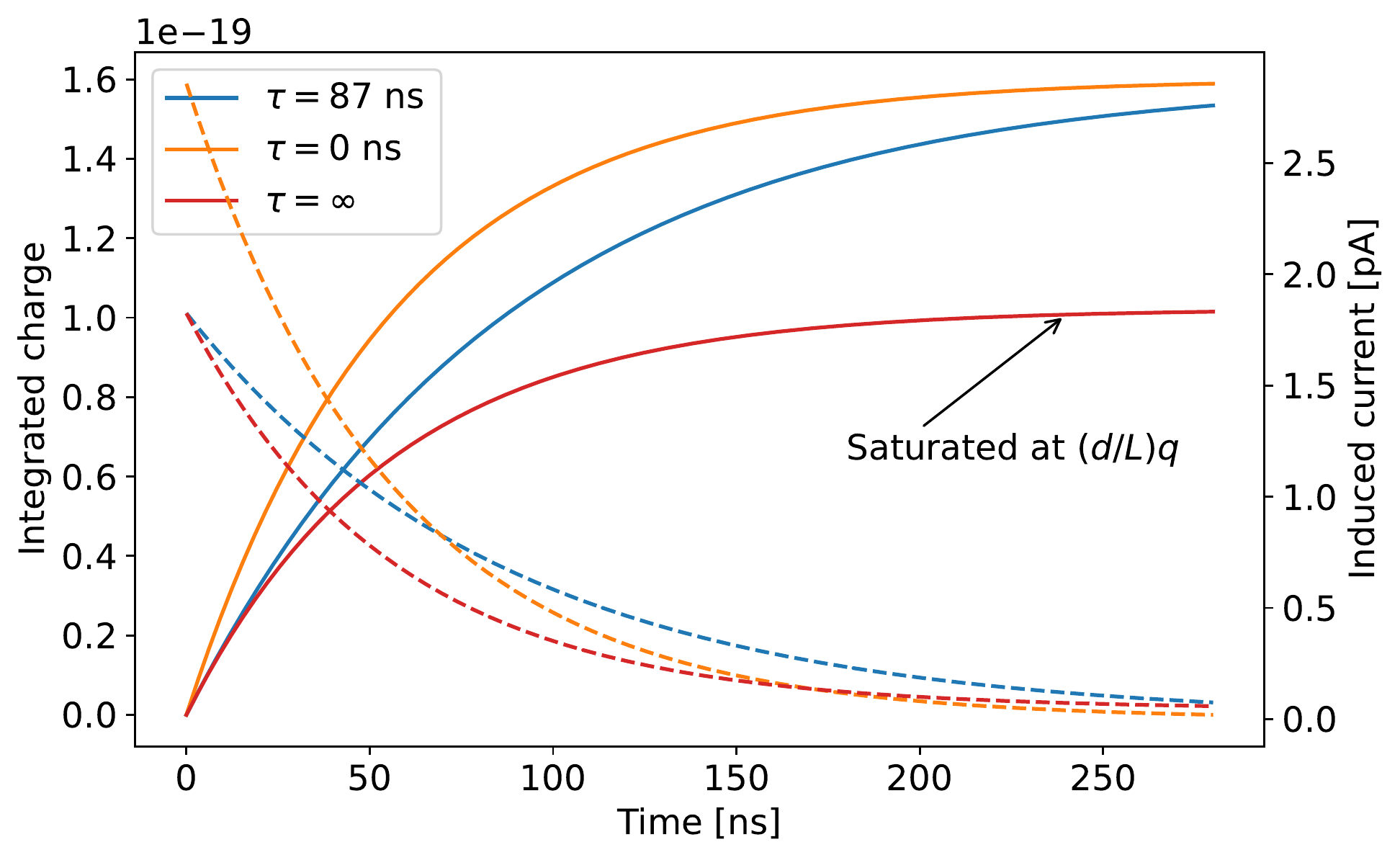}
    \caption{Integrated charge (solid) and current (dashed) for a single electron moving in a silicon device with thickness $2$ mm, impurity density of $4\times 10^{10}$ cm$^{-3}$ and applied bias voltage $V = 50$ V.}
    \label{fig:time_dependent_I_Q}
\end{figure}

The characteristic timescale of Eq. (\ref{eq:t_undepleted}) is not negligible relative to the transit time in the Nab detectors due to the increased mobility at low temperatures. Specifically, taking an operating temperature of $T = 120$ K and an impurity density concentration of $1\times 10^{10}$ cm$^{-3}$, one finds $\epsilon/\sigma \approx 50$ ns. The latter is comparable to the transit time of electrons starting at the front face of the detector and as such presents a significant difference in the predicted induced charge. Figure \ref{fig:time_dependent_I_Q} shows the effect on the induced charge as a function of time for different approximations of $\tau$. The static behaviour of Gunn's theorem with Eq. (\ref{eq:E_undepl_depl}) corresponds to $\tau = 0$, i.e. the undepleted region is a conductor with infinite conductivity. In cases where $\tau \gg \tau_e$, on the other hand, the response of the undepleted medium is extremely slow and finite integration times in the (pre)amplifying system will introduce a ballistic deficit and the pulse saturates at $(d/L)q$. Note that the time for the integrated charge to go from 10\% to 90\% of its maximal value is always longer when underdepleted medium effects are taken into account.

The parallel plate capacitor result of Eq. (\ref{eq:W_t_simple}) can be expected to hold throughout most of the detector volume, with the exception of the pixel isolation structures discussed in the previous section. As such, radial gradients in the impurity density concentration (see Sec. \ref{subsection: DopingProfile}) can potentially be studied by looking at the time response for individual pixels. We investigate this in greater detail in Sec. \ref{sec:analytical_baseline_pulse}.

\section{Carrier transport simulation}
\label{sec:carrier_transport_sim}
As carriers are created from ionization events along the track of an incident particle, their close proximity to neighbours will cause the charge cloud to expand due to electrostatic repulsion. Whereas the latter matters only when densities are high, additional broadening of the charge cloud occurs due to random thermal diffusion. Both effects will cause time-dependent perturbations to the charge transport process and the observed pulse shapes. Using the machinery developed for these effects, we additionally perform a study of the quasiparticle transport and charge collection efficiency in the entrance window.

\subsection{Thermal diffusion}
\label{sec:diffusion_sim}
\subsubsection{Analytical results}
\label{sec:diffusion_analytical}
In the assumption of isotropic diffusion, we may approximate the charge cloud as a Gaussian distribution rather than a point charge, where the charge distribution is then
\begin{equation}
    \rho(\bm{r}, t) = \frac{Q}{(4\pi Dt)^{3/2}}\exp\left(-\frac{(\bm{r}-\bm{r}_0(t))^2}{4Dt} \right)
    \label{eq:diffusion_concentration}
\end{equation}
centered at a position $\bm{r}_0(t)$ with the previously-defined diffusion constant, $D$ (Eq. (\ref{eq:einstein_diffusion})). In this case, we must modify the induced current relationship (Eq. (\ref{eq:I_SR}) and (\ref{eq:I_Gunns})) to integrate over the full charge carrier volume,
\begin{equation}
    I(t) =  \int d\bm{r} \rho(\bm{r}, t) \bm{v}(\bm{r}, t) \cdot \bm{W}(\bm{r}).
    \label{eq:I_charge_cloud}
\end{equation}


Upon some simplifying assumptions one may derive analytical results for the expected induced current and integrated charge \cite{Ruch1968}. Specifically, if one considers only the average motion of the charge cloud the results becomes insensitive to velocity variations from individual charge carriers but retains the effects of broadening. In this case, the current will decrease gradually as the charge cloud reaches the far electrode rather than stopping abruptly. In the case of a constant electric field, Ref. \cite{Ruch1968} derived an expressions for the time it takes for the current to drop from 95\% to 5\% of its maximal value,
\begin{equation}
    \tau_\mathrm{diff} = 2.32 \frac{\sqrt{4D(x_0+L)}}{v^{3/2}}
    \label{eq:tau_diff}
\end{equation}
where $x_0$ is the initial starting position of the charge cloud. Setting $E = 1$ kV/cm at $T=120$ K Eq. (\ref{eq:tau_diff}) results in $\tau_\mathrm{diff} = 2$ns. While the results of Ref. \cite{Ruch1968} were valid only for a constant electric field, we may generalize the result by solving
\begin{equation}
    \langle I\rangle (t) = \frac{Q v(t)}{\sqrt{4\pi D t}L}\int_0^L d\bm{r} \exp \left(-\frac{(\bm{r}-\bm{r}_0(t))^2}{4Dt} \right)
\end{equation}
for a detector thickness $L$, to find
\begin{equation}
    \langle I \rangle (t) = \frac{Qv(t)}{2L}\left[\mathrm{erf}\left(\frac{L-r_0(t)}{\sqrt{2Dt}}\right)+\mathrm{erf}\left(\frac{r_0(t)}{\sqrt{2Dt}}\right) \right],
    \label{eq:I_diffusion_erf}
\end{equation}
where $\mathrm{erf}$ is the error function. The average velocity and position can now be solutions to arbitrary field configurations, such as those for a linearly decreasing electric field (see Eq. (\ref{eq:x_v_linear_E})).

Whereas the average behaviour of the charge cloud gives rise to broadening features in the induced current, velocity fluctuations due to the thermal diffusion of individual charge carriers cause additional noise. The latter has been treated in depth in several works \cite{Reggiani1985}, but analytical results are available only when assuming a white spectrum. In that case, the noise current spectral density is found to be
\begin{equation}
    S_I(0) = \frac{4e^2N}{L^2}D
\end{equation}
where $N$ is the number of charge carriers. Similarly, one can define an equivalent noise temperature for quasiparticles \cite{Zimmermann1977}
\begin{equation}
    T_n = \frac{eD}{k_B\mu}
\end{equation}
that is dependent on the applied electric field and equilibrium temperature through both the mobility and diffusion constant. For fields larger than about 1 V/cm the noise temperature increases significantly (hence the name \textit{hot} electrons commonly used for charge carrier transport), and measurements performed at 77 K show good agreement with a parametrization \cite{Takagi1977}
\begin{equation}
    T_n = T_0(1+\beta E^2)
\end{equation}
where $T_0$ is the lattice temperature, and one find good agreement with data for $\beta = 2.5\cdot 10^{-7}$ cm$^2$ V$^{-2}$ for fields up to 10 kV/cm.

\subsubsection{Monte Carlo simulation}
\label{sec:diffusion_MC}
The discussion above was valid only for the average charge cloud behaviour and simple electric fields. In order to more generally describe the charge transport process, we create a standalone simulation using custom electric and weighting fields and perform a step-by-step simulation of charge carriers following the procedure of Ref. \cite{Brigida2004}. The equation of motion is integrated using the Runge-Kutta method, where the user specifies the simulation granularity using a parameter $\varepsilon$ used to control the time step, $\delta t$, using
\begin{equation}
    \delta t = \frac{\varepsilon}{|\bm{v}[\bm{r}(t)]|}
\end{equation}
where $\bm{v}[\bm{r}(t)] = \mu \bm{E}[\bm{r}(t)]$ is the velocity of the charge carrier at position $\bm{r}(t)$. A step, $\delta \bm{r}$, then consists of a drift component due to the electric field, $\delta \bm{r}_E = \delta t \bm{v}$, and a diffusion component
\begin{equation}
    \delta \bm{r}_D = \left(\begin{array}{ccc}
        c_\varphi c_\theta & -s_\varphi & -c_\varphi s_\theta \\
        s_\varphi c_\theta & c_\varphi & -s_\varphi s_\theta \\
        s_\theta & 0 & c_\theta
    \end{array} \right) \left(\begin{array}{c}
         \sigma_1 \\
         \sigma_2 \\
         \sigma_3
    \end{array} \right)
    \label{eq:diffusion_explicit_step}
\end{equation}
where $s(c)$ denotes the (co)sine and $\sigma_i$ is a random value chosen from a Gaussian distribution centered around 0 and standard deviation $\sigma = \sqrt{2D \delta t}$, with $D$ the diffusion constant\footnote{While this simulation is performed in two dimensions, the generalization of Eq. (\ref{eq:diffusion_explicit_step}) to three dimensions is trivial but does not influence our results.}. The induced current for charge carrier $k$ is evaluated using Gunn's theorem (Eq. (\ref{eq:I_Gunns})) using custom electric and weighting fields.

\begin{figure}[ht]
    \centering
    \includegraphics[width=0.48\textwidth]{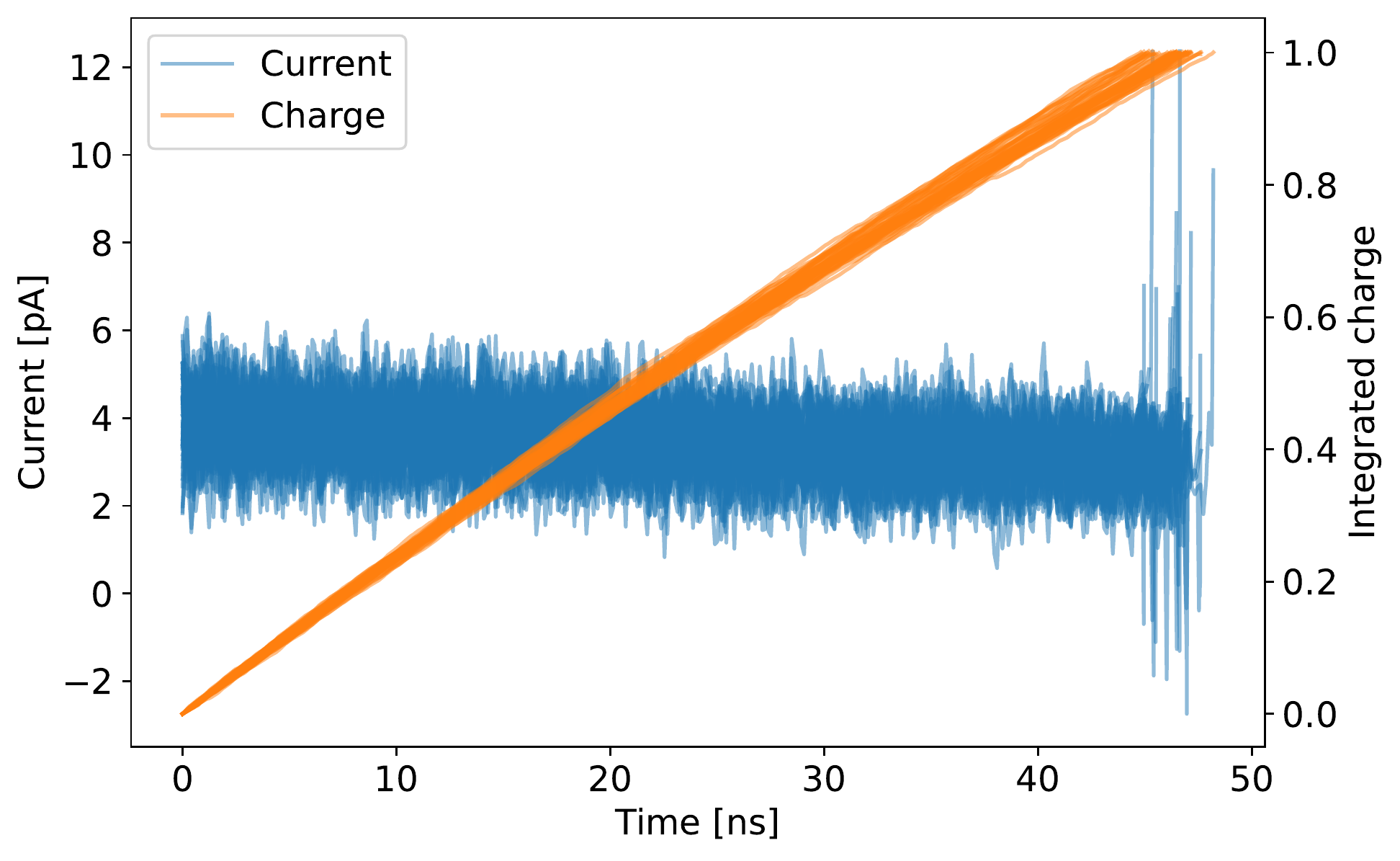}
    \caption{Induced current and integrated charge following charge carrier transport inside a linearly varying electric field with explicit diffusion as in Eq. (\ref{eq:diffusion_explicit_step}). The electric field decreases towards the end of the track, resulting in reduced average current and parabolic behaviour in the integrated charge. Large variations near the end of the collection time arise due to the small number of remaining charge carriers.}
    \label{fig:diffusion_charge_current}
\end{figure}

As a simple example, we first consider transport inside a linearly varying field. Parameters are set to similar conditions as the Nab experiment, using a parallel plate geometry. The weighting field is simply taken to be a constant (see Sec. \ref{sec:pixel_weighting_potential}), resulting in Figure \ref{fig:diffusion_charge_current}.

Figure \ref{fig:diffusion_charge_current} shows an example of the induced current and integrated charge for a large number of individual charge carriers moving in a linearly varying electric field. The random walk process introduces an additional noise source in the induced current and total transit time, similar to Eq. (\ref{eq:tau_diff}). The transit time distribution was studied for both a linearly varying and constant electric field. No significant difference is observed in the width of the distribution. Additionally, even though the width of the individual carrier arrival time distribution can exceed the nanosecond level, the total signal timing uncertainty (being the sum of the all individual carriers) is reduced by a factor $\sqrt{N}$, where $N$ is the total number of charge carriers for a single event. In the case of a 30 keV proton impinging upon a silicon detector, the resulting timing uncertainty due to carrier diffusion is reduced by almost a factor 100 and is rendered negligible. 

\begin{figure}[ht]
    \centering
    \includegraphics[width=0.48\textwidth]{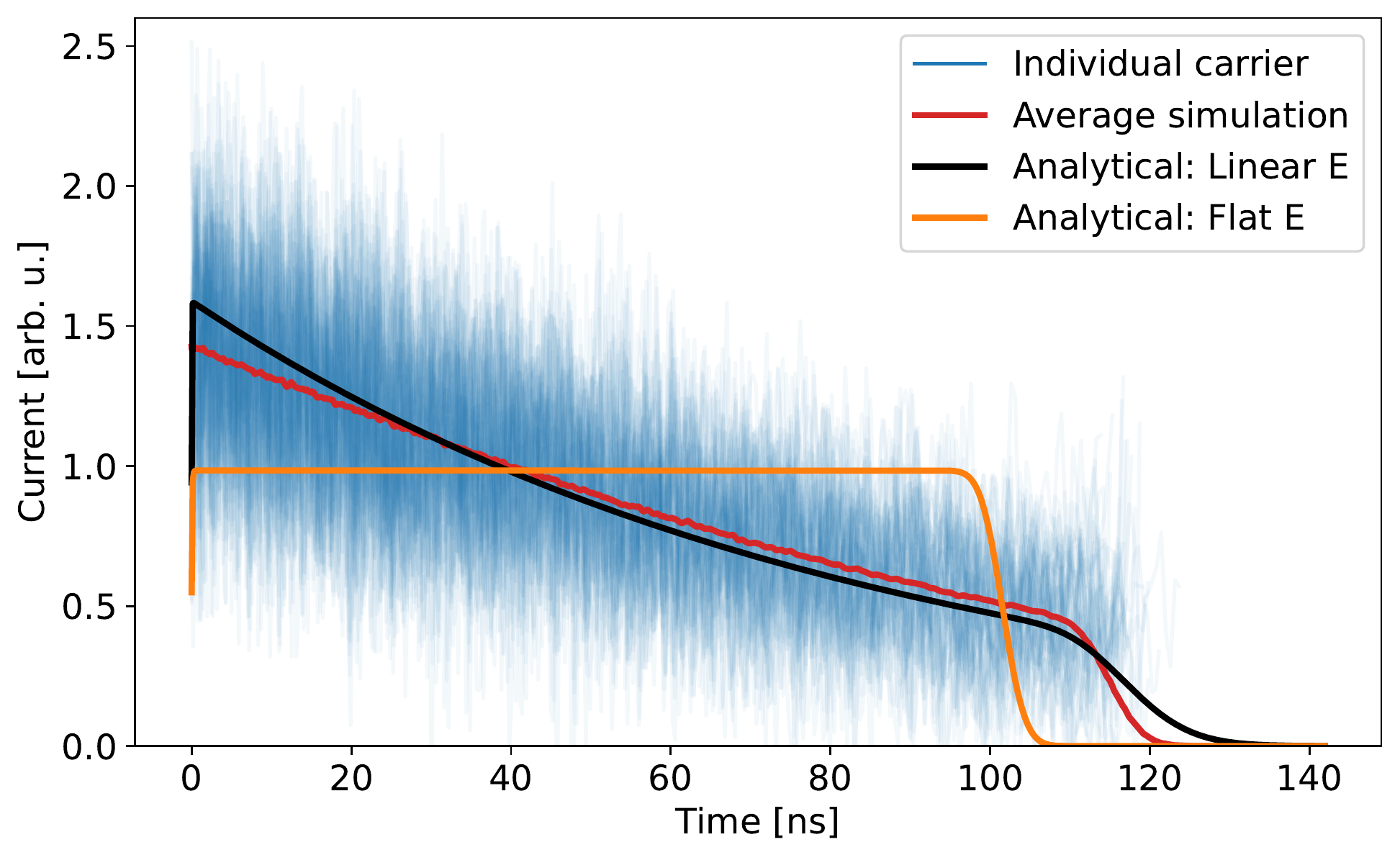}
    \caption{Comparison of the analytical results of Eq. (\ref{eq:I_diffusion_erf}) of the induced current when including diffusion with Monte Carlo simulations for a linearly varying electric field. }
    \label{fig:current_diffusion_analytical}
\end{figure}

While Eq. (\ref{eq:I_diffusion_erf}) is a general result, the non-linear behaviour of the mobility as a function of electric field means that a correct implementation becomes convoluted when the charge carriers approach saturation velocities. Figure \ref{fig:current_diffusion_analytical} shows a comparison of explicit simulation with the analytical results of Eq. (\ref{eq:I_diffusion_erf}) for a constant and linearly varying electric field corresponding to the simulation condition. Whereas the former behaves poorly both in amplitude and timing, the latter obtains very good agreement throughout the entire charge collection time. Minor differences arise due to velocity saturation at early times and an overestimation of the rampdown time near the end of the pulse.

Regardless, it is clear that charge carrier thermal diffusion gives rise to significant differences in the time profile of induced currents. While newly derived analytical results can give good descriptions of the average diffusive charge cloud behaviour in simplified geometries, explicit numerical simulation is required for more advanced geometries. We will use the machinery developed here to discuss more complicated field configurations (Sec. \ref{sec:chargeSharing}) and charge trapping phenomena (Sec. \ref{sec:charge_trapping_dl}). Diffusion is not the only process determining the charge cloud evolution, however, and we first treat plasma and self-repulsion effects.

\subsection{Self-repulsion and plasma effects}
\label{sec:self_repulsion_plasma}
The electron-hole pairs liberated by an incident charged particle creates a large local difference in charge density. When this difference is sufficiently high, the charge cloud becomes effectively a plasma that shields external fields \cite{Tove1967, Taroni1969, Finch1979, Finch1982, Finch1980}. The result is a delay in charge collection until the field is large enough for drift to dominate quasiparticle movement, denoted a plasma delay time. Previous analytical methods in the literature \cite{Neidel1980, Kanno1987, Kanno1990, Kanno1994, Kanno1999} rely on phenomenological factors calibrated to MeV/$u$ fission fragment data far outside the operational window for Nab, however, meaning extrapolation is unlikely to yield satisfactory results. Even so, performing such an extrapolation results in systematic plasma delay times in the window between 0.1 and 1 nanoseconds, which exceeds the Nab uncertainty budget (see Sec. \ref{sec:timing_requirement}). Numerical efforts \cite{Parlog2010, Sosin2012} were performed only for heavy fission fragment ions using dielectric theory and can similarly not be extrapolated to low proton energies. As such, below we describe an explicit simulation effort to quantify this effect as a function of field strength for 30 keV protons.

We construct an $N$-body simulation by explicitly taking into account individual Coulomb interactions between all electron and hole pairs, and take into account the dielectric, three-dimensional diffusion and drift response. The calculation proceeds as follows:

$(i)$ At the site of energy deposition, a number of electron-hole pairs, $n$, is generated such that each charge carrier contains an effective fractional charge $q_{eff} = N_p/n$ where $N_p = 30$ keV$/\varepsilon_{ph}(T)$ is the total number of charge carriers created by a 30 keV proton at temperature $T$.

$(ii)$ Newly created electrons and holes are distributed in space according to a Gaussian distribution with width $\sigma_0 = \sqrt{2D_{e/h}(T)\delta t}$, where $\delta t$ is the time step of the simulation and centered on the creation site. The initial velocity distribution is generated according to a Maxwell-Boltzmann distribution at the lattice temperature $T$.

$(iii)$ The effective electric field is calculated according to all individual Coulomb interactions between all charge carriers together with the external electric field. The Coulomb interaction is reduced by the dielectric strength of silicon to take into account polarization effects, but is bounded in $r^{2} \simeq 10^{-16}$ m$^2$ to avoid numerical instability. The length is physically motivated to correspond to the thermal average de Broglie wavelength, below which quantum mechanical effects are expected to become important. Following the discussion in Ref. \cite{Salas-Bacci2014}, however, we may neglect plasma recombination effects for typical conditions associated with neutron $\beta$ decay events.

$(iv)$ The effective electric field is used to calculate electron and hole velocities following Eq. (\ref{eq:mobility_definition}), and the transport step occurs as described in the diffusion case above, i.e. $\delta \bm{r} = \mu \bm{E}_{eff} \delta t + \delta \bm{r}_D$.

Using the above procedure, we may calculate the position distribution of both charge carriers and derive effective electric fields as a function of time. Positions and velocities may finally be translated into induced currents and integrated charges on electrodes. For all results discussed below we use $\delta t = 30$ ps and $n=300$ unless otherwise mentioned.

\begin{figure}[ht]
    \centering
    \includegraphics[width=0.48\textwidth]{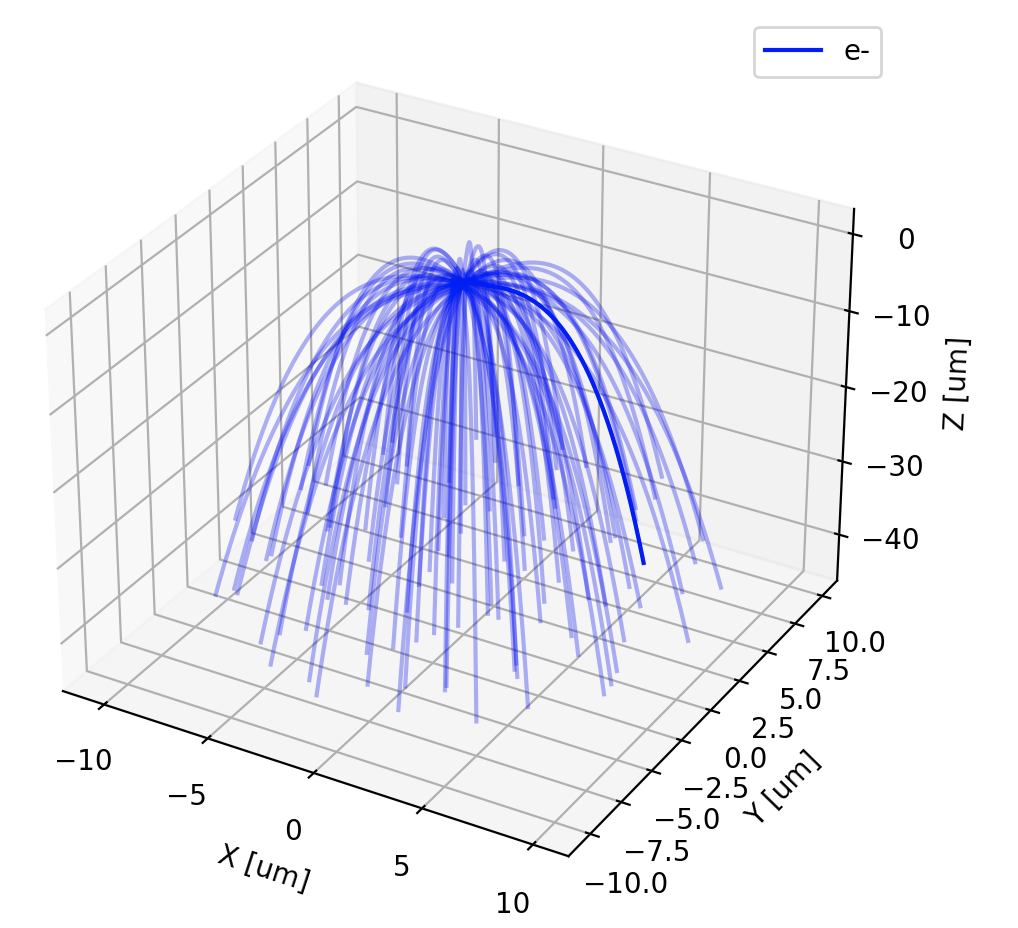}
    \caption{Effects of charge carrier self-repulsion through explicit simulation without diffusion with an external electric field applied along the $Z$-axis of 750 V/cm. The charge cloud undergoes rapid expansion which then slows significantly due to the $1/r^2$ behaviour of the Coulomb interaction.}
    \label{fig:electron_self_repulsion}
\end{figure}

\begin{figure}[ht]
    \centering
    \includegraphics[width=0.48\textwidth]{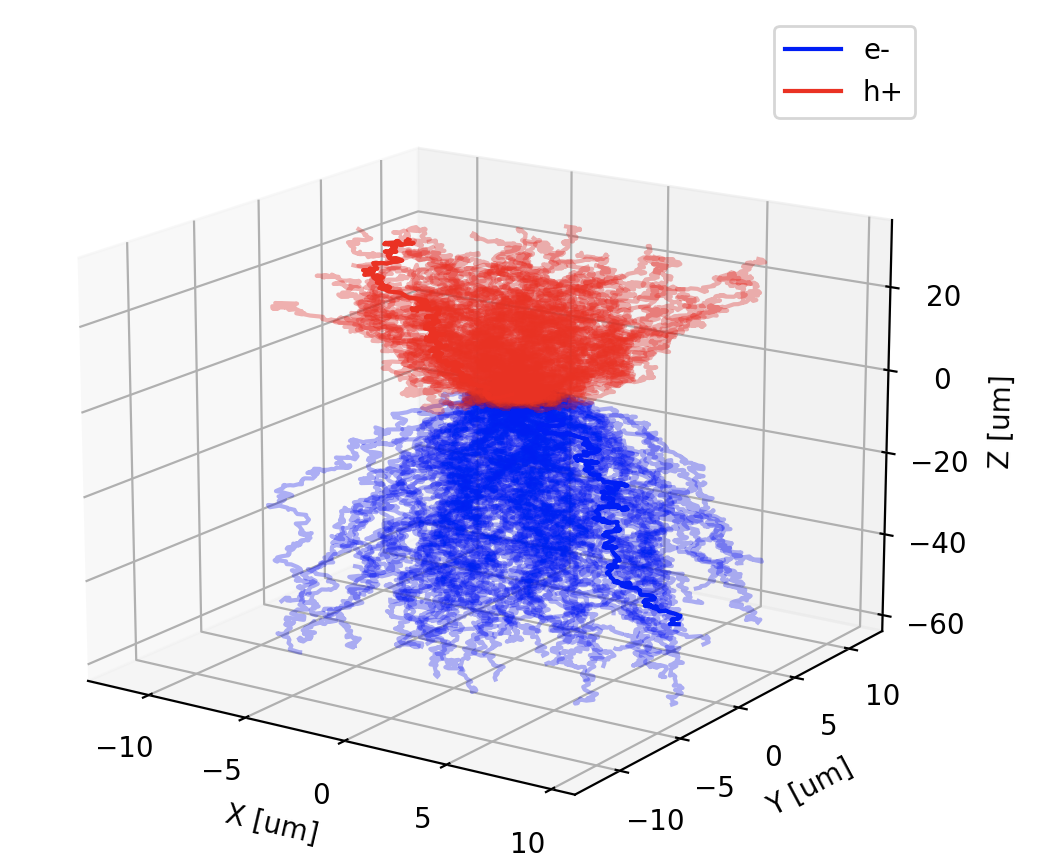}
    \caption{Charge cloud evolution of 300 electron-hole pairs interacting continuously while undergoing drift and diffusion. The final positions are obtained after 1 ns of simulation time. Electrons can be observed to drift and diffuse faster than holes as anticipated, whereas the charge cloud extension at this time scale is determined by self-repulsion as shown in Fig. \ref{fig:electron_self_repulsion}.}
    \label{fig:charge_cloud}
\end{figure}

Figure \ref{fig:electron_self_repulsion} shows the effect of self-repulsion on an electron cloud under the effect of an externally applied electric field after one nanosecond of simulation time. The charge cloud expands rapidly due to the Coulomb interaction and reaches a $7$ $\mu$m radius after $t = 1$ ns, compared to $\sigma \sim $ 2.6 $\mu$m from diffusion at $T=300$K. Figure \ref{fig:charge_cloud} shows the charge clouds of electrons and holes when interacting together including effects due to thermal diffusion after the same time. The extent of the charge cloud is similar to that without diffusion, corroborating the estimate above. 

\begin{figure}[ht]
    \centering
    \includegraphics[width=0.48\textwidth]{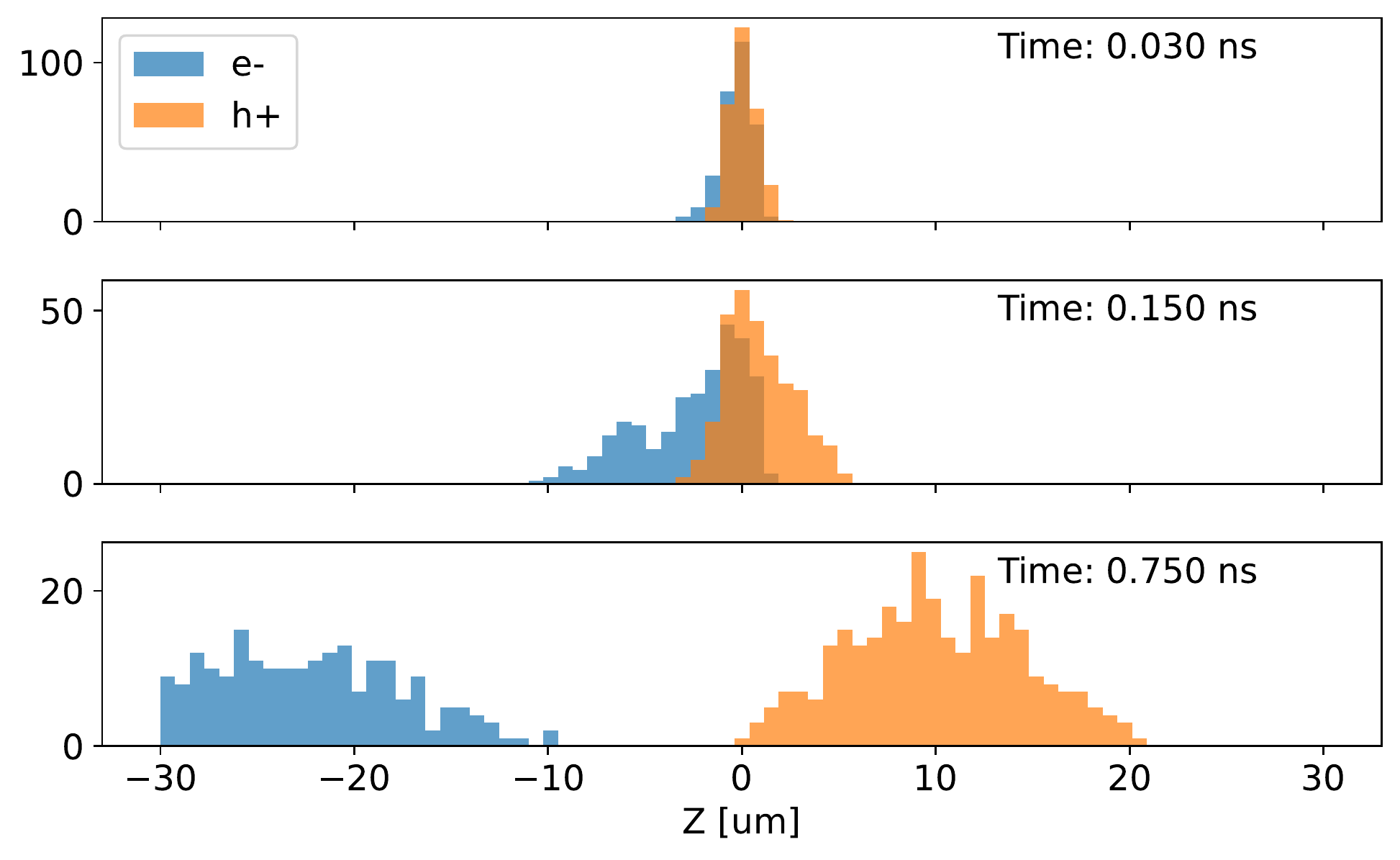}
    \caption{Carrier position distribution as a function of time inside an externally applied field of 750 V/cm at $T=$120 K. The strong asymmetry in position distribution in the middle panel is due to the screening of the external field due to the charge cloud, keeping charge carriers stationary for longer.}
    \label{fig:charge_cloud_vs_time}
\end{figure}

Looking at the time dependence of the charge cloud in Fig. \ref{fig:charge_cloud_vs_time}, it is clear that at $t=0.15$ ns parts of the charge distributions overlap significantly where the external electric field is largely cancelled. Aided by diffusion, the outer layers are swept away and thereby decrease the shielding felt in the center of the charge cloud. After $t=0.75$ ns both charge distributions are sufficiently separated, with the electrons moving out of the window due to their larger mobility relative to holes.

\begin{figure}[ht]
    \centering
    \includegraphics[width=0.48\textwidth]{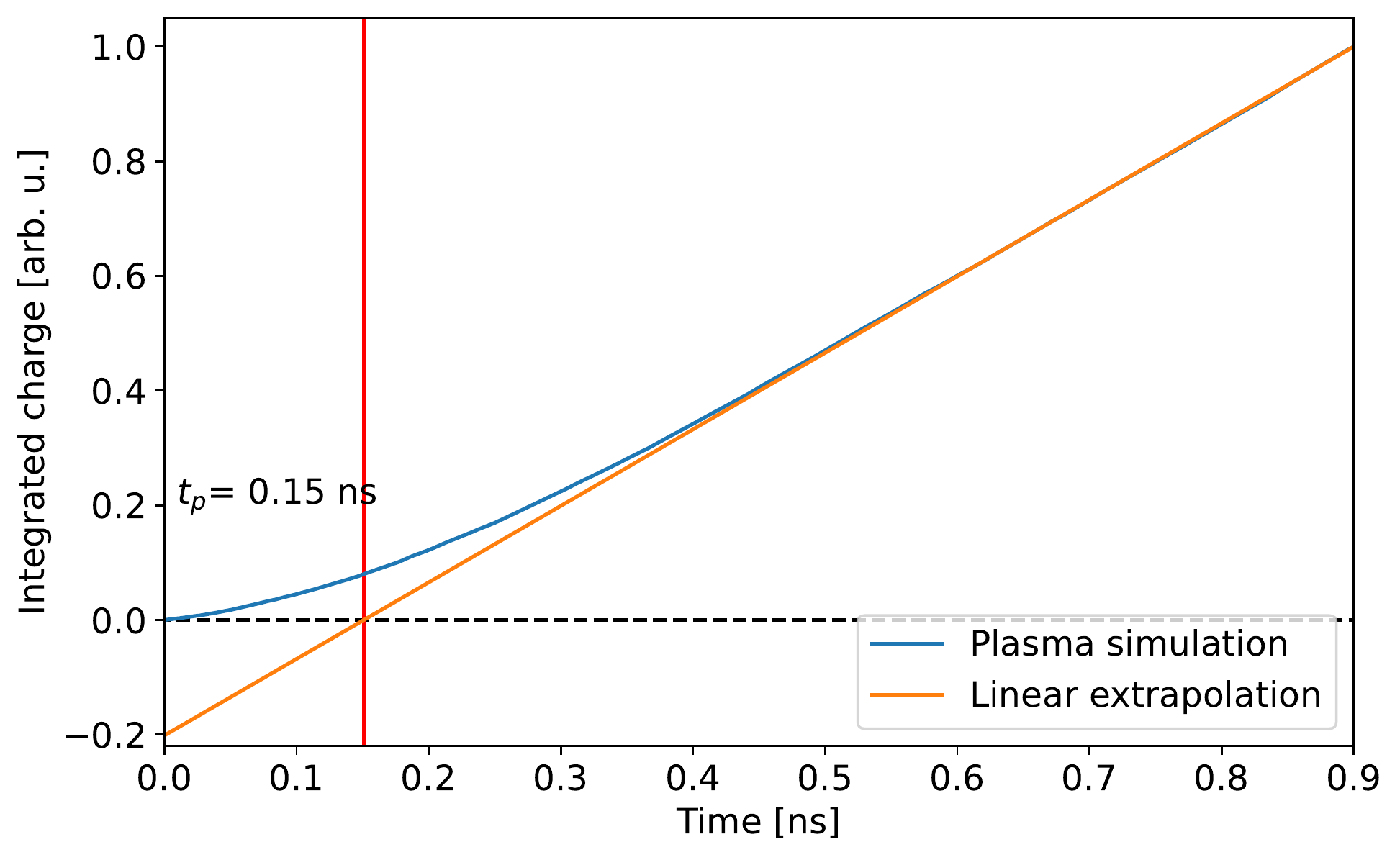}
    \caption{Comparison of the integrated charge when performing a full $N$-body simulation versus a linear extrapolation to $t_0$ in the later region of the pulse. The difference between the true start time and the extrapolated $t_0$ corresponds to a plasma delay time, $t_p$. Results are shown for $E = $ 750 V/cm and $T=$120 K.}
    \label{fig:plasma_time_extrapolation}
\end{figure}

The reduced electric field inside the charge cloud results in a portion of the charge carriers staying stationary for longer than they would in the absence of Coulomb interactions. Once the distributions have separated sufficiently, each group's velocity moves with an average velocity equal to $v_d = \mu E$ and the induced charge integrated on a contact increases linearly for constant electric fields. In the Nab experiment, the low signal-to-noise of 30 keV protons does not allow one to resolve the start time of the event, denoted $t_0$. While several strategies are being investigated to enable optimal $t_0$ extraction, we may quantify the effect of the plasma delay time at very early times using a linear extrapolation. As such, we may define the plasma delay time as the difference in zero-point crossings of a linear extrapolation and the actual start time of the interaction. Figure \ref{fig:plasma_time_extrapolation} shows the determination of the plasma delay time, $t_p$, for an electric field set to 750 V/cm (corresponding to a bias voltage of $\sim$150 V) and temperature $T=120$ K. These conditions correspond to an estimated lower bound in bias voltage and temperature for the Nab running conditions. Following the procedure outlined above, we find $t_p = 0.15$ ns. Imposing higher bias voltages, however, serve to break up the charge clouds sooner. Invoking instead an average electric field of 1.5 kV/cm, an anticipated upper bound for Nab, reduces the plasma delay time to $t_p = 0.08$ ns following the same analysis. More generally, the plasma delay time follows a $1/E$ behaviour, consistent with other approaches \cite{Finch1979, Kanno1987}. The procedure described here is not fully self-consistent, however, as the size of the delay depends on the square root of the time step through the initialization of the charge cloud density. Using a time step of $\delta t = 3$ ps, the plasma time is increased by a factor three. While other approaches of estimating the initial density have been explored \cite{Kanno1987, Sosin2012}, no consistent scheme has emerged. Our approach provides a result consistent with other literature estimates and performs well in a regime where no other simulation data exists. While results are not fully model-independent, the observed range appears to be sufficiently small for the timing requirements of the Nab experiment. Further research is needed, however, to perform a fully self-consistent calculation. 

\subsection{Transport inside the entrance window}
\label{sec:charge_trapping_dl}


As discussed in Sec. \ref{sec:charge_collection_efficiency}, the charge collection (defined as the fraction of quasiparticles that can escape the entrance window and get collected at an electrode) in the entrance window of the detector is poor due to the high doping concentration in the junction. Two models for this `dead layer' were put forward from the literature, Eqs. (\ref{eq:cce_hard_dead}) and (\ref{eq:cce_soft_dead}), but neither has been compared against explicit simulation. With the SIMS data as presented in Sec. \ref{subsection: DopingProfile}, we may construct an detailed model using the Monte Carlo transport approach as discussed above. At these short length- and timescales, the charge collection is determined by the quasiparticle lifetime and local electric fields, discussed below.

\subsubsection{Carrier lifetime}
We first consider charge loss and trapping mechanisms. In the highly doped layer of the entrance window, quasiparticle losses occur through radiative and Auger recombination \cite{Sze2007}. In indirect bandgap semiconductors such as silicon, the former is suppressed by a factor $10^4$ relative to a direct bandgap semiconductor such as GaAs and can be neglected \cite{Rein2005}. Auger recombination, on the other hand, is expected to be contribute significantly. In this process, a particle-hole pair recombines to give its excess energy to a third carrier which then thermalizes in the crystal through phonon emission. In the Boron-implanted entrance window, liberated electron quasiparticles are a low-level injection into a hole-dominated regime, causing the $ehh$ process to dominate. As a consequence, the carrier lifetime depends quadratically on the $p$-type doping concentration, $N_A$, as
\begin{equation}
    \tau_{\rm Auger}^{LLI} = \frac{1}{C_pN_A^2}, \qquad C_p(T) = C_p^0\left(\frac{T}{300 {\rm\, K}}\right)^{1.18}
\end{equation}
where $C_p^0 = 9.9\times 10^{-32}$ cm$^6$s$^{-1}$ \cite{Dziewior1977} with the temperature dependence by Klaassen \cite{Klaassen1992a}. The results are valid for extremely high doping ($N_A > 1\times 10^{18}$cm$^{-3}$), whereas at lower doping concentrations the minority lifetime is experimentally seen to be larger than anticipated \cite{Rein2005}. This is explained via the Coulomb-enhanced Auger recombination process, which modifies $C_p \to g_{ehh}(N_A) C_p $ where
\begin{equation}
    g_{ehh}(N_A) = \left(1+44\left\{1-\tanh\left[\frac{N_A}{5\times 10^{16}\, {\rm cm}^{-3}}\right]^{0.29}\right\}\right)
\end{equation}
is a multiplicative factor \cite{Altermatt1997} to the $ehh$ Auger recombination cross section. 

The other dominant effective charge loss mechanism is through capture at so-called trapping centers or defects. It is described using Shockley-Reed-Hall (SRH) statistics, and is most effective for defects with energies close to the middle of the bandgap \cite{Sze2007}. Depending on the type of defect, electrons or holes can become temporarily trapped at one of these sites, and if the detrapping time is longer than the signal integration time it counts effectively as a lost charge. The specific carrier lifetime depends linearly on the trap density, $N_t$, as
\begin{equation}
    \frac{1}{\tau_{\rm SRH}} = C_{\rm SRH}N_t, \quad C_{\rm SRH}(T) = C_{\rm SRH}^0 \left(\frac{300 {\rm\, K}}{T} \right)^{1.77}
\end{equation}
where $C_{\rm SRH}^0 = 3 \times 10^{-13}$ cm$^3$s$^{-1}$ \cite{Klaassen1992a}. The latter was derived in the assumption of a single trapping energy level and without explicit specification of the contribution of different contaminants \cite{Newman1982, Graff1995}. Finally, the bulk has a characteristic carrier lifetime, $\tau_0$, that is assumed to be independent of temperature so that the total carrier lifetime is
\begin{equation}
    \frac{1}{\tau_c(N_A, T)} = \frac{1}{\tau_0} + g_{ehh}C_pN_A^2 + C_{\rm SRH} N_t
\end{equation}
Typical values of $\tau_0$ for extremely pure silicon are in the millisecond range, which is what we will adopt here. At such carrier lifetimes, the charge collection efficiency is very close to 100\%. Effective carrier lifetimes for the Nab detectors using the SIMS data of Fig. \ref{fig:sims_data} are shown in Fig. \ref{fig:loss_time_E}.

\subsubsection{Local electric field}

Before capture, the transport of the particle is determined by the local electric field environment, its mobility and diffusion constant. The latter two were described already in Sec. \ref{sec:mobility} and are related via the Einstein diffusion relation (Eq. (\ref{eq:einstein_diffusion})). As a consequence, both depend strongly on temperature and impurity concentration (for $N_{A, D} > 1 \times 10^{15}$ cm$^{-3}$). The electric field in the presence of strong doping gradients is less straightforward to calculate, and in fact several standard methods suffer from conceptual issues \cite{Redfield1979, Pimbley1988}. In the usual depletion approximation, the electric potential is considered constant except for regions of depleted charge. For strongly asymmetric junctions typical in particle detectors, however, this assumption is not valid when looking at the electric potential at the scale of the entrance window. Specifically, the electric field must be non-zero when any doping gradient is present, even when that region is not considered depleted. This can easily be understood from the drift-diffusion equations (Eq. (\ref{eq:drift_diffusion})), where the net carrier current can be set to zero in thermal equilibrium so that
\begin{equation}
    \bm{E} = -\frac{D}{\mu_n}\frac{1}{n}\nabla n
    \label{eq:E_drift_diff_eq}
\end{equation}
where $D/\mu_n$ evaluates to $k_BT/q$ via Eq. (\ref{eq:einstein_diffusion}). For an exponentially graded implantation region \cite{Chawla1971}, $\nabla n/n$ reduces to a simple constant but, more generally, one writes \cite{Pimbley1988, Lanyon1981}
\begin{equation}
    E(x) = -\frac{k_BT}{q} \frac{\partial [\log \{N(x)\}]}{\partial x}
    \label{eq:E_log_N}
\end{equation}
where $N(x)$ is the impurity density. Using the SIMS data presented in Sec. \ref{subsection: DopingProfile} we might be tempted to evaluate Eq. (\ref{eq:E_log_N}) directly. At very high doping concentrations, however, ($N > 1 \times 10^{19}$ cm$^{-3}$), degeneracy and bandgap narrowing effects modify this relation substantially \cite{Lanyon1981, Redfield1981}. For electron quasiparticles, which are our main concern in this work for incoming protons, the effective electric field for doping concentrations above $N = 1\times 10^{19}$ cm$^{-3}$ is strongly suppressed. Below this concentration, we use the SIMS data to get an effective electric field using Eq. (\ref{eq:E_log_N}). To reduce the effect of statistical scatter at lower concentrations (see Fig. \ref{fig:sims_data}), we find that the Boron concentration from the SIMS data can be fit well using a double-exponential function beyond 10 nm,
\begin{align}
    N_B(x) &= 3.7\times 10^{20} {\rm\, cm}^{-3} \exp(-x/1.0 \times 10^{6} {\rm\, cm}) \nonumber \\
    &+ 1.5 \times 10^{19} {\rm\, cm}^{-3} \exp(-x/3.5\times 10^{5} {\rm\, cm}).
\end{align}
To represent the transition regime between zero electric field where $N_B > 1\times 10^{19}$ cm$^{-3}$ ($x \lesssim 50$ nm) and below, we add an additional Gaussian function of 5 nm width to provide a smooth connection between the two regimes.

\begin{figure}[ht]
    \centering
    \includegraphics[width=0.48\textwidth]{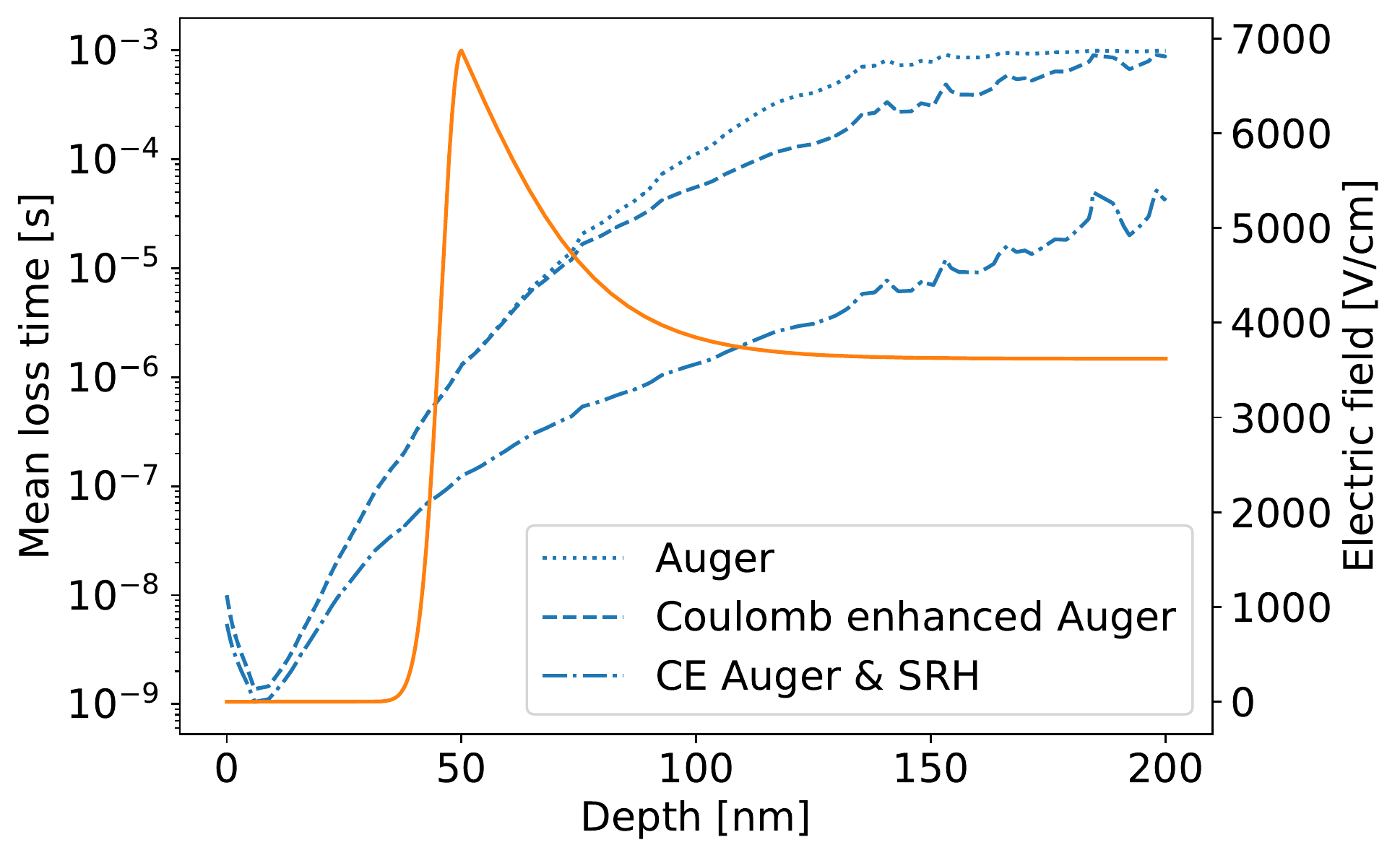}
    \caption{Effective carrier lifetime (blue, left axis) and electric field (orange, right axis) as a function of depth inside the entrance window. The electric field in the high doping regime ($\lesssim$ 50 nm) is set to zero due to degeneracy and bandgap narrowing \cite{Lanyon1981}. The SRH carrier lifetime is evaluated in a worst-case scenario where we set $N_t = N_A$.}
    \label{fig:loss_time_E}
\end{figure}

Figure \ref{fig:loss_time_E} shows the effective carrier lifetime and electric field as a function of depth from the front surface using all available information. Since we do not have prior information on the trapping density in the entrance window, the SRH lifetime was evaluated in a worst-case scenario where we set $N_t = N_A$. Contrary to the usual analysis, the electric field in the region beyond $50$ nm is substantially larger than that inside the bulk for typical bias voltages due to the strong gradients in impurity density (compare Eq. (\ref{eq:E_drift_diff_eq}) and Eq. (\ref{eq:E_undepl_depl})). Both mobility and drift coefficients are strongly suppressed in this regime, however, due to the high impurity concentration so that the increased field effect is largely mitigated.

\subsubsection{Monte Carlo charge collection simulation}

We perform a Monte Carlo simulation of individual quasiparticle transport in the entrance window using the carrier lifetime and electric fields discussed above using the procedure of Sec. \ref{sec:diffusion_MC}. The latter is modified to take into account the finite carrier lifetime, where at each step the local lifetime is calculated and the time step is forced to be smaller than 1\% of the lifetime. The probability for loss is then simply $\delta t/ \tau_c$. If the quasiparticle survives, its transport is performed using local effective diffusion constants and electric field drift to its next location. This process continues until either the quasiparticle is lost through recombination or capture, it is collected at the front electrode, or reaches an arbitrary distance of 500 nm away from the front face, in which case it is considered `safe' to move towards the back electrode and contribute to signal formation. We repeat the process $N$= 10000 times as a function of initial starting position, and define the charge collection efficiency as the fraction of $N$ that reach the 500 nm threshold.

The transport of quasiparticles is purely diffusive within the first 50 nm due to the absence of an effective electric field and swiftly proceeds via drift beyond this point. Since both drift and diffusion are regulated by the mobility, the drift-diffusion cross-over point is determined solely by the effective electric field caused by the impurity gradient rather than the absolute value of the mobility. Since diffusion depends linearly on temperature (see Eq. (\ref{eq:einstein_diffusion})), the cross-over point moves closer to 50 nm as the temperature decreases.

\begin{figure}[ht]
    \centering
    \includegraphics[width=0.48\textwidth]{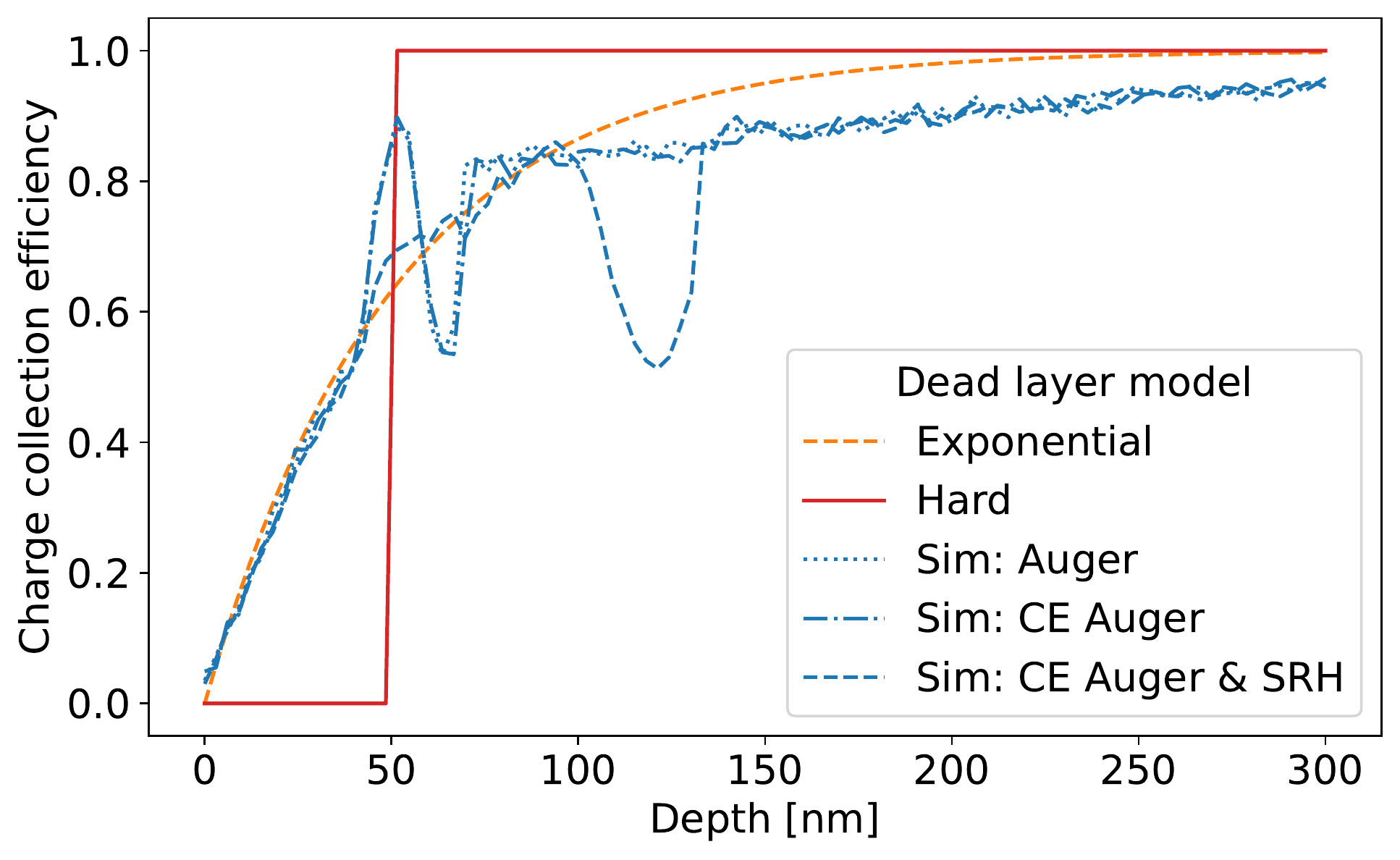}
    \caption{Simulated charge collection efficiency for electron quasiparticles inside the entrance window to the detector using the three carrier lifetime models shown in Fig. \ref{fig:loss_time_E}. Shown also are the `hard' dead layer and an exponential parametrization of Eq. (\ref{eq:cce_hard_dead}) and (\ref{eq:cce_soft_dead}), respectively.}
    \label{fig:CCE_sim_dl}
\end{figure}

Figure \ref{fig:CCE_sim_dl} shows the results of the Monte Carlo simulation using the three different carrier lifetime models shown in Fig. \ref{fig:loss_time_E}, together with the phenomenological models discussed in Sec. \ref{sec:charge_collection_efficiency}. The simulated data show strong local differences at $x \sim 50$ nm, where the electric field is suddenly turned on, and around $x\sim 125$ nm, where the SRH carrier lifetime is large enough for substantial losses to occur in the decreased electric field regime. Given that $\tau_{\rm SRH}$ was evaluated in a worst-case scenario where $N_t = N_A$, it is unlikely that such strong local effects will be experimentally observed. Outside of these local changes, all three models show remarkable convergence over the full range.

When comparing to the phenomenological models, we find that the `hard' dead layer model of Eq. (\ref{eq:cce_hard_dead}) is too crude an approximation to capture the simulated data, and indeed charge collection is finite inside the traditional `dead layer'. The exponential model of Eq. (\ref{eq:cce_soft_dead}) captures the general trend of the simulated data reasonably well but shows discrepancies due to local effects discussed above. Overall, however, through appropriate parameter choices it is able to satisfactorily capture the behaviour.

Charge collection inside the SiO$_2$ layer (see Sec. \ref{sec:detector_geometry}) is assumed to be absent. While effects have been studied in the literature \cite{Poehlsen2013, Jones1988}, the thermally grown oxide layer for the Nab geometry is much smaller than typically the case and we do not treat it in detail. Instead, we may simply shift the results of the CCE presented here by the estimated oxide thickness similar to the parametrization of Eq. (\ref{eq:cce_soft_dead}).

In summary, we have treated in detail several microscopic phenomena of quasiparticle transport. Depending on the local impurity and field environment, the first nanosecond of their movement is a complex interplay of different mechanisms. For several of these, we have derived new analytical and numerical results and found deviations relevant to the timing restrictions of the Nab experiment as outlined in Sec. \ref{sec:timing_requirement}. We now move on to the final part of the pulse shape simulation process, which connects the quasiparticle transport processes to a model pulse shape to be compared to experimental observation.

\section{Pulse shape simulation}
\label{sec:pulse_simulation}
The ingredients described above form parts of a simulation chain that come together into a realistic description of wave forms which allow for the extraction of detector parameters and perform sensitivity studies to a variety of scenarios. Specifically, for detectors as large as those of the Nab experiments, significant variations can occur in the doping uniformity along its radial directions whereas the high segmentation implies strong geometrical effects. As discussed above, a detailed simulation pipeline is able to differentiate aforementioned effects and disentangle experimental data. In this section, we present an overview of the individual components of the pulse shape simulation procedure after introducing analytical results to compare against.

\subsection{Analytical baseline}
\label{sec:analytical_baseline_pulse}
In order to provide a reference for the effect of the simulations we will describe an analytical baseline using some ingredients from the previous sections. Starting from an electric field linearly varying with position (i.e. for constant impurity density) as in Eq. (\ref{eq:E_undepl_depl}), we may write the electron position and velocity as
\begin{subequations}
\begin{align}
    x(t) &= \frac{b}{a}\left[\frac{E_0}{b}\exp(a\mu t) -1\right] \\
    v(t) &= \mu E_0 \exp(a\mu t)
\end{align}
\label{eq:x_v_linear_E}
\end{subequations}
where $E = ax+b$ and $E_0 = ax_0 + b$ is the electric field at the starting position $x_0$. The electron takes a time $t_\mathrm{max} = \ln(E_d/E_0)/a\mu$ to reach the end of the depletion zone at a distance $d$, so that the current is simply $qv(t)/d$ and induced charge is
\begin{equation}
    Q(t) = \frac{qE_0}{ad}[\exp(a\mu t)-1],
\end{equation}
for a uniform weighting field, $\nabla W = 1/d$. The generalization to an inhomogeneous weighting potential such as the analytical result of Eq. (\ref{eq:weighting_pot_circ_pixel}) can be done in a straightforward fashion by replacing $d^{-1}$ with $\nabla W(\rho, x(t))$. For a uniform weighting field the time taken for the integrated charge to go from $r_LQ_{\rm max}$ to $r_H Q_{\rm max}$ where $Q_{\rm max} = q(d-x_0)/d$ and $0 < r_{L, H} < 1$ is simply
\begin{equation}
    t_{L-H} = \frac{1}{a\mu}\log\left(\frac{1+r_H A}{1+r_L A}\right),
    \label{eq:t_L_H}
\end{equation}
where $A = a(d-x_0)/E_0$ or $A=a/b$ when setting $x_0=0$. Using the results from Eq. (\ref{eq:E_undepl_depl}) the behaviour of $t_{L-H}$ changes when the detector reaches full depletion, as $b \propto \sqrt{V}$ for $V < V_d$ and $b \propto V$ otherwise.

The electronic response may be treated similarly to Refs. \cite{Bruyneel2006, Brigida2004} by numerically integrating the time signal using expressions in the Laplace domain\footnote{While closed expressions for the inverse Laplace transform with the current as above can be obtained, they are not particularly insightful nor do they allow for easy noise insertion in the input signal.}. More specifically, if the transfer function may be written as a rational function where both numerator and denominator are polynomials of at most second degree, $H(s) = (a_0 + a_1 s + a_2s^2)/(b_0 + b_1 s + b_2 s^2)$, the time domain equations may be written as
\begin{equation}
    b_0 y(t) + b_1 \frac{d y(t)}{dt} + b_2 \frac{d^2 y(t)}{dt^2} = a_0 x(t)+ a_1 \frac{d x(t)}{dt} + a_2 \frac{d^2 x(t)}{dt^2}
\end{equation}
for input signal $x(t)$ and output signal $y(t)$. If the latter are written as functions of discretized time, $\{t_i\}$, derivatives may be replaced by differences and one obtains a recursion relation for $y(t_n)$ in terms of $\{x, y\}(t_{n-1, n-2})$.

In its simplest iteration, the preamplifier may be reduced to the impedance of a feedback resistor and capacitor in parallel \cite{Bruyneel2006}, i.e. $H_{p}(s) = R_{fb}/(1+sC_{fb}R_{fb})$, whereas the shaping amplifier can be written as an $RC$-$CR$ shaper with gain $G$
\begin{equation}
    H_{sh}(s) = G\frac{sC_1R_2}{(1+sC_1R_1)(1+sC_2R_2)}.
    \label{eq:H_sh_elec}
\end{equation}
For the circuit of interest we set $R_{fb} = 1$ M$\Omega$, $C_{fb} = 1$ pF, $G=20$, $C_1 = 10$ pF, $C_2 = 1$ pF, $R_1 = 150$ k$\Omega$ and $R_2 = 5$ k$\Omega$.

\begin{figure}[ht]
    \centering
    \includegraphics[width=0.48\textwidth]{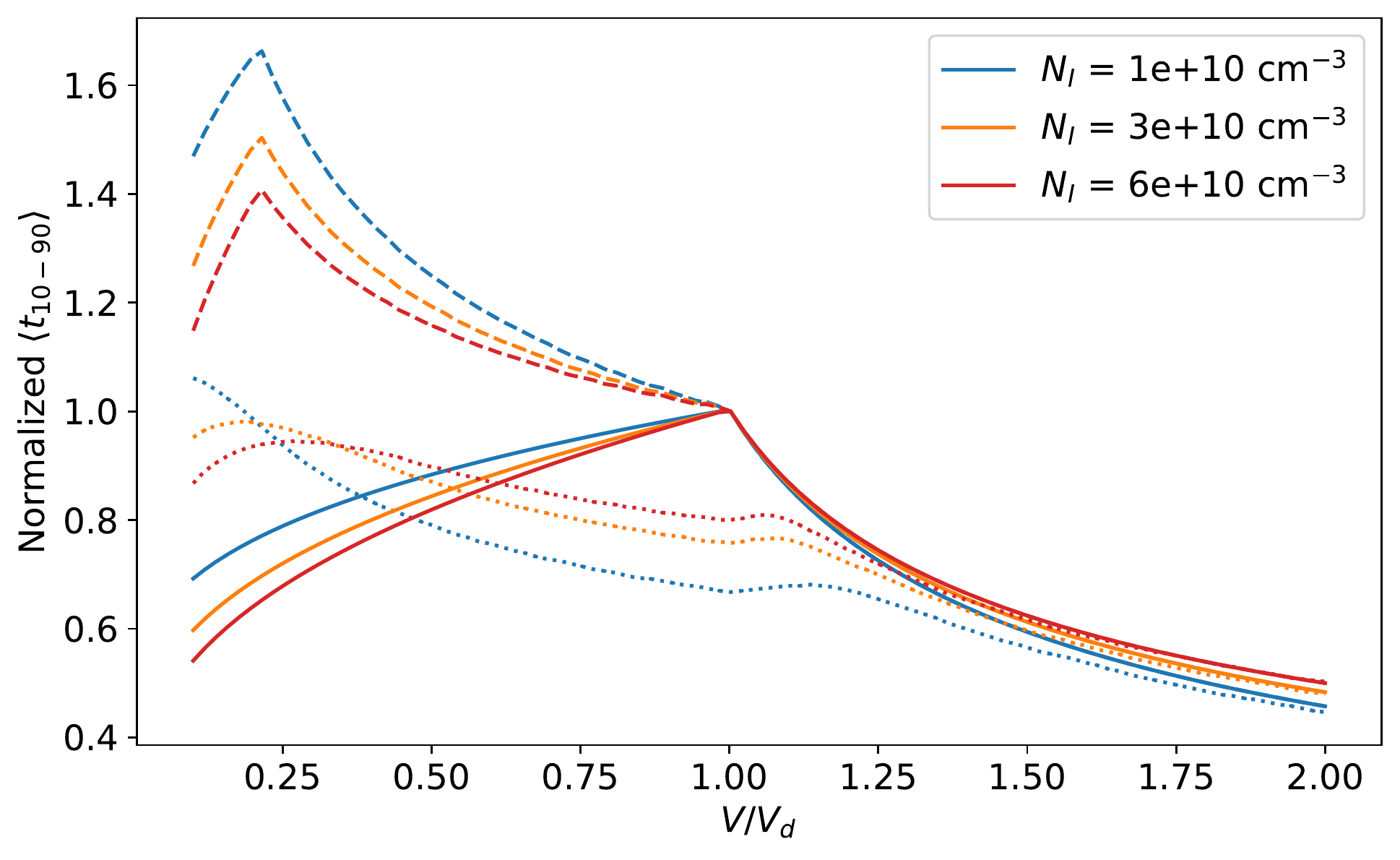}
    \caption{Elapsed time for the signal to reach 10\% to $90\%$ of the integrated charge normalized to its value at the depletion voltage according to Eq. (\ref{eq:t_L_H}) (solid line), taking into account time-dependent effects from Eq. (\ref{eq:W_t_simple}) (dashed line), and time-dependent effects after passing through the electronic filter of Eq. (\ref{eq:H_sh_elec}) (dotted line) for different base impurity concentrations at $T=120$ K.}
    \label{fig:normalized_t_10_90_analytical}
\end{figure}

Figure \ref{fig:normalized_t_10_90_analytical} shows the $10\%$ to $90\%$ rise time for different impurity concentrations assuming the static response of Eq. (\ref{eq:t_L_H}), taking into account the time dependence of the undepleted layer as discussed in Sec. \ref{sec:time_dependent_weighting_field}, and including the electronic response function. Even though closed expressions for induced currents including the electronic response and undepleted effects are available, the crossing points must be determined numerically. In all three cases, the behaviour below the depletion voltage is a sensitive function of the impurity concentration, and could be used as a way to obtain an accurate assessment. The behaviour is completely opposite, however, when considering static versus time-dependent results. Including the shaping electronics suppresses the rise time variation as long transit times will lead to ballistic deficit. Even so, while increased voltage noise due to the higher capacitance of an underdepleted detector might complicate precision measurements, the presence of time-dependent effects could be straightforward to observe.

\subsection{Monte Carlo transport}
\label{sec:MC_sim}
A necessary input in constructing simulated pulse shapes produced in our detectors is the simulation of proton interactions with the detector material. In order to construct detector signals, it is imperative to know the the location and magnitude of ionizing interactions. We used two different software packages to achieve this. The first was SRIM \cite{Ziegler2010} that was designed to calculate the stopping and range of ions in matter. The second was Geant4 \cite{Agostinelli2003} which is designed to simulate the passage of particles through matter.

\subsubsection{SRIM, Geant4 inter \& intracomparison}
\label{sec:SRIM_Geant4_comp}
\begin{figure}[ht]
    \centering
    \includegraphics[width=0.48\textwidth]{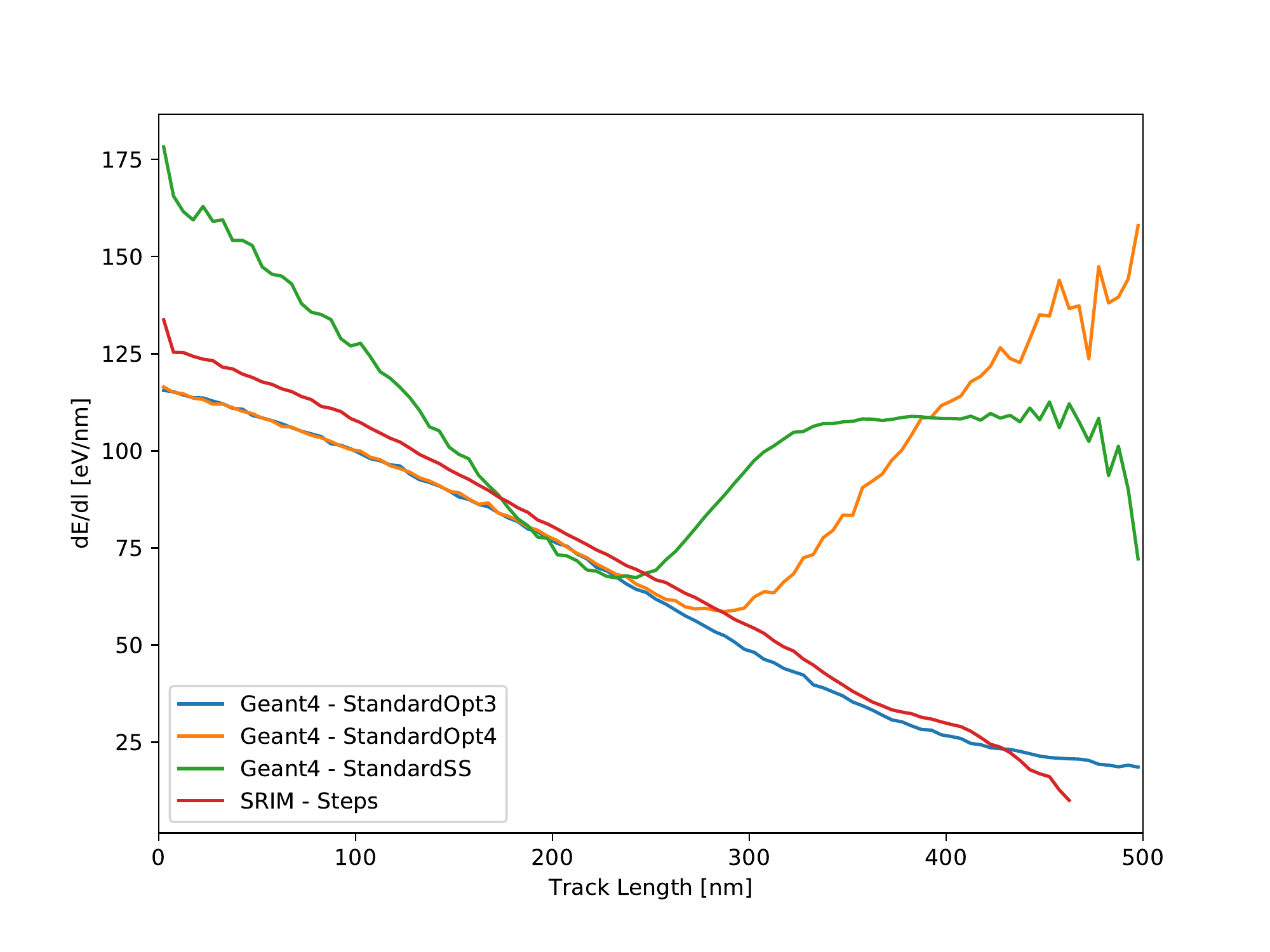}
    \caption{Comparison of extracted $dE/dl$ versus track length for $30$ keV protons impinging on pure silicon using different Geant4 electromagnetic physics lists and SRIM. For the latter, 'Steps' refers to procedure in the text, compared to standard output by SRIM.}
    \label{fig:dEdl_comp}
\end{figure}

Both SRIM and Geant4 were used to simulate 30 keV protons normally incident on silicon and the results were compared. The stopping power of silicon was compared in SRIM and Geant4 with several different low energy electromagnetic physics lists. The Geant4 physics lists used in the comparisons included the standard \texttt{option 3}, \texttt{option 4} and \texttt{single scattering}, which differ in their use condensed history scattering algorithm. The stopping power was extracted by taking the ratio of the energy loss to step length for steps in the silicon and then plotted versus the total length of the track at the step. This generates a 2D histogram which is then averaged along the $x$-axis. The results are shown in Fig. \ref{fig:dEdl_comp}. Other than a maximal step length of 5 nm, all physics constructors used default parameters. All Geant4 models shown here use the Bragg model for ionization losses along the proton path length but differ in the way (multiple) Coulomb scattering is treated. In the case of option 4, Geant4 uses the \texttt{WentzelVI} multiple scattering algorithm instead of the \texttt{Urban} model for option 3. Single Coulomb scatters are treated only for large angular deviations, except for the single scattering algorithm where each is treated individually. The Bragg model used for the ionization stopping power is based on the Lindhard theory \cite{Lindhard1963, Lindhard1965} and is very similar to what is used in SRIM \cite{Brandt1982, Ziegler2010}. Even so, substantial changes are observed between the different options and SRIM, particularly with the single scattering result overemphasizing backscattered protons and losses early in the track. The standard option 3 physics list was found to give the best agreement with SRIM, and was used for the remainder of this work.

\begin{figure}[ht]
    \centering
    \includegraphics[width=0.48\textwidth]{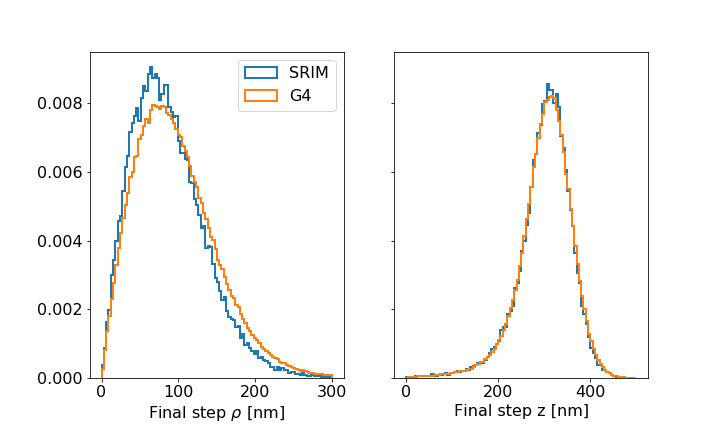}
    \caption{Comparison of the range distributions of 30 keV protons normally incident on silicon using Geant4 and SRIM using the coordinates of the final track step. Here, $\rho = \sqrt{x^2+y^2}$ is the lateral displacement. }
    \label{fig:range_comp}
\end{figure}

A comparison of the distribution of the proton range in silicon was made betweeen SRIM and Geant4. The results are shown in Fig. \ref{fig:range_comp}. The mean of the distributions in z agree to within 0.3\%.  The lateral range for the Geant4 simulation using the standard option 3 physics list was 10\% wider than that found from SRIM. As both packages use very similar electronic stopping power models to describe the proton's ionization losses, both simulations generate distributions of track lengths that are in good agreement. In the case of normally incident source particles, the $z$ component of the final step is closely correlated with the overall track length. The discrepancy in the radial component of the range distribution is attributed to differences in the treatment of small angle scattering in the two simulations. 

\begin{figure}[ht]
    \centering
    \includegraphics[width=0.48\textwidth]{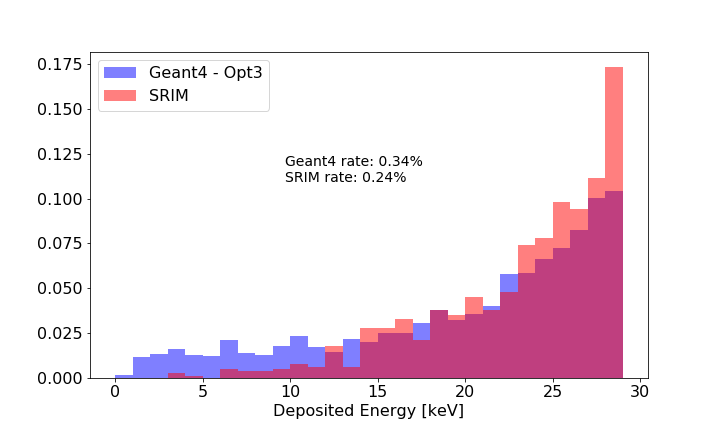}
    \caption{Comparison of backscattering rate and deposited energy spectrum of 30 keV protons normally incident on silicon using Geant4 with the Standard Option 3 electromagnetic physics list and SRIM. Here, a backscattered particle is defined as leaving the silicon detector and therefore depositing less than its full kinetic energy. Note that the spectrum shows the deposited rather than detected energy, i.e. before charge collection losses and thresholds.}
    \label{fig:bs_comp}
\end{figure}

Additionally, we compared the overall rate and energy spectrum of partially deposited energy from proton back scattering. We define a backscatter as any proton that leaves the silicon and therefore deposits less than its full energy. The results are shown in Fig. \ref{fig:bs_comp}. The backscattering probability was found to be higher in Geant4 with a softer spectrum. The latter is consistent with the lateral displacement difference as coming from the treatment of (multiple) Coulomb scattering. A higher average angular deflection from multiple Coulomb scattering results both in a widened lateral displacement and softer backscatter spectrum.

\subsubsection{Simulated NIEL}
\begin{figure}[ht]
    \centering
    \includegraphics[width=0.48\textwidth]{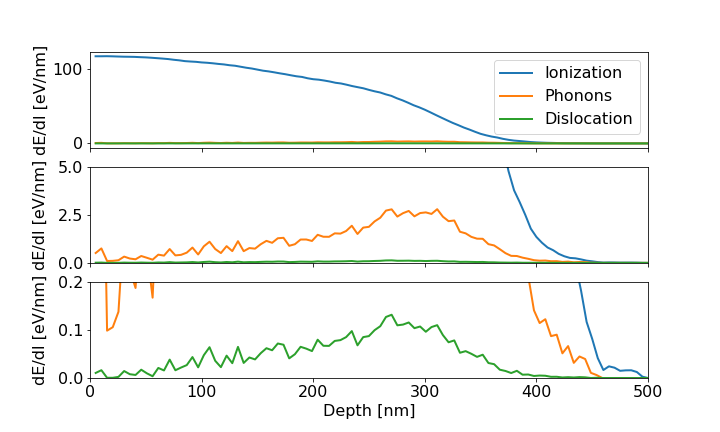}
    \caption{Simulated non-ionizing energy loss and its breakdown into different processes, obtained using SRIM for 30 keV protons impinging upon pure silicon.}
    \label{fig:NIEL_SRIM}
\end{figure}

Since only the movement of quasiparticles through an electric and weighting field contribute to a signal induced in electrodes of the detector (see Sec. \ref{sec:electric_field}), any process that involves a particle losing energy without creating particle-hole pairs will not contribute. The collective effect of these processes is denoted Non-Ionizing Energy Loss (NIEL), and has received substantial interest from both collider and space exploration research programs \cite{Valentin2012a, Jakubek2013, Huhtinen2002, Hopf2008, Inguimbert2009, Srour2003}. Previously \cite{Salas-Bacci2014}, predictions for NIEL within the Nab experiment were studied using analytical methods. Here, we use SRIM simulation to find the stopping power of the different processes for 30 keV protons incident on silicon. The predominant NIEL processes occur via phonon emission and dislocation of silicon atoms. In the binary collision approximation \cite{Robinson1994a}, phonon emission can be understood as a Coulomb interaction with an energy transfer which is smaller than the energy required for a silicon atom to create a vacancy, i.e. the displacement energy $E_{\rm disp} \sim 21$ eV. Large-angle Coulomb scatters, on the other hand, eject a silicon atom that can itself go on and create particle-hole pairs or dissipate energy through phonon emission. We use SRIM to obtain a distribution of the proportion of total energy lost to non-ionizing processes, shown in Fig. \ref{fig:NIEL_SRIM}. On average the non-ionizing loss is less than 2\% of the total energy, consistent with earlier results \cite{Salas-Bacci2014}. As expected, the proportion of NIEL attributable to phonon losses is substantially higher than dislocations. Relative energy losses due to NIEL are substantially higher towards the end of  the track, however. Combined with the poor charge collection in the entrance window (see Sec. \ref{sec:charge_trapping_dl}), it is feasible for events to fall below a detection threshold due to large NIEL effect near the end of a track. 

\subsection{Electronic response function: SPICE simulation}
\label{sec:spice}

\begin{figure*}[ht]
    \centering
    \includegraphics[width=\textwidth]{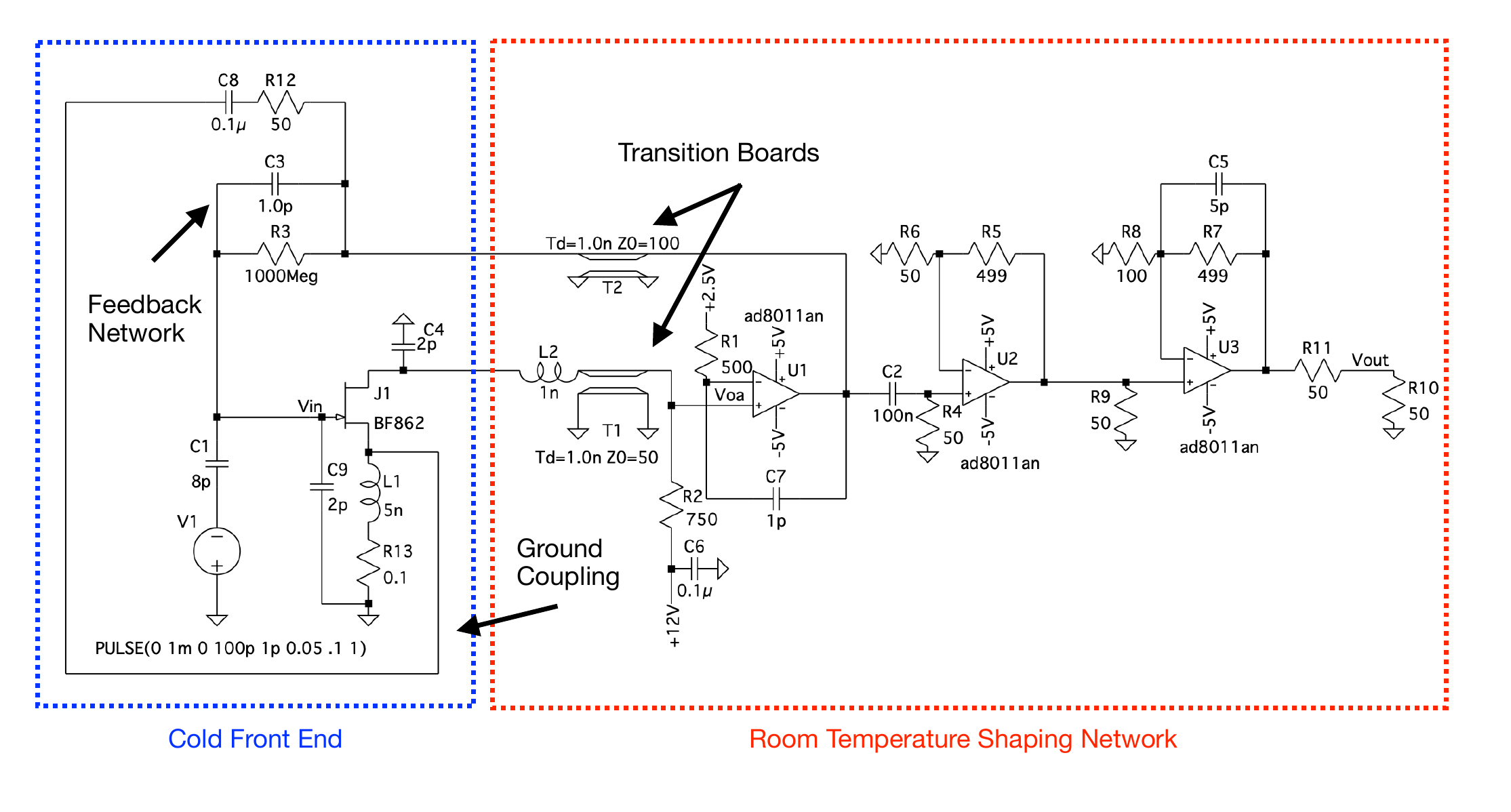}
    \caption{Overview of the amplification stage in the Nab experiment as modeled using the SPICE simulation. The FET has an internal capacitance of 8.3 pF between the source and gate for the frequency range of expected detector signals. The ground coupling denotes an imperfect connection to ground through resistive and inductive loads, as well as feedback effects due to shared grounds. Parasitic parameters were tuned to reproduce experimentally observed responses.}
    \label{fig:spice_circuit}
\end{figure*}

The charges induced on individual electrodes are read out in a standard charge-sensitive configuration, meaning a negative feedback FET + integrating capacitor scheme with a number of shaping networks following the initial amplification. The FET capacitance is chosen such that it corresponds closely to the total pixel capacitance for a fully depleted detector. As discussed above, the anticipated total capacitance is on the order of 10 pF after accounting for additional parasitic capacitance before reaching the FET amplification stage. 

\begin{figure}[ht]
    \centering
    \includegraphics[width=0.48\textwidth]{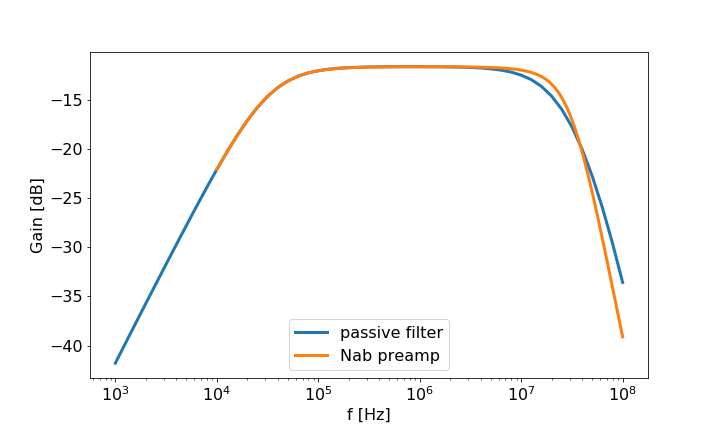}
    \caption{Frequency response of the preamplifier circuit from the SPICE simulation (orange), and the analytical $CR$-$(RC)^2$ approximation (blue). Both filters contain the same corner frequencies, but the Nab system's behaviour at the time scale of charge collection ($t_{\rm coll} \sim 50$ ns) is designed to have a steeper roll-off.} 
    \label{fig:passband}
\end{figure}

A SPICE model of the Nab detector preamplifier was constructed based on the components used in their design as-built. The circuit schematic is shown in Fiq.~\ref{fig:spice_circuit}. The first stage of the preamplifier circuit is composed of a FET amplifier in the common source configuration. The FET used in the circuit is a BF862 n-channel junction FET. It has an internal capacitance of 8.3 pF between the source and gate. It was chosen in part because this capacitance is comparable to the capacitance of the detector when depleted. This FET and the feedback resistor (R3) and capacitor (C3) are housed near the detector and kept at cryogenic temperatures. The first stage is connected to the subsequent stages through so-called "transition boards" represented in the circuit as transmission lines with nominal impedances of 50 $\Omega$ in the signal path (T1) and 100 $\Omega$ in the feedback path (T2) with 1 ns of delay for both. The "transition boards" create a thermal break between the cold FET and the rest of the circuit which is kept at room temperature. Immediately following the transition board in the signal path is an operational amplifier (U1) used as a unity gain buffer with low output impedance to drive the shaping network. The shaping network consists of a high pass filter with an RC time constant of 5 $\mu$s followed by an active low pass filter with an RC time constant of 7 ns. The frequency response of the preamplifier circuit was simulated in SPICE and the results are shown in Fig. \ref{fig:passband}. The 3 dB corner frequencies are found at 31 kHz and 20 MHz.

\begin{figure}[ht]
    \centering
    \includegraphics[width=0.48\textwidth]{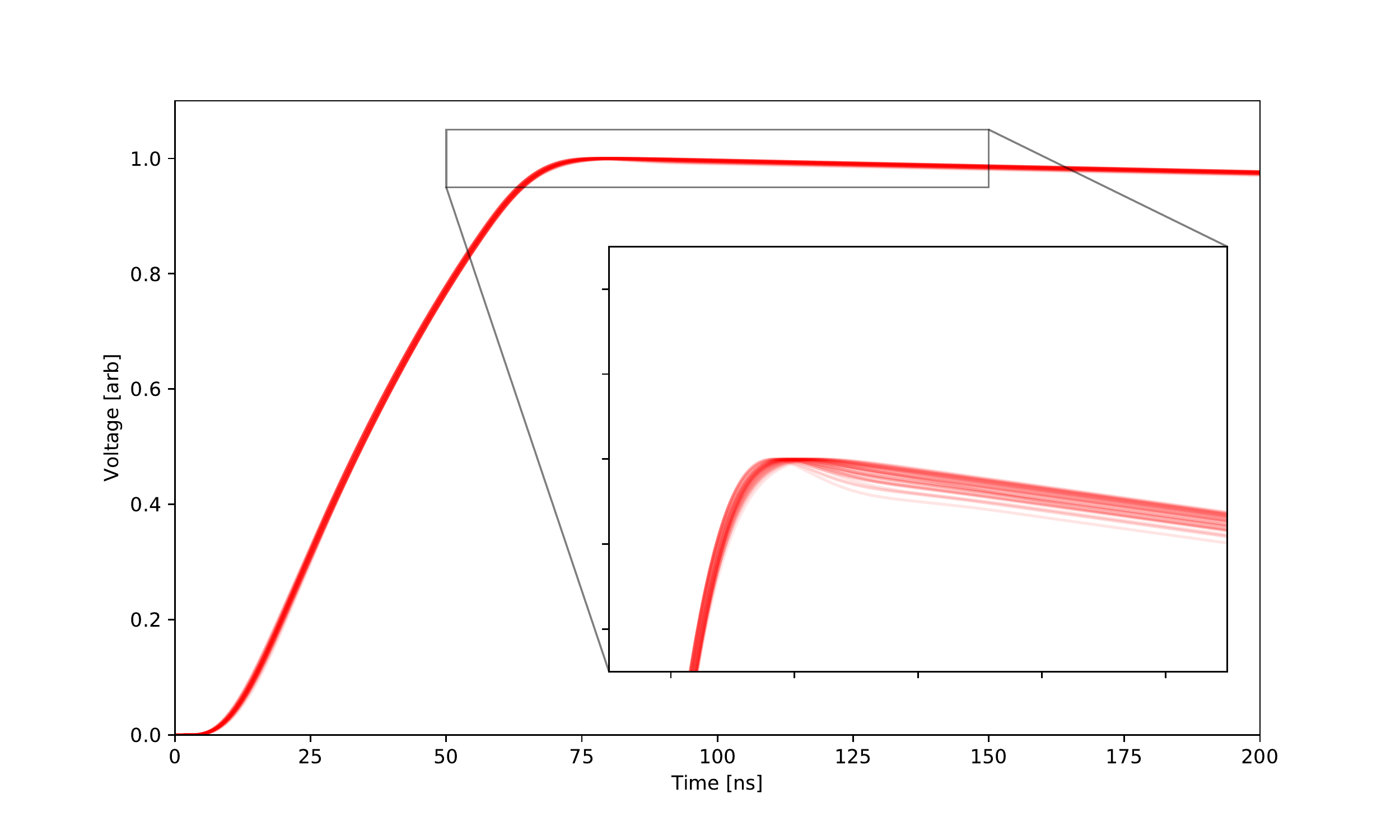}
    \caption{Pulse shapes created by taking simulated preamp response from SPICE varying the transmission line impedance and capacitance convoluted with the current pulse from drifting charge as simulated in SSD. The top and bottom figure are the same but shown at different $y$-axis scales.}
    \label{fig:spice_pulse}
\end{figure}

In order to take into account potential distortions, we add a number of parasitic elements to the circuit. One such modification is the coupling between the ground at the FET source and at the feedback resistor and capacitor, represented by a connection between R12 and the FET source. Additionally, a small resistance (R13), inductance (L1) and a parasitic capacitance (C9) were introduced to model a realistic ground connection. Finally, a capacitance (C4) and inductance (L2) were added in front of the transmission line. The capacitance and inductance were allowed to vary up the point that strong oscillations occurred. The impulse response for each set of values was generated in SPICE and convolved with a detector signal generated in SSD to create simulated pulses. The resulting pulses are shown in Fig. \ref{fig:spice_pulse}. A comparison with data and bench testing of the electronics boards will determine the final simulation parameters of these parasitic elements.

\subsection{Precision pulse shapes}
\label{sec:precision_pulse_shapes}
\begin{figure}[ht]
    \centering
    \includegraphics[width=0.48\textwidth]{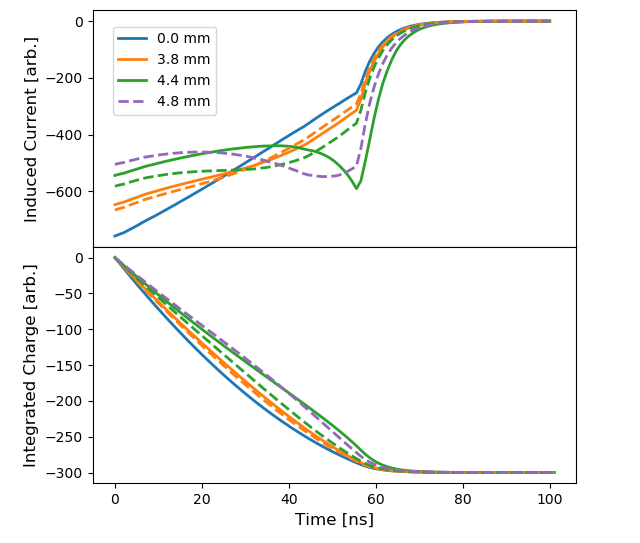}
    \caption{Induced current and integrated charge on the central pixel from the transit of the quasiparticles within the silicon detector at 4 different proton impact radii using the electric fields of Fig. \ref{fig:Efield_z}. Solid line represents the impact position moving towards edge of the hexagon, and the dashed line represents the impact position moving towards corner of the hexagon (See Fig. \ref{fig:EdgeWP}).}
    \label{fig:SSDQI}
\end{figure}

\begin{figure}[ht]
    \centering
    \includegraphics[width=0.48\textwidth]{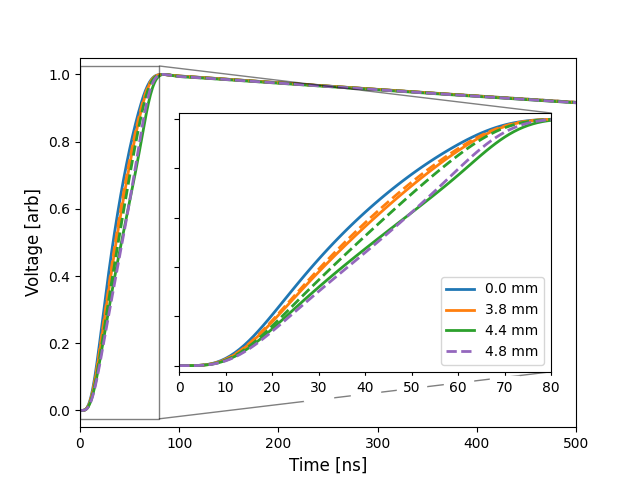}
    \caption{Pulse shape created by convolving the induced current of protons impacting 4 different radii of the pixel with the preamp impulse current response from SPICE. Solid line represents the impact position moving towards edge of the hexagon, and the dashed line represents the impact position moving towards corner of the hexagon.}
    \label{fig:SSDI_SPICE}
\end{figure}

Combining the Monte Carlo simulation of interactions between the impinging proton and pure silicon, propagation of created e-h pairs in the electric fields of a hexagonal pixel detector, explicit charge collection losses, calculation of the induced current on the central pixel contact, and finally convolution of the induced current with the impulse current response of the preamplifier shaping electronics, we now present the resulting simulation of the precision pulse shape for the 30 keV proton interacting with the Nab Detector System. As further discussed in Sec. \ref{sec:sensitivity_study}, the details of the simulation model impacts the simulated pulse shape. However, to highlight the impact of high precision pulse shape simulation, nominally expected parameters have been used to assess the effect of pulse shape on the systematic bias in the proton impact timing extraction. For this end, the impurity concentration of the silicon bulk was set to $n_0 = 4\times10^{10}\textrm{ cm}^{-3}$, resulting in the depletion voltage of $V_d \approx -120$ V. These settings correspond to the discussion in Sec. \ref{sec:electric_field} and the electric fields shown in Fig. \ref{fig:Efield_z}. The silicon detector temperature is set to 110 K and we simulate protons impacting at 18 different locations within a pixel moving both towards the flat edge of the hexagon as well as one of the corners. These locations correspond to 11 different radii, and induced current and the integrated charge on the pixel contact for 4 different impact radii is shown in Fig. \ref{fig:SSDQI}. Close to the edge of the hexagon, strong deviations occur due to weighting potential effects as discussed in Secs. \ref{sec:pixel_weighting_potential} and \ref{sec:pixel_isolation}. After convolving the induced current with the impulse current response of the preamplifier electronics, we obtain the resulting pulse shapes shown in Fig. \ref{fig:SSDI_SPICE}. 

As described in the introduction, an accurate determination of the proton time-of-flight lies at the heart of the Nab experiment. The variation in pulse shapes in Fig. \ref{fig:SSDI_SPICE}, however, would lead to a systematic timing bias of several nanoseconds when using standard techniques such as leading edge or constant fraction triggering. Instead, one might opt to use pulse shape information. Using the results from the detailed simulations presented here, we may quantify an introduced timing bias when instead using only a single pulse shape as a fitting template for each hit location inside the location. We determine the timing bias by overlaying Gaussian noise with a signal-to-noise ratio of 36, on each of the simulated pulse shapes and performing a 2-parameter curve fit (amplitude and $t_0$) using the pulse shape with an impact position at the center of the pixel.

\begin{figure}[ht]
    \centering
    \includegraphics[width=0.48\textwidth]{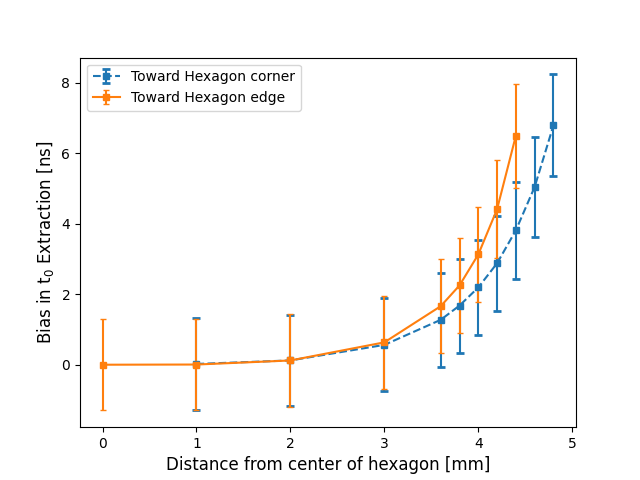}
    \caption{Mean timing bias arising from utilizing only the pulse shape of central impact position in extraction of the impact timing. The error bars show the width of the distribution rather than the uncertainty on the average shift, and as such are highly correlated.}
    \label{fig:t0bias}
\end{figure}

\begin{figure}[ht]
    \centering
    \includegraphics[width=0.48\textwidth]{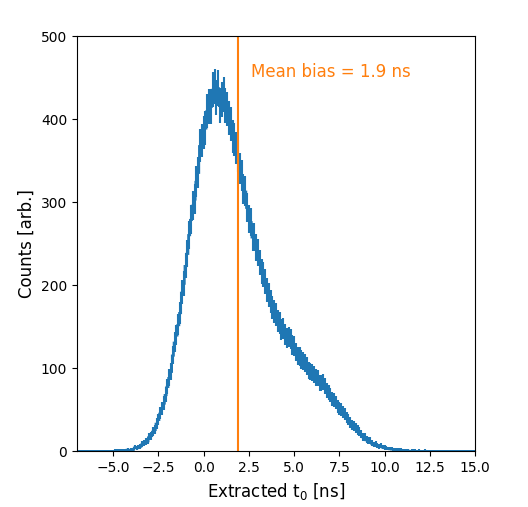}
    \caption{Cumulative histogram of extracted $t_0$ bias on an individual waveform when fitting every hit event with a central hit pulse shape. Individual histograms at every radial position - see Fig. \ref{fig:t0bias} - were weighted with their geometrical surface contribution to form the cumulative distribution shown here. Using this procedure, a mean bias of 1.9 ns is obtained.}
    \label{fig:t0biasDistribution}
\end{figure}

Results for a homogeneous electric field are shown in Fig. \ref{fig:t0bias}. Within the first 3 mm of the pixel center, the pulse shape does not vary significantly. As a consequence, the mean timing bias from utilizing "incorrect" pulse shapes does not result in any substantial mean timing bias. Near the outer perimeter of the pixel, however, effects due to the weighting potential (see Sec. \ref{sec:pixel_weighting_potential}) change the pulse shape significantly and one obtains a mean bias between 6.5 and 6.8 ns depending on the direction. As shown in Fig. \ref{fig:t0biasDistribution}, weighting by the surface area ratio of the different impact positions, the mean timing bias across the entire pixel is 1.9 ns. While this may be corrected for \textit{a posteriori}, the scale of the latter significantly exceeds the required precision for the Nab experiment as discussed in Sec. \ref{sec:timing_requirement}. Depending on the signal-to-noise ratio of the experimental data, however, pulse shape fitting may discriminate between the different hit positions on an event-by-event basis. Should this not be feasible, an average correction may be applied on a pixel-by-pixel basis after appropriate determinations of the model parameters as mentioned above.

\begin{figure}[ht]
    \centering
    \includegraphics[width=0.48\textwidth]{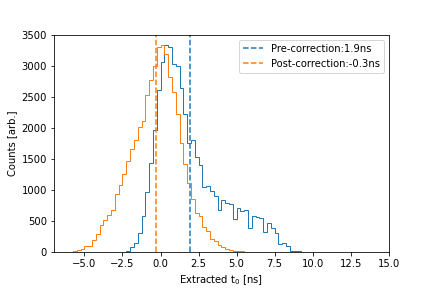}
    \caption{Cumulative histogram of extracted $t_0$ bias on an individual waveform before and after corrections from pulse shape discrimination. Similarly to Figure \ref{fig:t0biasDistribution}, the hit positions were weighted geometrically. At a signal to noise ratio of 100:1 (where SNR is taken as the square of the ratio of signal and noise amplitudes, respectively) a reduction in the mean bias from $1.9$ ns to $-0.3$ ns was achieved.}
    \label{fig:t0BiasPostCorrection}
\end{figure}

\begin{figure}[ht]
    \centering
    \includegraphics[width=0.48\textwidth]{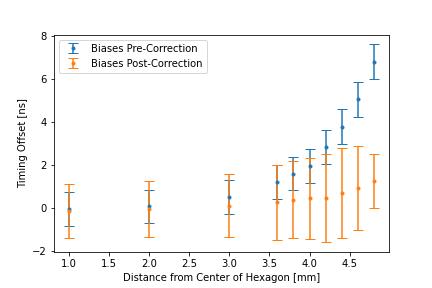}
    \caption{Comparison of the timing extraction on a per-waveform basis before and after corrections from pulse shape discrimination. Note that this test was performed on the detector positions along the axis towards the hexagonal pixel corner, or corresponding to the blue dataset in Figure \ref{fig:t0bias}. The increase in timing uncertainty for each hit position arises from under/over-correcting the timing bias when the hit position is misidentified.}
    \label{fig:t0BiasPerShapePostCorrection}
\end{figure}

If it were possible to consistently identify which pulse shape a detected signal most closely matched, then the timing bias could be corrected using the known biases from Figure \ref{fig:t0bias}. One effective way to perform this identification is through fitting each measured waveform to a series of template functions representative of the expected waveform shape at different hit positions in the pixel as seen in Figure \ref{fig:SSDI_SPICE}. Through comparison of the $\chi^2$ values returned from these fits, the most closely matching template shape can be identified and the bias corrected for \cite{Mathews2022}. Figure \ref{fig:t0BiasPostCorrection} shows the results of this method on the overall timing bias and Figure \ref{fig:t0BiasPerShapePostCorrection} shows the per-waveform bias and uncertainty before and after corrections. The performance of this method varies highly with the signal to noise ratio of the measured data. Below a ratio of around 36:1 this method shows no improvements over the default performance as the uncertainty in the identification of the waveform shape is simply too large making the cleanliness of the measured data critically important for the application of a method such as this.

\section{Probing model parameters}
\label{sec:sensitivity_study}

The previous sections treated several aspects of the pulse shape formation using models either extracted from the literature or constructed in this work. While many of the parameters in these models can be well-motivated and have been verified experimentally, the behaviour of the Nab detectors will depend on several macroscopic observables that are specific to their production. 

First among these is the bulk impurity density profile and potential radial gradients. As discussed in Sec. \ref{subsection: DopingProfile}, the large diameter of the Nab detectors makes substantial radial variation likely. We investigate the effects of radial gradients on the signal rise time and discuss potential measurement schemes of extracting the local impurity density. Additionally, we show how capacitance-voltage curves can be instructive in determining large-scale impurity density variations as a way of obtaining complementary information.

Charge collection losses in the entrance window were discussed in Sec. \ref{sec:charge_trapping_dl} and determine the fraction of sub-threshold proton events. An energy dependence in the latter - as protons arrive with kinetic energies between 30 and 30.8 keV - causes a disturbance in the reconstructed spectrum akin to a false $a_{\beta\nu}$. As such, an accurate reconstruction of the charge collection is required, and we show how the energy spectrum of backscattered protons can provide useful information.

Finally, the pixel isolation structure and its simulation in Sec. \ref{sec:pixel_isolation} determines the fraction and occurrence of physical charge sharing, where carriers get collected on different electrodes on either side of the pixel isolation structure. The latter is similarly important for sub-threshold proton detection efficiency effects.

\subsection{Bulk doping profile}

\subsubsection{Rise time distributions}
\label{sec:RiseTimeDistributions}
As discussed in Section \ref{sec:Measurement_principle} the proton time of flight ($t_p$) must be determined precisely in the Nab experiment using digitized pulse shapes. The signal rise time for protons is determined primarily by the time it takes for the electron quasiparticles to travel from the front of the detector to the back electrode, which depends on the local impurity density. Following the electric field calculations of Sec. \ref{sec:electric_field} for different radial gradients, we characterize the radial dependence of signal rise times using \texttt{SolidStateDetectors.jl} for quasiparticle transport.

As a starting point we take individually simulated events from Geant4 as discussed in Section \ref{sec:SRIM_Geant4_comp}. Using the same 7 pixel geometry and impurity density profiles described in Sections \ref{sec:pixel_weighting_potential} and \ref{sec:electric_field}, we then simulated pulse shapes at a range of radial positions within a single pixel. We define a rise time, $t_{10-90}$, as the time it takes a signal to reach 10\% and 90\% of its maximal value, respectively, and analogously for 0\% to 10\%. The former can be similarly extracted from experimental data whereas the limited signal to noise ratio renders the latter inaccessible.

\begin{figure}[ht]
    \centering
        \includegraphics[width=0.5\textwidth, left]{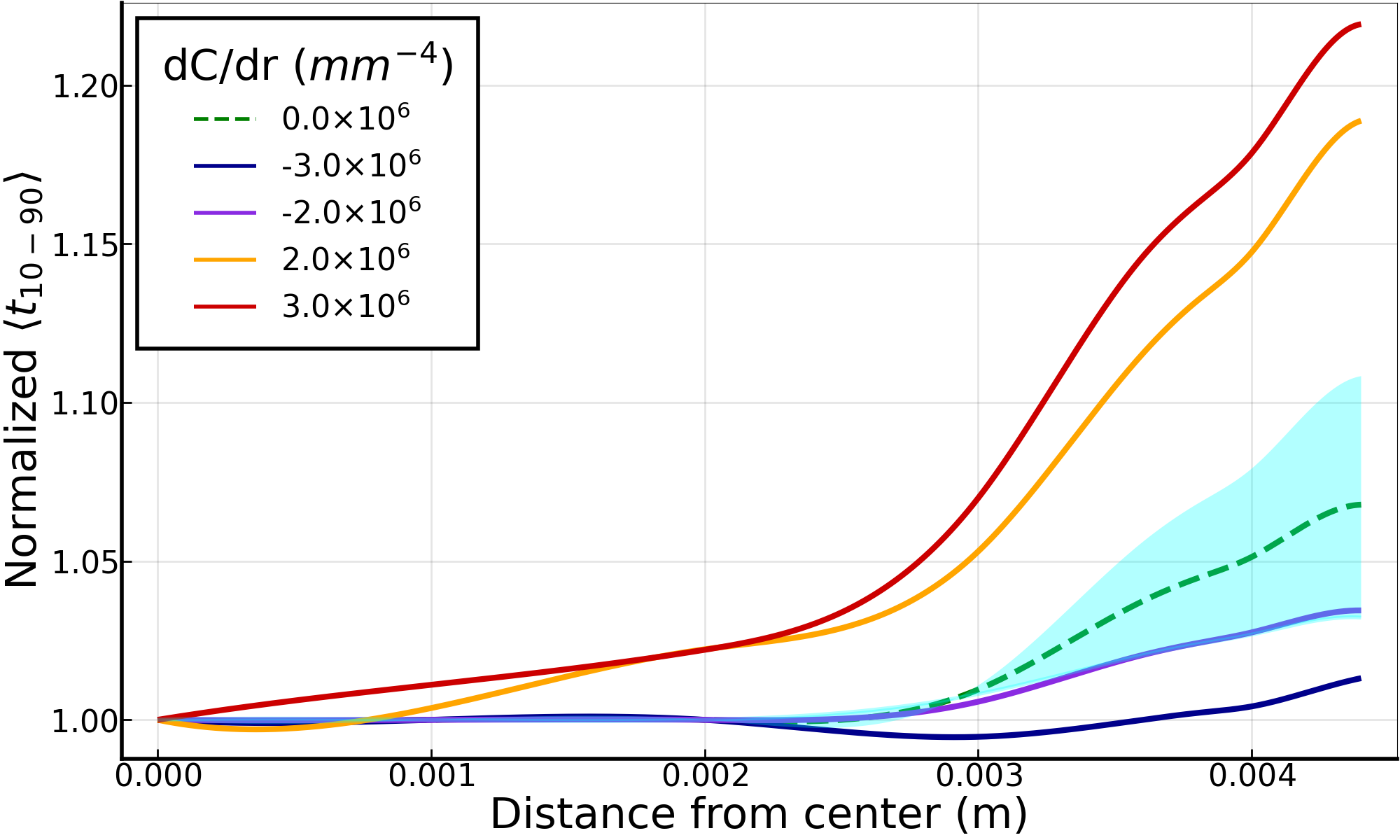}
    \caption{Average 10-90\% rise times normalized to center of pixel rise time for $4\times10^{10}\textrm{ cm}^{-3}$ detector at various impurity radial gradients. Results for 150V are represented by solid lines and the range of 180V results are shown by the shaded cyan region. The dashed line is the zero gradient result.}
    \label{fig:risetimes_150v180}
\end{figure}

In Fig. \ref{fig:risetimes_150v180} we show the results for a single pixel with base impurity concentration $n_0 = 4\times10^{10}\textrm{ cm}^{-3}$ at 150 and 180V (i.e. $V >V_d \sim 120$ V). The detector needs to be significantly over-depleted so that the edge of the detector is still depleted for positive impurity concentration gradients. Near the pixel edge differences in impurity concentrations have a significant affect on rise times due to the combination of the radial dependence of the electric field slope and the weighting potential (see Figs. \ref{fig:EdgeWP} and \ref{fig:Efield_z} and Eq. (\ref{eq:I_SR})). For negative impurity density gradients, both effects largely cancel and the rise time stays constant within a few percent over the entire pixel surface. Positive gradients, on the other hand, have a large impact on rise times as both lower field strengths and later rise in weighting potential contribute to differences of up to 20\% in the signal rise time. Differences are obviously more pronounced for bias voltages close to the depletion voltage.

\begin{figure}
    \centering
    \includegraphics[width=0.5\textwidth, left]{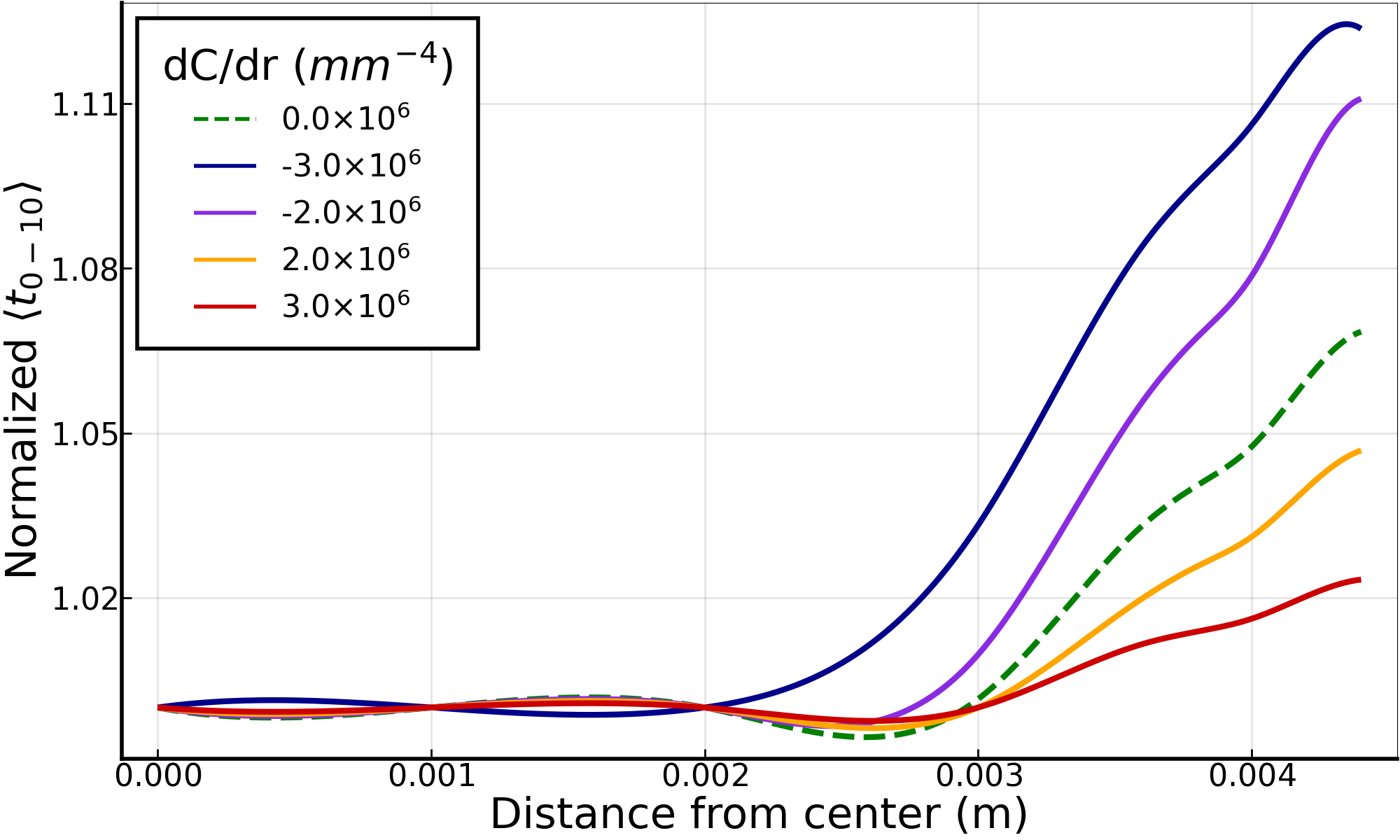}
    \caption{Average 0-10\% rise times normalized to center of pixel rise time for a detector with $n_0 = 4\times10^{10}\textrm{ cm}^{-3}$ at various impurity radial gradients. The dashed line is the zero gradient result.}
    \label{fig:risetimes_010}
\end{figure}

Similarly, Fig. \ref{fig:risetimes_010} shows the average normalized 0-10\% rise times for single pixel with $n_0 = 4\times10^{10}\textrm{ cm}^{-3}$ impurity density at 150 V. Within a 2 mm radius, no effect on the 0-10\% rise time can be observed, consistent with expectation. Unlike the behaviour discussed above, however, positive radial gradients leave the 0\% to 10\% rise time largely unaffected while negative gradients show substantial changes. The latter can be understood through field differences close to the front contact where higher impurity concentrations have larger field strengths (see Fig. \ref{fig:Efield_z}).  Note that the Nab experiment will utilize pixels with much larger radial offsets than those presented here, producing systematic variation in the pulse shape response which depends on the pixel "ring" in which the signal originates.  These effects are the subject of a separate experimental program based at the University of Manitoba \cite{Harrison2013}.

\begin{figure}
    \centering
    \includegraphics[width=0.5\textwidth, left]{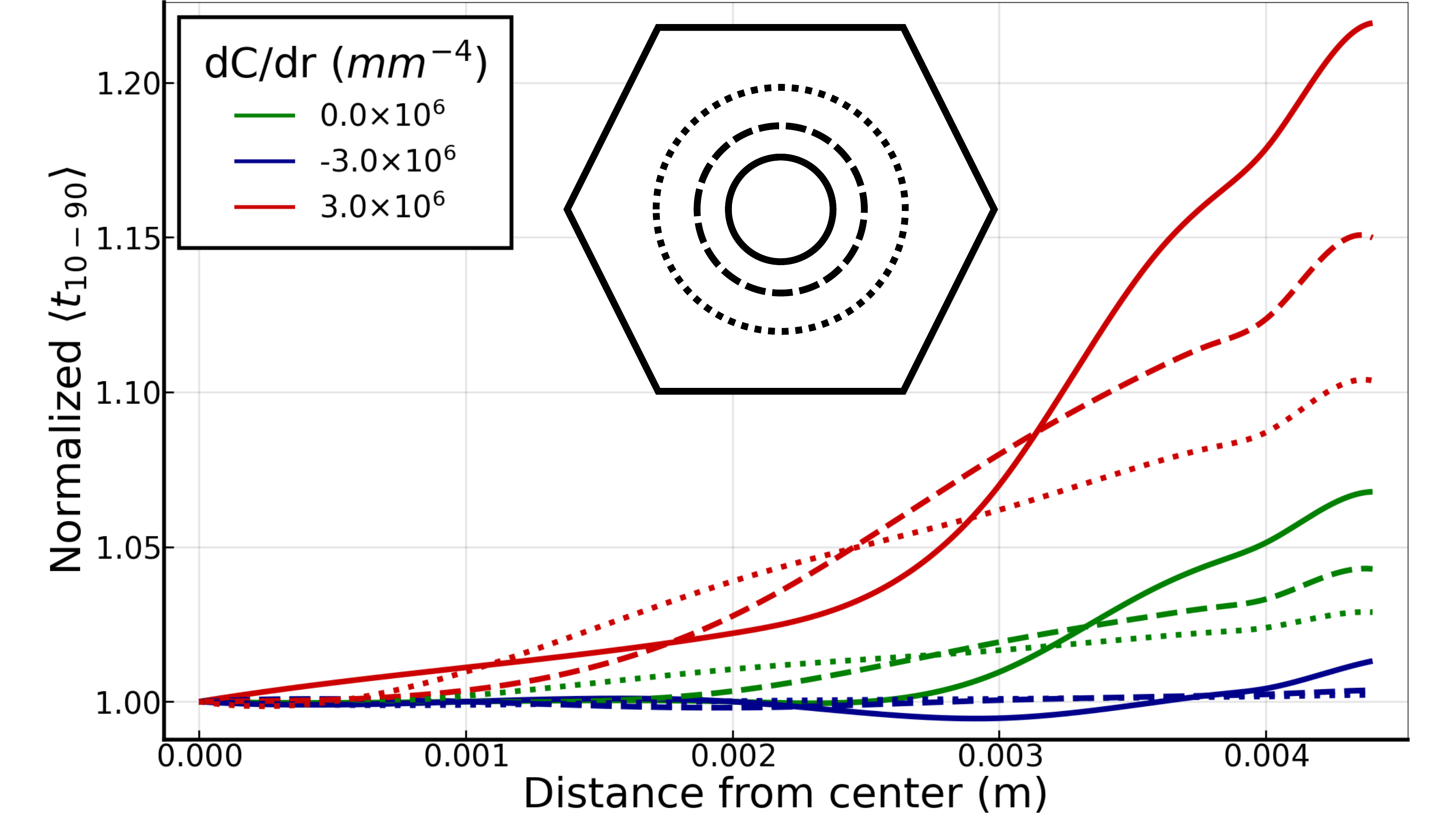}
    \caption{Average 10-90\% rise times normalized to center of pixel rise time for $4\times10^{10}\textrm{ cm}^{-3}$ detector at 150V various impurity radial gradients. Solid, dashed, and dotted lines are used to represent 1, 2, and 3mm beam sizes.}
    \label{fig:risetimes_150_radii}
\end{figure}

Radial impurity density gradients might be determmined using a collimated proton beam that is swept radially across the detector and analysing the 10-90\% rise times. In Fig. \ref{fig:risetimes_150_radii} we show the results of simulating such an experiment with a beam of  radius 1,2, and 3 mm. While a smaller beam size is obviously more sensitive, even with a 3 mm beam radius differences of up to 10\% can be observed towards the pixel edge. As observed in Fig. \ref{fig:risetimes_150v180}, maximal sensitivity occurs for bias voltages just above depletion voltage. This points both towards running conditions substantially above depletion voltage for regular data taking and diagnostic studies near depletion.

We briefly comment on the possible existence of longitudinal gradients in the impurity density. From the manufacturer, impurity density gradients along the boule symmetry axis can range from vanishingly small to as large as about $3\times10^{10}$cm$^{-4}$. From the Poisson equation, such a linear longitudinal impurity density gradient creates a quadratic electric field which diverges from the standard result by at most ~10\%. This difference in electric field was numerically found to give differences in the average 10-90\% rise time of $<1\%$, i.e. substantially smaller than shifts from radial gradients and edge effects (up to 20\%), and we do not further consider its effects.

\subsubsection{$C$-$V$ curves}
Often, measurements of the capacitance-voltage curve of the entire detector at once are performed during the manufacturing phase and can be repeated afterwards by chaining all pixels together. Besides the plateau in a typical $C$-$V$ curve showing full depletion, the shape as it approaches the latter can be sensitive to bulk properties such as radial gradients. As discussed in Sec. \ref{subsection: DopingProfile}, radial gradients in the impurity density profile can be significant in geometries as large as those used in Nab. As a consequence, inner regions may be depleting faster or slower than outer regions depending on the sign of the gradient.

\begin{figure}[ht]
    \centering
    \includegraphics[width=0.48\textwidth]{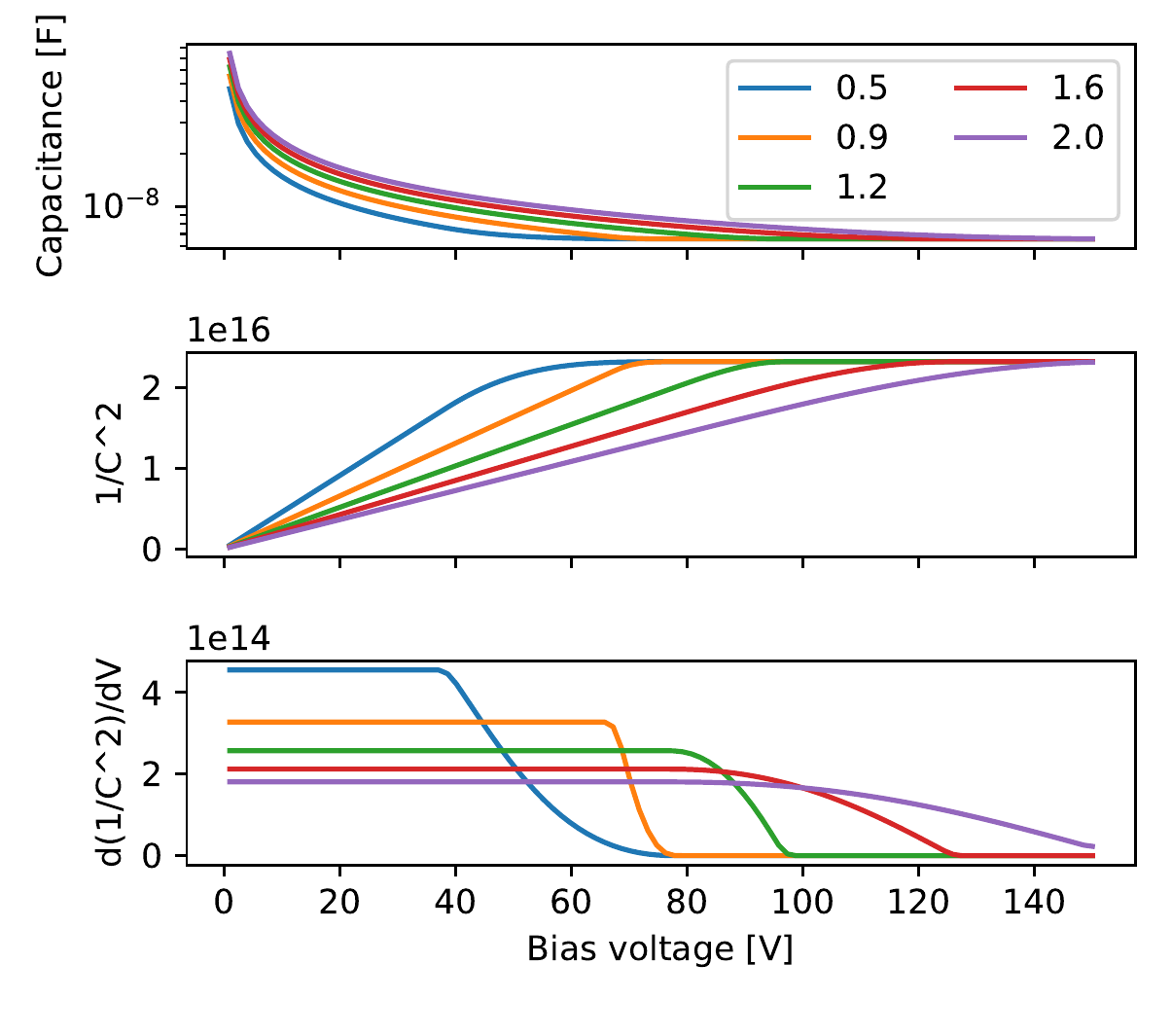}
    \caption{Comparison of $C$-$V$ curves with differing radial gradients in the impurity density profile using Eq. (\ref{eq:d_approximation}). The label denotes the relative change in impurity density at the outer edge compared to the center.}
    \label{fig:C_V_analytical}
\end{figure}

A simple analytical approximation to investigate potential effects can be constructed when restricting to purely radial gradients. In this case one may, in a first approximation, consider the total detector capacitance as a construction of concentric rings where each is a parallel plate capacitor with depletion thickness $d(r)$. The total detector capacitance is then simply
\begin{equation}
    C_\mathrm{det} = \int_0^R dr 2\pi r \frac{\varepsilon}{d(r)}
\end{equation}
where $d(r)$ is approximated as
\begin{equation}
    d(r) \approx \sqrt{\frac{2\varepsilon_r\varepsilon_0V}{qN(r)}}
    \label{eq:d_approximation}
\end{equation}
when less than the detector thickness, $t$, and $t$ otherwise. These simple approximations neglect, e.g., transverse fields but can shed some light on anticipated changes in $C$-$V$ curves. Taking the impurity density to depend linearly on the radius, i.e. $N(r) = N_0+gr$, results in $C$-$V$ curves shown in Fig. \ref{fig:C_V_analytical} for a number of different values of $g$.

Besides the pure $C$-$V$ result, often-quoted variables such as $1/C^2 \propto d^2 \propto V/N$ and $d(1/C^2)/dV \propto 1/N$ are also shown in the figure. In the absence of any radial gradients, the former shows a discrete kink indicating full depletion. As anticipated, the presence of radial gradients serve to smooth out this kink. The sign of the gradient, however, changes the second derivative as can be observed in the bottom panel of Fig. \ref{fig:C_V_analytical}. This is a feature that persists in the more detailed simulation as discussed below, and can be a useful diagnostic tool for the bulk behaviour.

\subsection{Charge collection in the entrance window}
\begin{figure}[ht]
    \centering
    \includegraphics[width=0.48\textwidth]{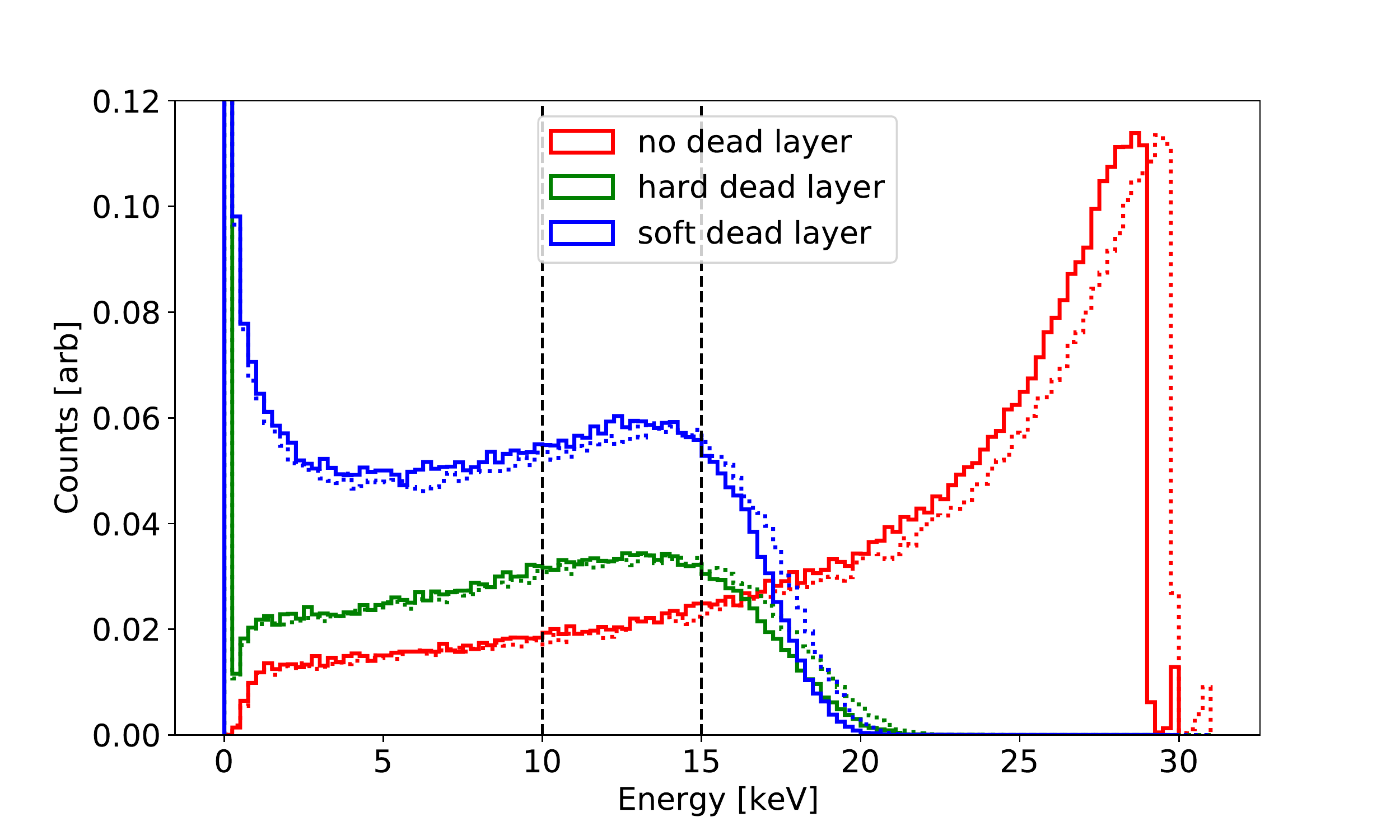}
    \caption{Spectra of deposited energy for 30 (solid) and 30.8 keV (dots) protons scattering out of silicon using different dead layer models. The black vertical lines represent the range of detector thresholds.}
    \label{fig:CCE_comparison}
\end{figure}

Following the discussion of Sec. \ref{sec:charge_trapping_dl}, we perform a comparison of different dead layer models using the Geant4 simulation as described in Sec. \ref{sec:SRIM_Geant4_comp}. We study both phenomenological models discussed in Sec. \ref{sec:charge_collection_efficiency}, i.e. the `hard' and `soft' models of Eqs. (\ref{eq:cce_hard_dead}) and (\ref{eq:cce_soft_dead}), respectively. In the Nab experiment, protons are accelerated by a 30 kV potential before striking the silicon detector (see Fig. \ref{fig:nab_setup_overview}). As such, protons emerging from neutron $\beta$ decay range between 30.0 keV and 30.8 keV by the time they reach the upper detector. The `hard' dead layer was implemented with a depth of 70 nm and similarly the `soft' dead layer was implemented with a characteristic length $l$ of 70 nm. The spectra for \textit{detected} energy of 30.0 and 30.8 keV protons that scatter out of the silicon with each dead layer model applied are shown in Fig. \ref{fig:CCE_comparison}. Substantial differences are observed in the rate and shape for each charge collection model, meaning the study of the backscattered proton spectrum can be a valuable tool for distinguishing between them.

A simple diagnostic tool can be simply the number of over-threshold backscattered events. For this purpose, we define thresholds of 10 and 15 keV assuming a detector resolution that is Gaussian with a width of 2 keV. The fraction of backscattered events that cross the threshold out of the total number of events is shown in Table \ref{tab:CCE_threshold} for both 30 and 30.8 keV protons. As anticipated from Fig. \ref{fig:CCE_comparison}, the expected difference between fractional missed backscatter rates for hard and soft charge collection models is substantial, such that discrimination between the two models using the same characteristic difference should be straightforward. 


\begin{table}[ht]
\begin{ruledtabular}
\begin{tabular}{ p{1cm}p{2cm}p{2cm}p{2cm}  }
 & \multicolumn{1}{l}{Energy} &\multicolumn{2}{c}{Threshold}\\
 \hline
 & & \multicolumn{1}{c}{10 keV} & \multicolumn{1}{c}{15 keV}\\
 \hline
 \multirow{2}{*}{soft} &  30.0 keV & 13.33(5)e-4 & 4.52(3)e-4\\

  & 30.8 keV & 13.72(5)e-4 & 5.24(3)e-4\\
\multirow{2}{*}{hard} &  30.0 keV & 8.10(4)e-4 & 3.02(2)e-4\\
  & 30.8 keV & 8.49(4)e-4 & 3.53(2)e-4
\end{tabular}
\end{ruledtabular}
\caption{Fraction of backscattered proton events above threshold out of total number of events for different thresholds, energies and dead layer models.}
\label{tab:CCE_threshold}
\end{table}

\subsection{Charge sharing}
\label{sec:chargeSharing}

Section \ref{sec:pixel_isolation} discussed the local electric field environment near pixel boundaries in great detail. Here, we will use those results to investigate the effects of physical charge sharing. The latter occurs when energy is deposited close to a pixel boundary and freed charge carriers may be due to a combination of drift and diffusion get collected on either side of the pixel boundary. 

\begin{figure}[ht]
    \centering
    \includegraphics[width=0.48\textwidth]{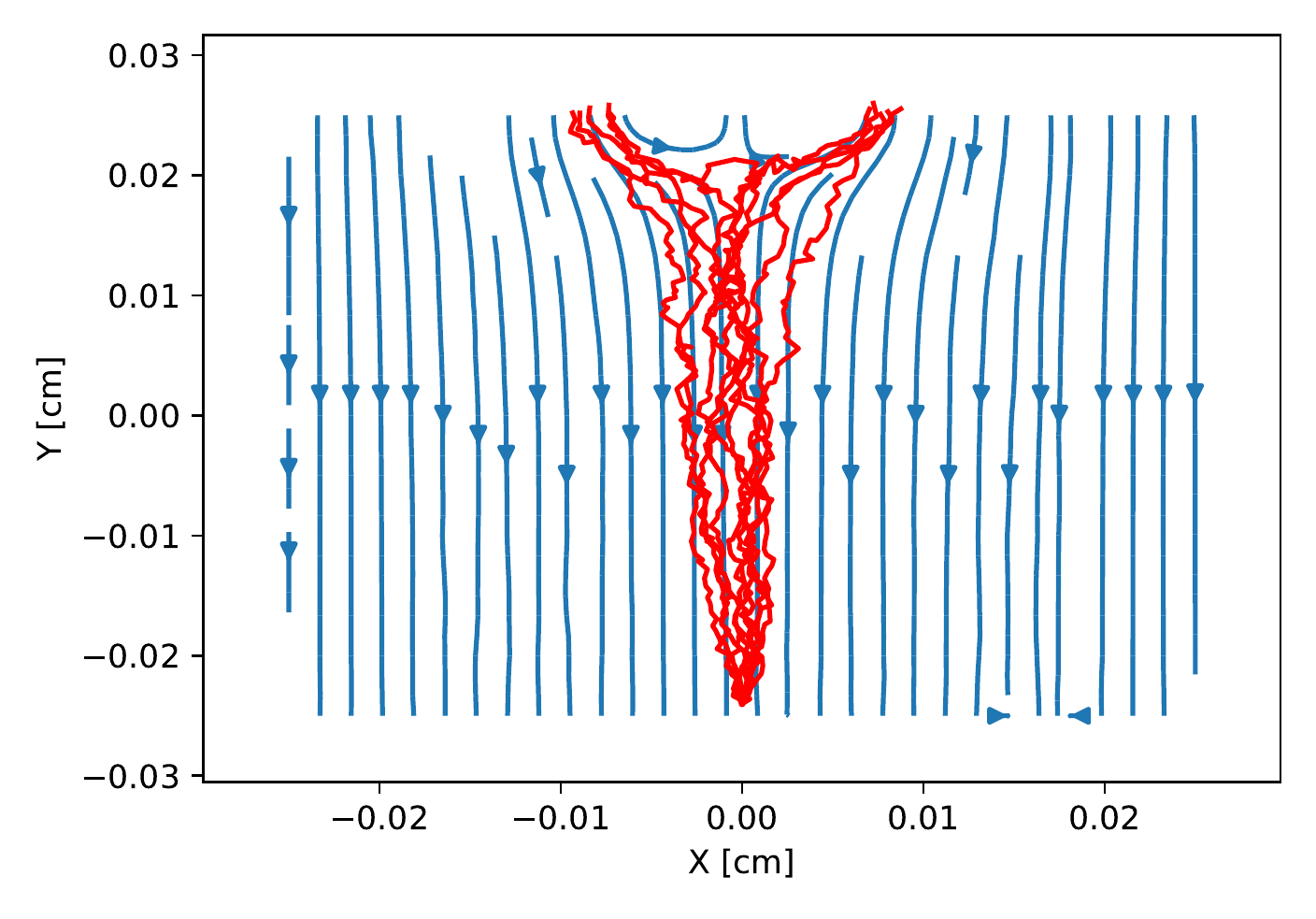}
    \caption{Example of explicit free carrier drift including diffusion close to the pixel isolation structure.}
    \label{fig:particle_drift_pspray}
\end{figure}

An example of charge carrier motion in this configuration is shown in Fig. \ref{fig:particle_drift_pspray}, where a cloud of 10 electrons are released at the geometrical center of a pixel boundary and are free to drift and diffuse. As discussed above, the p-spray and p-stop configurations serve to repel charges from the pixel boundary so that they are collected in either electrode. In this region, however, the electric field magnitude drops significantly, so that their motion is dominated by diffusion.

In order to quantify the effects of charge sharing, we define a charge asymmetry according to
\begin{equation}
    A = \frac{Q_R-Q_L}{Q_L+Q_R}
    \label{eq:charge_asym}
\end{equation}
where $Q_{L(R)}$ is the total collected charge on the contact left (right) of the boundary. By varying the initial position of charge carriers relative to the boundary, we may map the behaviour of $A$. The relevant scale in this problem is the relative distance traversed by diffusion and drift, i.e. $\mathcal{S} \propto \sqrt{Dt}/\langle t \rangle \propto \sqrt{T/\langle E \rangle}$ with temperature $T$ and average electric field $\langle E \rangle$, as losses are negligible.

\begin{figure}[ht]
    \centering
    \includegraphics[width=0.48\textwidth]{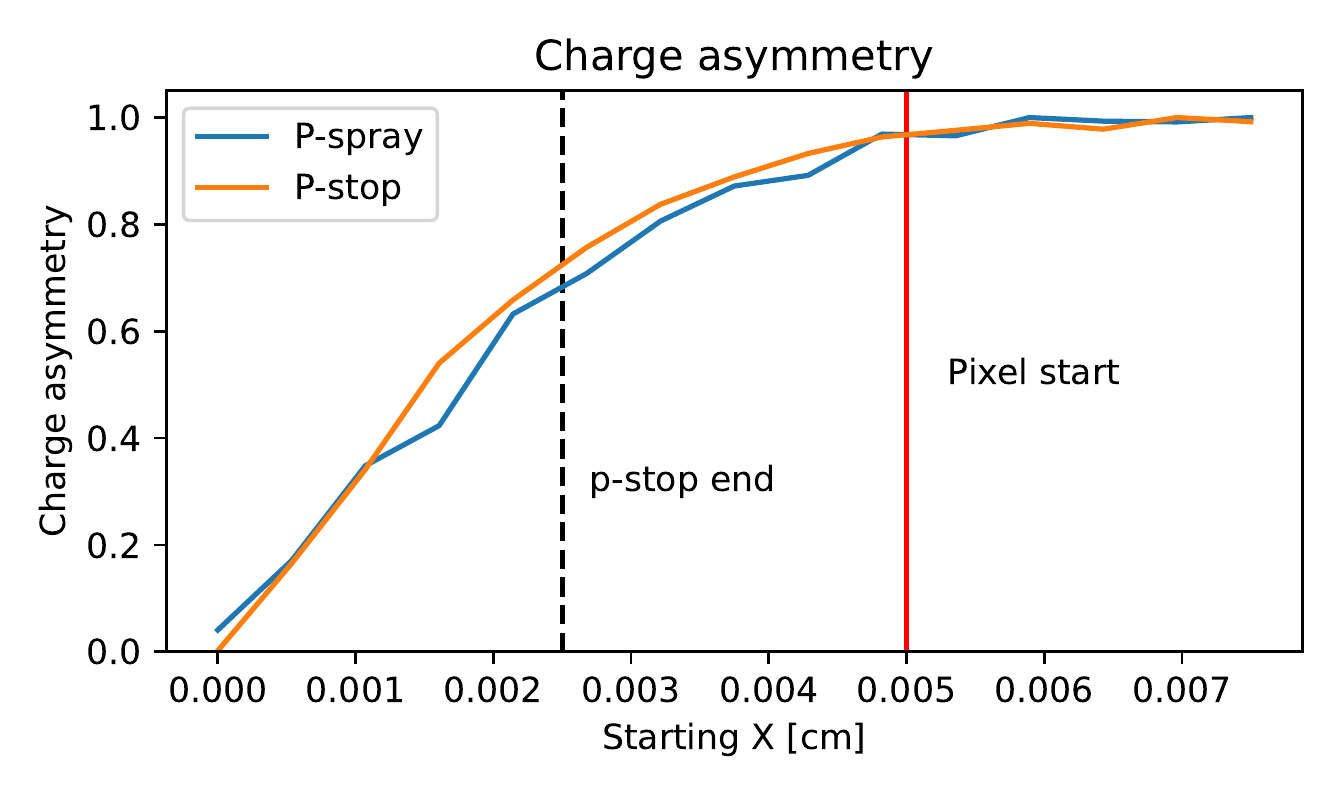}
    \caption{Charge asymmetry, defined in Eq. (\ref{eq:charge_asym}), when varying the initial position of the energy deposition relative to the pixel isolation structure for both p-stop and p-spray.}
    \label{fig:charge_asymmetry_isolation}
\end{figure}

Fig. \ref{fig:charge_asymmetry_isolation} shows $A$ as a function of distance from the pixel boundary center for both p-stop and p-spray configurations, with $T = 150$ K and biased at twice the depletion voltage. Within statistical uncertainty, there are no systematic differences between p-stop and p-spray. Both configurations reach full asymmetry (i.e. complete collection in just one contact) close to the physical boundary. 

\begin{figure}[ht]
    \centering
    \includegraphics[width=0.48\textwidth]{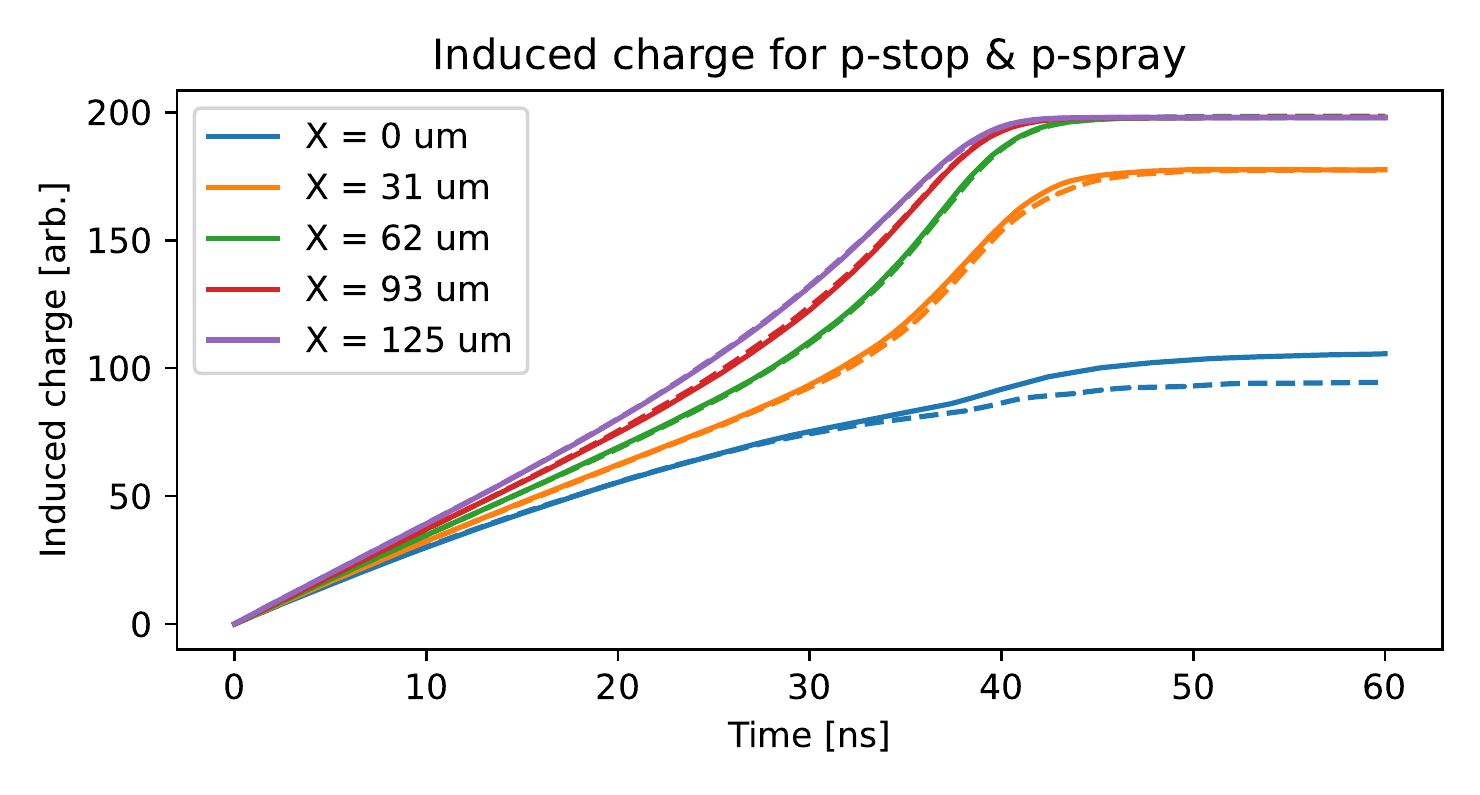}
    \caption{Average induced charge versus time in a contact for events close to the pixel boundary using detailed electric and weighting field simulations for p-stop (solid) and p-spray (dashed). Charge sharing is explicitly taken into account resulting in lowered amplitude pulses in agreement with Fig. \ref{fig:charge_asymmetry_isolation}. Strong deviations in the pulse shape for fully collected events arise from near-perpendicular weighting fields on the physical boundary, as shown in Fig. \ref{fig:overview_pspray_COMSOL}.}
    \label{fig:induced_charge_pixel_boundary}
\end{figure}

The induced charge as a function of time is shown in Fig. \ref{fig:induced_charge_pixel_boundary} for different starting positions relative to the pixel boundary center. In the center of the inter-pixel gap, charge collection at either electrode is halved, resulting a pulse shape with half amplitude. Slightly off-center, however, interesting pulse shapes emerge as the shape of the weighting field surrounding the pixel insulation (Fig. \ref{fig:pixel_isolation}) becomes important. Moving along a straight line from the front face, the weighting field become progressively more perpendicular to the electric field so that it's possible for a moving quasiparticle to induce zero net charge (see Eq. \ref{eq:I_Gunns}). As the quasiparticle approaches the p-stop or p-spray region, the electric field lines change direction to the closest contact in a small region and diffusion determines the fraction of charge collected on either pixel. The last part of the transport results in a swift increase of induced charge, similar to the edge effects discussed in Sec. \ref{sec:pixel_weighting_potential}. 

\section{Conclusion}
\label{sec:conclusion}

An accurate timing reconstruction of the proton time-of-flight in the Nab experiment at the nanosecond level is required to measure the beta-antineutrino angular correlation at the 0.1\% level. We have focused here on how detector effects, through a variety of different mechanisms, produce differences in the pulse shape and extracted start time. Using detailed electric and weighting field calculations, we have shown how the hexagonal pixel geometry gives rise to complex weighting potentials and derived an analytical expression that gives excellent agreement except for the sharp corners. Additionally, we have studied the effect of radial gradients in the bulk impurity density and how they cause a substantial rise time spread even within a single pixel. Finite element analysis was performed on a detailed simulation of inter-pixel isolation technologies using p-stop and p-spray, where one must use Gunn's theorem rather than the standard Shockley-Ramo approach to obtain correct results.

Using Monte Carlo methods, we presented detailed studies of collective effects in quasiparticle transport through thermal diffusion and plasma effects. Our approach for the latter is the first microscopic treatment of the effect for low energy protons and shows delays in the charge collection onset between 0.1 and 0.5 ns. Additionally, we have used Secondary Ion Mass Spectroscopy results to establish a detailed charge collection efficiency function inside the entrance window. As many experiments in low-energy particle and nuclear physics are sensitive to the details of this `dead layer' but typically consider only simplified models, our results show a potential avenue for a more detailed understanding.

Finally, we presented ways in which the remaining free parameters in the model description can be probed using auxiliary experiments. These are predominantly concerned with establishing the radial impurity variations throughout the large crystals used for the Nab experiment. We find that, using collimated beams with a diameter on the order of one to a few millimeters may be sufficient to establish the bulk behaviour within the required specifications for the Nab experiment.

In summary, we have provided an overview and in-depth treatment of precise pulse shape prediction for high purity silicon detectors used in nuclear and particle physics. The current work represents an improvement in the state of the art which might prove fruitful in the efforts to constrain Beyond the Standard Model physics using neutron and nuclear $\beta$ decay, and points to several areas of interest for further research.

\begin{acknowledgements}
This article was supported through the Department of Energy (DOE), Low Energy Physics [contracts DE-FG02-ER41042, DE-AC05-00OR22725, DE-FG02-03ER41258, DE-SC0008107 and DE-SC0014622] and National Science Foundation (NSF) [contracts PHY-1914133, PHY-2209590, PHY-2111363]. This research was sponsored by the U.S. Department of Energy, Office of Science, Office of Workforce Development for Teachers and Scientists (WDTS) Graduate Student Research (SCGSR) program, and the Science Undergraduate Laboratory Internship (SULI) program.
\end{acknowledgements}

\bibliography{library}

\end{document}